\documentclass{article}%
\usepackage{amssymb}
\usepackage{amsmath}
\usepackage{makeidx}
\usepackage{cite}
\usepackage{amsfonts}
\usepackage{graphicx}
\begin{document}

\author{Hartmut Wachter\thanks{E-Mail: Hartmut.Wachter@gmx.de}\\An der Schafscheuer 56\\D-91781 Wei\ss enburg, Federal Republic of Germany}
\title{Zero-Point Energy of a Scalar Field in $q$-De\-\-formed Euclidean Space}
\maketitle
\date{}

\begin{abstract}
We examine the energy of a scalar field in its ground state within
$q$-de\-formed Euclidean space. Specifically, we compute the total vacuum
energy of the entire $q$-de\-formed Euclidean space, originating from the
scalar field's ground-state energy. Our results show that, for a massless
scalar field, the total vacuum energy vanishes. In contrast, when evaluating
the average ground-state energy over finite, localized regions of the
$q$-de\-formed Euclidean space, we find that the vacuum energy density can
assume significant values.

\end{abstract}
\tableofcontents

\section{Introduction}

\paragraph{Zero-Point Energy - Historical and Theoretical Background}

When Max Planck originated quantum theory with his famous radiation law, the
concept of zero-point energy emerged shortly thereafter. In his
\textquotedblleft second quantum theory\textquotedblright, Planck described
black-body radiation using a harmonic oscillator of frequency $\omega$, which
absorbs radiant energy continuously but emits energy in discrete quanta. For
an oscillator in radiative equilibrium at temperature $T$, the average energy
contains a tem\-per\-a\-ture-in\-de\-pen\-dent contribution $\hbar\omega/2$:%
\begin{equation}
E(\omega)=\frac{\hbar\omega}{\operatorname{e}^{\hbar\omega/kT}-1}+\hbar
\omega/2.
\end{equation}
As $T\rightarrow0$, the mean energy of the quantum oscillator approaches
$\hbar\omega/2$. For this reason, the term $\hbar\omega/2$ is referred to as
the zero-point energy. In modern quantum theory, zero-point energy originates
from the Heisenberg uncertainty principle, representing the minimum energy
permitted for the quantum harmonic oscillator by this principle
\cite{milonni1994quantum}.

Difficulties arise when zero-point energy is considered in the context of free
quantum fields. A quantum field can be expressed as a superposition of modes
with all possible frequencies $\omega$, such that each mode corresponds to a
harmonic oscillator at every point in space. Consequently, each spatial point
holds an energy equal to the sum of the zero-point energies of all harmonic
oscillators located there. In the absence of an upper bound on the allowed
frequencies, this sum diverges.

In most areas of quantum physics, this divergence poses no problem: the total
zero-point energy merely constitutes a constant offset that can be subtracted
by redefining the energy reference point. However, in the presence of gravity,
this procedure is not valid. In general relativity, the stability of
space-time requires an upper limit to the oscillator frequencies.

Furthermore, as shown by Nernst, Lenz, and Pauli, even a finite zero-point
energy may produce significant gravitational effects, potentially leading to
an unrealistic large-scale structure of the universe \cite{Gomez:2017}.
Consequently, it remains unclear whether zero-point energy contributes to
gravitation at all \cite{rugh2000quantum}.

In the late 1990s, observations by Perlmutter, Riess, and collaborators
revealed that the expansion of the universe is accelerating
\cite{Riess_1998,Perlmutter_1999}. This phenomenon can be explained by the
cosmological constant $\Lambda$. When the universe is modeled as an ideal
fluid, a non-zero $\Lambda$ acts as a constant pressure driving accelerated
expansion \cite{Maggiore:2010}, and the value of $\Lambda$ determines the rate
of this expansion. The cosmological constant $\Lambda$ is related to the
vacuum energy density $\rho_{0}$ via%
\begin{equation}
\Lambda=\frac{8\pi G}{c^{\hspace{0.01in}2}}\rho_{0}.
\end{equation}
Astronomical observations give a value for the vacuum energy density of
approximately%
\begin{equation}
\rho_{0}\thickapprox10^{-9}~\text{J/m}^{3}.
\end{equation}
In quantum field theory, vacuum energy arises from the zero-point energy of
quantum fields. However, predictions for $\rho_{0}$ from quantum field theory
exceed the observed value by many orders of magnitude
\cite{Weinberg:1989,akhmedov2002vacuum}.

\paragraph{Issues in Modern Quantum Field Theory}

Since its inception, the development of quantum field theory (QFT) has
repeatedly encountered fundamental challenges, many of which could only be
addressed through the introduction of new theoretical concepts. The problem of
vacuum energy may be another instance where such innovations are required.

QFT arose from the synthesis of quantum mechanics and group theory. By
employing Poincar\'{e} symmetry, Dirac successfully predicted several key
properties of the electron. Despite these achievements, it soon became clear
that QFT corrections are finite only at the lowest perturbative order. The
early pioneers of quantum electrodynamics (QED) recognized that higher-order
corrections influence physical phenomena at energy scales beyond the theory's
original applicability.

The introduction of renormalization allowed QED to be formulated without
producing infinite or otherwise ill-defined expressions. Nevertheless,
predictions from renormalized QFT remain valid only within a restricted energy
range, and the theory itself cannot predict the empirical values substituted
for the divergent quantities. As long as renormalization remains the sole
procedure for handling divergences, new conceptual frameworks within QFT will
be necessary.

One of the central aims of QFT is to describe all fundamental interactions
within a unified theoretical framework. This has long motivated the
expectation that a comprehensive QFT of all interactions might eliminate
divergences entirely. A key strategy for unification has been the exploitation
of internal symmetries. Gauge symmetries, for example, have enabled the
unification of the electromagnetic and weak interactions into a single
electroweak theory. However, no existing QFT provides a complete and
empirically consistent description of all known interactions.

The principal obstacle is likely the quantization of gravity. The
gravitational interaction is so weak that quantum corrections become relevant
only at the Planck scale. At these energies, quantum gravitational effects are
expected to modify the structure of space-time itself, implying that the
Planck length constitutes a fundamental limit to position-measurement accuracy
\cite{Garay:1995,Hagar:2014}. If such a minimal length exists, standard
Poincar\'{e} symmetry may fail to describe the geometry of space-time at the
Planck scale. Thus, a unified QFT of all fundamental forces may require not
only an internal symmetry group governing interactions between elementary
particles, but also a modified space-time symmetry.

\paragraph{Motivation for $q$-Deformation}

Non-com\-mu\-ta\-tive coordinate algebras and their associated symmetry
structures provide a mathematical framework for modeling modified space-time
geometries
\cite{Snyder:1947a,Doplicher:1994zv,Chu:1998qz,Schomerus:1999ug,Lukierski:1991pn,Majid:1994cy}%
. In a non-com\-mu\-ta\-tive space-time, the measurement of one spatial
coordinate necessarily affects the precision with which another can be
measured, owing to the non-commutative nature of the position operators. This
feature induces an intrinsic positional uncertainty, implying a discretized
space-time structure. Since momentum and wavelength of plane waves are inversely related, spatial discretization imposes an upper bound on momentum. As a result, momentum-space
integrals in high\-er-order quan\-tum-field-the\-o\-ret\-ic corrections -
otherwise divergent - would become finite.

A particularly relevant example in physics is the three-di\-men\-sion\-al
$q$-de\-formed Euclidean space \cite{Faddeev:1987ih,Lorek:1997eh}. Its
co\-or\-di\-nate-gen\-er\-a\-tor relations arise from a continuous deformation
of those in standard three-di\-men\-sion\-al Euclidean space. For small
deformation parameters, deviations from standard Euclidean geometry occur only
at extremely short distances. At low energies, the effects of $q$%
-de\-for\-ma\-tion are negligible, and established physical theories remain
valid for describing such processes with high accuracy.

In the context of the zero-point energy problem, conventional quantum field
theory fails to yield an accurate prediction for the vacuum energy density. If
this discrepancy originates from an incomplete description of space-time
geometry at small scales, it becomes natural to investigate whether
$q$-de\-formed space-time symmetries could help resolve these challenges.

\paragraph{Organization of the Paper}

Developing a quantum theory with $q$-de\-formed space-time symmetries,
requires specialized mathematical tools. In particular, we employ a
non-com\-mu\-ta\-tive product, the \textit{star product}, for multiplying
functions on $q$-de\-formed coordinate algebras (cf. App.~\ref{KapQuaZeiEle}).
Ordinary derivatives and integrals are replaced by \textit{Jackson
derivatives} and \textit{Jackson integrals} on $q$-de\-formed spaces (cf.
App.~\ref{KapParDer}). These tools enable the formulation and solution of
Klein-Gordon equations in the $q$-de\-formed Euclidean space (cf.
App.~\ref{KapPlaWavSol}).\footnote{The $q$-de\-formed Klein-Gordon equation
presented here does not exhibit $q$-deformed Poincar\'{e} symmetry
\cite{ogievetsky1992}, as the time coordinate is not included in the
deformation. Analyzing Klein-Gordon equations in the $q$-de\-formed Minkowski
space is a more complex task and lies beyond the present scope
\cite{CarowWatamura:1990nk,Pillin:1993,Meyer:1994wi,Podles:1996,Blohmann:2001ph,Bachmaier:2003}%
.}

We adapt the standard procedure for calculating the zero-point energy of a
scalar field to the $q$-de\-formed Euclidean setting. To this end, we first
introduce and discuss delta functions specific to the $q$-de\-formed Euclidean
space [cf. Chap.~\ref{KapDreiDimDel}]. The expressions for these
$q$-de\-formed delta functions are derived from results on
one-di\-men\-sion\-al $q$-de\-formed Fourier transforms
\cite{olshanetsky1998q.alg,Wachter:2024}, summarized in Chap.~\ref{KapPrel}.
Readers may skip this introductory chapter and proceed directly to
Chap.~\ref{KapDreiDimDel}, consulting the relevant subsections of
Chap.~\ref{KapPrel} as needed.

In Chap.~\ref{KapVakEne}, we begin with the conventional calculation of the
vacuum energy density.\footnote{The zero-point energy is computed using the
\textit{momentum cutoff regularization method} rather than \textit{dimensional
regularization} \cite{akhmedov2002vacuum}.} We then derive an expression for
the vacuum energy arising from the ground state energy of a $q$-de\-formed
scalar field. Two scenarios are considered: the average ground state energy of
the $q$-de\-formed Klein-Gordon field over the nearest neighborhood of a
quasipoint (\textbf{localized case}),\footnote{A quasipoint is a compact
region with the smallest physically admissible volume.} the vacuum energy of
the entire $q$-de\-formed Euclidean space (\textbf{global case}). In the
localized case, the resulting vacuum energy density is found to be extremely
large, consistent with the corresponding classical result. In the global case,
the vacuum energy due to the ground state energy of a massless $q$-de\-formed
scalar field vanishes.

\section{Mathematical Preliminaries\label{KapPrel}}

\subsection{$q$-Derivatives and $q$-Exponential
Functions\label{KapqAnaExpTriFkt}}

The \textit{Jackson derivative} of a real function $f$ is defined as
\cite{Jackson:1910yd}:%
\begin{equation}
D_{q}f(x)=\frac{f(x)-f(qx)}{x-qx}. \label{DefJacAbl}%
\end{equation}
In the limit $q\rightarrow1$, the Jackson derivative converges to the ordinary
derivative \cite{Klimyk:1997eb}:%
\begin{equation}
\lim_{q\hspace{0.01in}\rightarrow1}D_{q}f(x)=\frac{\text{d}f(x)}%
{\text{d\hspace{0.01in}}x}. \label{KlaGreJacWdh}%
\end{equation}
From the definition, the Jackson derivative of a power function $x^{\alpha}$,
where $\alpha\in\mathbb{R}$, follows as:%
\begin{equation}
D_{q}\hspace{0.01in}x^{\alpha}=[[\alpha]]_{q}\hspace{0.01in}x^{\alpha-1}.
\end{equation}
The \textit{antisymmetric }$q$\textit{-num\-bers} are defined as:%
\begin{equation}
\lbrack\lbrack\alpha]]_{q}=\frac{1-q^{\alpha}}{1-q}. \label{DefAntSymQZah2}%
\end{equation}
The $q$\textit{-fac\-to\-ri\-al}, analogous to the classical factorial for
$n\in\mathbb{N}$, is given by:%
\begin{align}
\lbrack\lbrack0]]_{q}!  &  =1,\nonumber\\
\lbrack\lbrack\hspace{0.01in}n]]_{q}!  &  =[[1]]_{q}\hspace{0.01in}%
[[2]]_{q}\ldots\lbrack\lbrack\hspace{0.01in}n-1]]_{q}\hspace{0.01in}%
[[\hspace{0.01in}n]]_{q}. \label{DefQFakJacBas}%
\end{align}
Using the $q$-fac\-to\-ri\-al, the $q$\textit{-bi\-no\-mi\-al} coefficients
are defined for $k\in\mathbb{N}$ as:%
\begin{equation}%
\genfrac{[}{]}{0pt}{}{n}{k}%
_{q}=\frac{[[\hspace{0.01in}n]]_{q}!}{[[\hspace{0.01in}n-k]]_{q}%
!\hspace{0.01in}[[k]]_{q}!}. \label{qBinKoeBas}%
\end{equation}
The Jackson derivative satisfies the \textit{product rule}:%
\begin{align}
D_{q}\left(  f(x)\hspace{0.01in}g(x)\right)   &  =f(qx)\hspace{0.01in}%
D_{q}\hspace{0.01in}g(x)+g(x)\hspace{0.01in}D_{q}f(x)\nonumber\\
&  =f(x)\hspace{0.01in}D_{q}\hspace{0.01in}g(x)+g(qx)\hspace{0.01in}D_{q}f(x).
\label{ProRegJacAblBas}%
\end{align}
High\-er-or\-der Jackson derivatives can be calculated using the following
formulas \cite{Klimyk:1997eb}:%
\begin{align}
D_{q}^{n}f(x)  &  =\frac{1}{(1-q)^{n}x^{n}}\sum_{m\hspace{0.01in}%
=\hspace{0.01in}0}^{n}%
\genfrac{[}{]}{0pt}{}{n}{m}%
_{q^{-1}}(-1)^{m}q^{-m(m\hspace{0.01in}-1)/2}f(q^{m}x)\nonumber\\
&  =(-1)^{n}\frac{q^{-n(n\hspace{0.01in}-1)/2}}{(1-q)^{n}x^{n}}\sum
_{m\hspace{0.01in}=\hspace{0.01in}0}^{n}%
\genfrac{[}{]}{0pt}{}{n}{m}%
_{q}(-1)^{m}q^{m(m\hspace{0.01in}-1)/2}f(q^{n-m}x). \label{MehAnwJacAblFkt}%
\end{align}

The $q$-\textit{ex\-po\-nen\-tial} is introduced as an eigenfunction of the
Jackson derivative \cite{Majid:1993ud}, satisfying:%
\begin{equation}
D_{q}\exp_{q}(x)=\exp_{q}(x) \label{Bed1qExp}%
\end{equation}
with the normalization condition%
\begin{equation}
\exp_{q}(0)=1. \label{Bed2qExp}%
\end{equation}
Its power series expansion is:%
\begin{equation}
\exp_{q}(x)=\sum_{k\hspace{0.01in}=\hspace{0.01in}0}^{\infty}\hspace
{0.01in}\frac{1}{[[k]]_{q}!}\,x^{k}. \label{AusQExpEin}%
\end{equation}
From the $q$-ex\-po\-nen\-tial, the $q$\textit{-trigono\-met\-ric functions}
are defined as:%
\begin{align}
\sin_{q}(x)  &  =\frac{1}{2\text{i}}\left(  \exp_{q}(\text{i}\hspace
{0.01in}x)-\exp_{q}(-\text{i}\hspace{0.01in}x)\right)  ,\nonumber\\
\cos_{q}(x)  &  =\frac{1}{2}\left(  \exp_{q}(\text{i}\hspace{0.01in}%
x)+\exp_{q}(-\text{i}\hspace{0.01in}x)\right)  . \label{DefqTriSinCosExp1}%
\end{align}

The $q$-ex\-po\-nen\-tial in Eq.~(\ref{AusQExpEin}) can be used to define
$q$\textit{-trans\-la\-tion operators}
\cite{Chryssomalakos:1993zm,Majid:1993ud,rogov2000q.alg}:%
\begin{equation}
\exp_{q}(a|D_{q})=\sum_{k\hspace{0.01in}=\hspace{0.01in}0}^{\infty}\frac
{a^{k}}{[[k]]_{q}!}\,D_{q}^{k}. \label{qTraOpeDef}%
\end{equation}
Applied to a function $f(x)$, this yields the $q$-ana\textit{\-}log of
Taylor's formula:%
\begin{equation}
f(a\,\bar{\oplus}\,x)=\exp_{q}(a|D_{q})\triangleright f(x)=\sum_{k\hspace
{0.01in}=\hspace{0.01in}0}^{\infty}\frac{a^{k}}{[[k]]_{q}!}\,D_{q}^{k}%
\hspace{0.01in}f(x), \label{qTraVerGerUnKon}%
\end{equation}
and similarly:%
\begin{equation}
f(x\,\bar{\oplus}\,a)=\sum_{k\hspace{0.01in}=\hspace{0.01in}0}^{\infty}%
\frac{1}{[[k]]_{q}!}\hspace{0.01in}(D_{q}^{k}\hspace{0.01in}f(x))\,a^{k}.
\label{qTraVerGerKonUnk2}%
\end{equation}
An alternative translation operator is:%
\begin{equation}
\exp_{q^{-1}}(-a|D_{q})=\sum_{k\hspace{0.01in}=\hspace{0.01in}0}^{\infty}%
\frac{(-a)^{k}}{[[k]]_{q^{-1}}!}\,D_{q}^{k}=\sum_{k\hspace{0.01in}%
=\hspace{0.01in}0}^{\infty}\frac{q^{k(k\hspace{0.01in}-1)/2}}{[[k]]_{q}%
!}\,(-a)^{k}D_{q}^{k}. \label{AusQExpEinInv}%
\end{equation}
Applied to $f(x)$, this gives:%
\begin{equation}
f((\bar{\ominus}\,a)\,\bar{\oplus}\,x)=\exp_{q^{-1}}(-a|D_{q})\triangleright
f(x)=\sum_{k\hspace{0.01in}=\hspace{0.01in}0}^{\infty}\frac{q^{k(k\hspace
{0.01in}-1)/2}}{[[k]]_{q}!}\,(-a)^{k}D_{q}^{k}\hspace{0.01in}f(x),
\label{QTayForTyp2}%
\end{equation}
and similarly:%
\begin{equation}
f(x\,\bar{\oplus}\,(\bar{\ominus}\,a))=\sum_{k\hspace{0.01in}=\hspace
{0.01in}0}^{\infty}\frac{q^{k(k\hspace{0.01in}-1)/2}}{[[k]]_{q}!}\,(D_{q}%
^{k}\hspace{0.01in}f(x))(-a)^{k}. \label{QTayForTyp3}%
\end{equation}
The operators in Eqs.~(\ref{qTraOpeDef}) and (\ref{AusQExpEinInv}) are
inverses of each other \cite{Wachter:2024}:%
\begin{align}
\exp_{q^{-1}}(-a|D_{q})\exp_{q}(a|D_{q})\triangleright f(x)  &  =\exp
_{q}(a|D_{q})\exp_{q^{-1}}(-a|D_{q})\triangleright f(x)\nonumber\\
&  =f(x). \label{NacInvTra}%
\end{align}
For $q$-trans\-la\-tions of the $q$-ex\-po\-nen\-tial, the following
\textit{addition theorem} holds:\cite{Chryssomalakos:1993zm}:%
\begin{equation}
\exp_{q}(x\,\bar{\oplus}\,a)=\exp_{q}(x)\exp_{q}(a). \label{AddTheqExp1Dim}%
\end{equation}

Using Eq.~(\ref{QTayForTyp2}), one obtains $q$\textit{-in\-ver\-sions}
\cite{Wachter:2024}:%
\begin{equation}
f((\bar{\ominus}\,x))=\sum_{k\hspace{0.01in}=\hspace{0.01in}0}^{\infty}%
\frac{q^{k(k\hspace{0.01in}-1)/2}}{[[k]]_{q}!}\,(-x)^{k}(D_{q}^{k}%
\hspace{0.01in}f)(0).
\end{equation}
For powers of $x$, one finds (with $n\in\mathbb{N}$):%
\begin{equation}
(\bar{\ominus}\,x)^{n}=q^{n(n\hspace{0.01in}-1)/2}\,(-x)^{n}. \label{KonAntXn}%
\end{equation}
Moreover, it holds%
\begin{equation}
\exp_{q}(\bar{\ominus}\,x)=\sum_{k\hspace{0.01in}=\hspace{0.01in}0}^{\infty
}\frac{1}{[[k]]_{q^{-1}}!}\,(-x)^{k}=\exp_{q^{-1}}(-x) \label{ExpForInvQExp}%
\end{equation}
and%
\begin{equation}
\exp_{q}(\bar{\ominus}\,x)\exp_{q}(x)=\exp_{q}(x)\exp_{q}(\bar{\ominus}\,x)=1.
\end{equation}

\subsection{$q$-Integrals\label{KapQIntTrig}}

For $z>0$ and $0<q<1$, the one-di\-men\-sion\-al \textit{Jackson integral} is
defined as follows \cite{Jackson:1908}:%
\begin{align}
\int_{z}^{\hspace{0.01in}z.\infty}\text{d}_{q}x\hspace{0.01in}f(x)  &
=(1-q)\hspace{0.01in}z\sum_{j\hspace{0.01in}=1}^{\infty}q^{-j}f(q^{-j}%
z),\nonumber\\
\int_{0}^{\hspace{0.01in}z}\text{d}_{q}x\hspace{0.01in}f(x)  &  =(1-q)\hspace
{0.01in}z\sum_{j\hspace{0.01in}=\hspace{0.01in}0}^{\infty}q^{\hspace{0.01in}%
j}f(q^{\hspace{0.01in}j}z). \label{QInt0klQkl1}%
\end{align}
Accordingly, we have:%
\begin{align}
\int_{0}^{\hspace{0.01in}z.\infty}\text{d}_{q}x\hspace{0.01in}f(x)  &
=\int_{0}^{\hspace{0.01in}z}\text{d}_{q}x\hspace{0.01in}f(x)+\int_{z}%
^{\hspace{0.01in}z.\infty}\text{d}_{q}x\hspace{0.01in}f(x)\nonumber\\
&  =(1-q)\hspace{0.01in}z\hspace{-0.02in}\sum_{j\hspace{0.01in}=\hspace
{0.01in}-\infty}^{\infty}\hspace{-0.02in}q^{-j}f(q^{-j}z). \label{ImpJacInt}%
\end{align}
For $z<0$ and $0<q<1$, $q$-in\-te\-grals with a negative integration range are
defined as%
\begin{align}
\int_{z.\infty}^{\hspace{0.01in}z}\text{d}_{q}x\hspace{0.01in}f(x)  &
=(q-1)\,z\sum_{j\hspace{0.01in}=1}^{\infty}q^{-j}f(q^{-j}z),\nonumber\\
\int_{z}^{\hspace{0.01in}0}\text{d}_{q}x\hspace{0.01in}f(x)  &  =(q-1)\,z\sum
_{j\hspace{0.01in}=\hspace{0.01in}0}^{\infty}q^{\hspace{0.01in}j}%
f(q^{\hspace{0.01in}j}z), \label{QInt0klQkl1Neg}%
\end{align}
and%
\begin{equation}
\int_{z.\infty}^{\hspace{0.01in}0}\text{d}_{q}x\hspace{0.01in}%
f(x)=(q-1)\,z\hspace{-0.02in}\sum_{j\hspace{0.01in}=\hspace{0.01in}-\infty
}^{\infty}\hspace{-0.02in}q^{\hspace{0.01in}j}f(q^{\hspace{0.01in}j}z).
\label{JacIntMinInfZerNeg}%
\end{equation}
The $q$-in\-te\-grals in Eqs.~(\ref{ImpJacInt}) and (\ref{JacIntMinInfZerNeg})
can be combined to obtain the $q$-in\-te\-gral over the interval $\left(
-\infty,\infty\right)  $:%
\begin{align}
\int_{-z.\infty}^{\hspace{0.01in}z.\infty}\text{d}_{q}x\hspace{0.01in}f(x)  &
=\int_{0}^{\hspace{0.01in}z.\infty}\text{d}_{q}x\hspace{0.01in}f(x)+\int
_{-\infty.z}^{\hspace{0.01in}0}\text{d}_{q}x\hspace{0.01in}f(x)\nonumber\\
&  =\left\vert (1-q)\hspace{0.01in}z\right\vert \hspace{-0.02in}\sum
_{j\hspace{0.01in}=\hspace{0.01in}-\infty}^{\infty}\hspace{-0.02in}%
q^{\hspace{0.01in}j}\left[  f(q^{\hspace{0.01in}j}z)+f(-q^{\hspace{0.01in}%
j}z)\right]  . \label{UneJackIntAll}%
\end{align}

From Eqs.~(\ref{ImpJacInt}) and (\ref{JacIntMinInfZerNeg}), it is clear that
only the discrete set of points $\left\{  \pm\hspace{0.01in}q^{k}%
z\,|\,k\in\mathbb{Z}\right\}  $ contributes to the $q$-in\-te\-gral over
$\left(  0,\infty\right)  $ or $\left(  -\infty,0\right)  $. Therefore, for
convergency of the $q$-in\-te\-gral over these domains, the integrand $f$ must
satisfy the boundary condition (with $q<1$)%
\begin{equation}
\lim_{k\hspace{0.01in}\rightarrow-\infty}f(\pm\hspace{0.01in}q^{k}z)=0.
\label{RanBedFunEin}%
\end{equation}
If $f$ is continuous at $x=0$, this condition implies%
\begin{align}
\int\nolimits_{0}^{\hspace{0.01in}z.\infty}\text{d}_{q}x\hspace{0.02in}%
D_{q}f(x)  &  =(1-q)\hspace{-0.02in}\sum_{j\hspace{0.01in}=\hspace
{0.01in}-\infty}^{\infty}\hspace{-0.02in}zq^{\hspace{0.01in}-j}\,\frac
{f(q^{\hspace{0.01in}-j}z)-f(q^{\hspace{0.01in}-j+1}z)}{(1-q)\hspace
{0.01in}q^{\hspace{0.01in}-j}z}\nonumber\\
&  =\lim_{k\hspace{0.01in}\rightarrow\infty}\sum_{j=-k}^{k}\left[
f(q^{\hspace{0.01in}-j}z)-f(q^{\hspace{0.01in}-j+1}z)\right] \nonumber\\
&  =\lim_{k\hspace{0.01in}\rightarrow\infty}\left[  f(q^{-k}z)-f(q^{k+1}%
z)\right]  =-f(0). \label{PosParIntAblq1Dim}%
\end{align}
A similar reasoning shows%
\begin{equation}
\int\nolimits_{-\infty.z}^{\hspace{0.01in}0}\text{d}_{q}x\hspace{0.02in}%
D_{q}f(x)=f(0). \label{NegParIntAblq1Dim}%
\end{equation}
Combining Eqs.~(\ref{PosParIntAblq1Dim}) and (\ref{NegParIntAblq1Dim}) yields,
for $m\in\mathbb{N}$,%
\begin{align}
\int\nolimits_{-z.\infty}^{\hspace{0.01in}z.\infty}\text{d}_{q}x\hspace
{0.02in}D_{q}^{m}f(x)  &  =\int\nolimits_{0}^{\hspace{0.01in}z.\infty}%
\text{d}_{q}x\hspace{0.02in}D_{q}^{m}f(x)+\int\nolimits_{-\infty.z}%
^{\hspace{0.01in}0}\text{d}_{q}x\hspace{0.02in}D_{q}^{m}f(x)\nonumber\\
&  =-D_{q}^{m-1}f(0)+D_{q}^{m-1}f(0)=0. \label{UnIntJacAbl}%
\end{align}

If the functions $f$ and $g$ vanish at infinity,\footnote{A weaker condition,
namely $\lim_{k\rightarrow-\infty}f(\pm q^{k}a)\hspace{0.01in}g(\pm q^{k}%
a)=0$, is also sufficient.} Eq.~(\ref{UnIntJacAbl}) together with
Eq.~(\ref{ProRegJacAblBas}) in\ Chap.~\ref{KapqAnaExpTriFkt} gives the
$q$-de\-formed \textit{in\-te\-gra\-tion-by-parts formula}:%
\begin{equation}
\int\nolimits_{-z.\infty}^{\hspace{0.01in}z.\infty}\text{d}_{q}x\hspace
{0.01in}f(x)\hspace{0.01in}D_{q}\hspace{0.01in}g(x)=-\int\nolimits_{-z.\infty
}^{\hspace{0.01in}z.\infty}\text{d}_{q}x\hspace{0.01in}\left[  D_{q}%
f(x)\right]  g(qx).
\end{equation}
Iterating this relation leads to the following identity
\cite{olshanetsky1998q.alg} , valid for $k\in\mathbb{N}$:%
\begin{equation}
\int\nolimits_{-z.\infty}^{\hspace{0.01in}z.\infty}\text{d}_{q}x\hspace
{0.01in}f(x)\hspace{0.01in}D_{q}^{k}g(x)=(-1)^{k}q^{-k(k\hspace{0.01in}%
-1)/2}\int\nolimits_{-z.\infty}^{\hspace{0.01in}z.\infty}\text{d}_{q}%
x\hspace{0.01in}\left[  D_{q}^{k}f(x)\right]  g(q^{k}x).
\label{RegParIntBraLinK}%
\end{equation}

Using Eq.~(\ref{UnIntJacAbl}) together with Eq.~(\ref{qTraVerGerKonUnk2}) in
Chap.~\ref{KapqAnaExpTriFkt}, one finds that the $q$-in\-te\-gral over
$\left(  -\infty,\infty\right)  $ is invariant under $q$-trans\-la\-tion
\cite{Chryssomalakos:1993zm,Kempf:1994yd,1996q.alg.....8008K}:%
\begin{align}
&  \int\nolimits_{-z.\infty}^{\hspace{0.01in}z.\infty}\text{d}_{q}%
x\hspace{0.02in}f(x\,\bar{\oplus}\,a)=\int\nolimits_{-z.\infty}^{\hspace
{0.01in}z.\infty}\text{d}_{q}x\sum_{k\hspace{0.01in}=\hspace{0.01in}0}%
^{\infty}\frac{1}{[[k]]_{q}!}\left[  D_{q}^{k}f(x)\right]  a^{k}\nonumber\\
&  \qquad=\sum_{k\hspace{0.01in}=\hspace{0.01in}0}^{\infty}\frac{1}%
{[[k]]_{q}!}\int\nolimits_{-z.\infty}^{\hspace{0.01in}z.\infty}\text{d}%
_{q}x\left[  D_{q}^{k}f(x)\right]  a^{k}=\int\nolimits_{-z.\infty}%
^{\hspace{0.01in}z.\infty}\text{d}_{q}x\hspace{0.02in}f(x).
\label{qTraInvEinJacIntUnk}%
\end{align}
Similarly, it holds:%
\begin{equation}
\int\nolimits_{-z.\infty}^{\hspace{0.01in}z.\infty}\text{d}_{q}x\hspace
{0.02in}f(a\,\bar{\oplus}\,x)=\int\nolimits_{-z.\infty}^{\hspace
{0.01in}z.\infty}\text{d}_{q}x\hspace{0.02in}f(x).
\label{qTraInvEinJacIntUnk2}%
\end{equation}

Using the Jackson integral from Eq.~(\ref{ImpJacInt}), we define, for $0<q<1$,
the function%
\begin{equation}
\Theta_{q}(z)=\int\nolimits_{0}^{\hspace{0.01in}z.\infty}\text{d}_{q}%
x\,x^{-1}\sin_{q}(x)=(1-q)\sum_{m\hspace{0.01in}=\hspace{0.01in}-\infty
}^{\infty}\sin_{q}(q^{m}z). \label{DefTheZ}%
\end{equation}
From the definition of the improper $q$-in\-te\-gral in Eq.~(\ref{ImpJacInt}),
it follows that for $k\in\mathbb{Z}$:%
\begin{equation}
\Theta_{q}(q^{k}z)=\Theta_{q}(z). \label{PerTheFkt}%
\end{equation}
The function $\Theta_{q}(z)$ is related to%
\begin{equation}
\mathbf{Q}(z\hspace{0.01in};q)=(1-q)\sum_{m\hspace{0.01in}=\hspace
{0.01in}-\infty}^{\infty}\frac{1}{z\hspace{0.01in}q^{m}+z^{-1}q^{-m}}
\label{DefQzqFkt}%
\end{equation}
with%
\begin{equation}
\lim_{q\hspace{0.01in}\rightarrow1^{-}}\mathbf{Q}(z\hspace{0.01in}%
;q)=\frac{\pi}{2}. \label{KlaGreQFkt}%
\end{equation}
Concretely, it holds \cite{olshanetsky1998q.alg}%
\begin{equation}
\Theta_{q}(z)=\mathbf{Q}((1-q)z;q) \label{IdeTheQUmr}%
\end{equation}
with%
\begin{equation}
\lim_{q\hspace{0.01in}\rightarrow1^{-}}\Theta_{q}(z)=\lim_{q\hspace
{0.01in}\rightarrow1^{-}}\mathbf{Q}((1-q)z\hspace{0.01in};q)=\frac{\pi}{2}.
\label{TheQFktKlaGre}%
\end{equation}

\subsection{$q$-Distributions\label{KapQDisEin}}

Let $\mathcal{M}_{q}$ denote the set of functions defined on the
$q$-\textit{lat\-tice}%
\begin{equation}
\mathbb{G}_{q,\hspace{0.01in}x_{0}}\mathbb{=}\left\{  \pm\hspace{0.01in}%
x_{0}\hspace{0.01in}q^{m}|\,m\in\mathbb{Z}\right\}  \cup\{0\}.\label{DefQLat}%
\end{equation}
Using the $q$-in\-te\-gral over $\left(  -\infty,\infty\right)  $ [cf.
Eq.~(\ref{UneJackIntAll}) of Chap.~\ref{KapQIntTrig}], we define a
$q$-\textit{scalar product} on $\mathcal{M}_{q}$ \cite{Ruffing:1996ph}:%
\begin{equation}
\left\langle f,\hspace{0.01in}g\right\rangle _{q}=\int\nolimits_{-x.\infty
}^{\hspace{0.01in}x.\infty}\text{d}_{q}z\,\overline{f(z)}\,g(z)=\sum
_{\varepsilon\hspace{0.01in}=\hspace{0.01in}\pm}\,\sum_{j\hspace
{0.01in}=\hspace{0.01in}-\infty}^{\infty}(1-q)\hspace{0.01in}q^{\hspace
{0.01in}j}x\,\overline{f(\varepsilon q^{\hspace{0.01in}j}x)}\,g(\varepsilon
q^{\hspace{0.01in}j}x).\label{SkaProqDefEinBraLin}%
\end{equation}
Here, $\bar{f}$ denotes the complex conjugate of $f$. The $q$-scalar product
induces a $q$\textit{-norm} on $\mathcal{M}_{q}$%
\ \cite{Jambor:2004ph,Lavagno:2009vg}:%
\begin{equation}
\left\Vert f\right\Vert _{q}^{2}=\left\langle f,f\right\rangle _{q}%
=\sum_{\varepsilon\hspace{0.01in}=\hspace{0.01in}\pm}\,\sum_{j\hspace
{0.01in}=\hspace{0.01in}-\infty}^{\infty}(1-q)\hspace{0.01in}q^{\hspace
{0.01in}j}x\left\vert f(\varepsilon q^{\hspace{0.01in}j}x)\right\vert ^{2}.
\end{equation}
The set of functions with a finite $q$-norm forms the $q$\textit{-ana\-logue
of the Hilbert space of square-integrable functions}
\cite{Cerchiai:1999,Hinterding:2000ph,Wess:math-ph9910013}:%
\begin{equation}
L_{q}^{2}=\{f\,|\,\left\langle f,f\right\rangle _{q}<\infty\}.
\end{equation}

Next, we consider those functions in $\mathcal{M}_{q}$ that are continuous at
the origin of the lattice $\mathbb{G}_{q,\hspace{0.01in}x_{0}}$ (with
$0<q<1$):%
\begin{equation}
\lim_{m\hspace{0.01in}\rightarrow\hspace{0.01in}\infty}f(\pm\hspace
{0.01in}x_{0}\hspace{0.01in}q^{m})=f(0).\label{SteUrsGit}%
\end{equation}
Additionally, we assume the existence of positive constants $C_{k,\hspace{0.01in}l}(q)$ for
$k,l\in\mathbb{N}_{0}\ $such that%
\begin{equation}
\left\vert \,x^{k}D_{q}^{\hspace{0.01in}l}f(x)\right\vert \leq C_{k,\hspace{0.01in}l}(q)\quad\text{for}\quad
x\in\mathbb{G}_{q,\hspace{0.01in}x_{0}}.\label{SchAbfFktBed}%
\end{equation}
These conditions imply the asymptotic behavior%
\begin{equation}
\lim_{m\hspace{0.01in}\rightarrow\hspace{0.01in}\infty}(D_{q}^{\hspace{0.01in}l}f)(\pm
\hspace{0.01in}x_{0}\hspace{0.01in}q^{-m})=0.\label{LimQTesFkt}%
\end{equation}
The functions in $\mathcal{M}_{q}$ satisfying these properties form the set
$\mathcal{S}_{q}$, which plays the role of a $q$-\textit{ana\-logue}
\textit{of the space of test functions for tempered distributions}. For such
$q$-de\-formed test functions, we introduce a family of seminorms%
\begin{equation}
\left\Vert f\right\Vert _{k,l}=\max_{x\hspace{0.01in}\in\hspace{0.01in}%
\mathbb{G}_{q,x_{0}}}\left\vert x^{k}D_{q}^{\hspace{0.01in}l}f(x)\right\vert ,
\end{equation}
which equips $\mathcal{S}_{q}$ with a natural
topology\ \cite{olshanetsky1998q.alg}.

A continuous linear functional $l:\mathcal{S}_{q}\rightarrow\mathbb{C}$ is
called a $q$\textit{-dis\-tri\-bu\-tion}. Such a $q$-dis\-tri\-bu\-tion $l$ is
called \textit{regular} if there exists a function $f\in\mathcal{M}_{q}$ such
that%
\begin{equation}
l(\hspace{0.01in}g)=\left\langle f,g\right\rangle _{q}=\int\nolimits_{-\infty
}^{\hspace{0.01in}\infty}\text{d}_{q}x\,\overline{f(x)}\,g(x).
\label{DefRegDisqDef}%
\end{equation}
Using Eq.~(\ref{RegParIntBraLinK}) from the previous chapter, one obtains a
formula for the Jackson derivatives of a regular $q$-dis\-tri\-bu\-tion (for
$k\in\mathbb{N},$ $g\in\mathcal{S}_{q}$) \cite{Wachter:2024}:%
\begin{equation}
(D_{q}^{k}\hspace{0.01in}l)(\hspace{0.01in}g)=\int\nolimits_{-\infty}%
^{\hspace{0.01in}\infty}\text{d}_{q}x\,\overline{f(x)}\big((-q^{-1}D_{q^{-1}%
})^{k}g\big)(x)=l\big((-q^{-1}D_{q^{-1}})^{k}g\big). \label{JacAblKOrdDis}%
\end{equation}
New $q$-dis\-tri\-bu\-tions can be obtained as limits of sequences of regular
ones. Let $f_{n}(x)$ be a sequence of functions defining regular
$q$-dis\-tri\-bu\-tions. If the limit%
\begin{equation}
l(\hspace{0.01in}g)=\lim_{n\hspace{0.01in}\rightarrow\hspace{0.01in}\infty
}\left\langle f_{n}\hspace{0.01in},g\right\rangle _{q}%
\end{equation}
exists for all $g\in\mathcal{S}_{q}$, then the mapping $l:g\mapsto
l(\hspace{0.01in}g)$ defines a $q$-dis\-tri\-bu\-tion
\cite{olshanetsky1998q.alg}.

We now introduce a $q$-\textit{ana\-logue of the delta distribution}. For a
test function $g$ that is continuous at the origin, we define%
\begin{equation}
\delta_{q}(\hspace{0.01in}g)=\lim_{m\hspace{0.01in}\rightarrow\hspace
{0.01in}\infty}\frac{g(x_{0}\hspace{0.01in}q^{m})+g(-x_{0}\hspace{0.01in}%
q^{m})}{2}=g(0).\label{DefDelQEinDim}%
\end{equation}
Since $g$ is continuous at the origin, the $q$-delta distribution can be
decomposed as%
\begin{equation}
2\hspace{0.01in}\delta_{q}(\hspace{0.01in}g)=\delta_{q}^{+}(\hspace
{0.01in}g)+\delta_{q}^{-}(\hspace{0.01in}g),\label{ZerQDelDis}%
\end{equation}
with%
\begin{equation}
\delta_{q}^{\hspace{0.01in}\varepsilon}(\hspace{0.01in}g)=\lim_{m\hspace
{0.01in}\rightarrow\hspace{0.01in}\infty}g(\varepsilon x_{0}\hspace
{0.01in}q^{m})=g(0^{\varepsilon}),\qquad\varepsilon\in\{+,-\}.
\end{equation}
By introducing the localized basis functions%
\begin{equation}
\phi_{m}^{\hspace{0.01in}\varepsilon}(x)=\left\{
\begin{array}
[c]{ll}%
1 & \text{if\quad}x=\varepsilon q^{m}x_{0},\\
0 & \text{otherwise},
\end{array}
\right.  \label{BasVekFktBraLin}%
\end{equation}
each component of the $q$-delta distribution can be represented as the limit
of regular $q$-dis\-tri\-bu\-tions \cite{Wachter:2024}:%
\[
\delta_{q}^{\hspace{0.01in}\varepsilon}(\hspace{0.01in}g)=\lim_{n\hspace
{0.01in}\rightarrow\hspace{0.01in}\infty}\int\nolimits_{-\infty}%
^{\hspace{0.01in}\infty}\text{d}_{q}x\,\frac{\varepsilon\hspace{0.01in}%
\phi_{n}^{\hspace{0.01in}\varepsilon}(x)}{(1-q)\hspace{0.01in}x}\,g(x).
\]

We assume the existence of a function $\delta_{q}(x)$ such that%
\begin{equation}
\delta_{q}(\hspace{0.01in}g)=\int\nolimits_{-\infty}^{\hspace{0.01in}\infty
}\text{d}_{q}\hspace{0.01in}x\,\delta_{q}(x)\,g(x)=g(0).
\label{ChaIdeqDelQKal1dim}%
\end{equation}
Using the $q$-Taylor formula [cf. Eqs.~(\ref{QTayForTyp2}) and
(\ref{QTayForTyp3}) in Chap.~\ref{KapqAnaExpTriFkt}], one obtains%
\begin{align}
\delta_{q}(x\,\bar{\oplus}\,(\bar{\ominus}\,a))  &  =\sum_{k\hspace
{0.01in}=\hspace{0.01in}0}^{\infty}\frac{q^{k(k\hspace{0.01in}-1)/2}%
}{[[k]]_{q}!}\left(  D_{q}^{k}\hspace{0.01in}\delta_{q}(x)\right)
\hspace{0.01in}(-a)^{k},\nonumber\\
\delta_{q}((\bar{\ominus}\,a)\,\bar{\oplus}\,x)  &  =\sum_{k\hspace
{0.01in}=\hspace{0.01in}0}^{\infty}\frac{q^{k(k\hspace{0.01in}-1)/2}%
}{[[k]]_{q}!}\,(-a)^{k}\hspace{0.01in}D_{q}^{k}\hspace{0.01in}\delta_{q}(x).
\label{DefVerDelFktQ1Dim}%
\end{align}
Using Eq.~(\ref{RegParIntBraLinK}) from the previous chapter and
Eq.~(\ref{qTraVerGerUnKon}) from Chap.~\ref{KapqAnaExpTriFkt}, one verifies
that these expressions satisfy the following identities \cite{Wachter:2024}:%
\begin{equation}
\int_{-\infty}^{\hspace{0.01in}\infty}\text{d}_{q}\hspace{0.01in}x\,\delta
_{q}((\bar{\ominus}\,qa)\,\bar{\oplus}\,x)\,f(x)=\int_{-\infty}^{\hspace
{0.01in}\infty}\text{d}_{q}x\hspace{0.01in}f(x)\,\delta_{q}(x\,\bar{\oplus
}\,(\bar{\ominus}\,qa))=f(a). \label{HerChaIdeDel1Dim}%
\end{equation}

The relation in Eq.~(\ref{ChaIdeqDelQKal1dim}) holds for functions continuous
at the origin, in particular those admitting a pow\-er-se\-ries expansion
around the origin. To extend this property to functions with a Laurent
expansion around the origin, i.e.%
\begin{equation}
\int\nolimits_{-\infty}^{\hspace{0.01in}\infty}\text{d}_{q}x\,\delta
_{q}(x)\hspace{0.01in}g(x)=g_{0},\quad\text{where}\quad g(z)=\sum
_{k\hspace{0.01in}=\hspace{0.01in}-\infty}^{\infty}g_{k}\hspace{0.01in}z^{k},
\end{equation}
it suffices to establish (for $n\in\mathbb{Z)}$%
\begin{equation}
\int\nolimits_{-\infty}^{\hspace{0.01in}\infty}\text{d}_{q}x\,\delta
_{q}(x)\hspace{0.02in}x^{n}=\left\{
\begin{tabular}
[c]{ll}%
$1,$ & if $n=0,$\\
$0,$ & if $n\neq0.$%
\end{tabular}
\right.  \label{ErwChaIdeEindQDelt}%
\end{equation}
To prove Eq.~(\ref{ErwChaIdeEindQDelt}), we compute the $n$-th Jackson
derivative of $\delta_{q}(x)$. Using Eq.~(\ref{MehAnwJacAblFkt}) from
Chap.~\ref{KapqAnaExpTriFkt}, one finds \cite{Wachter:2024}:%
\begin{equation}
D_{q}^{n}\hspace{0.01in}\delta_{q}(x)=(-1)^{n}q^{-n(n\hspace{0.01in}%
+1)/2}\hspace{0.01in}[[\hspace{0.01in}n]]_{q}!\,x^{-n}\hspace{0.01in}%
\delta_{q}(x). \label{JacAblQDelN}%
\end{equation}
Combining this result with Eq.~(\ref{RegParIntBraLinK}) of
Chap.~\ref{KapQIntTrig} yields%
\begin{align}
\int_{-x_{0}.\infty}^{\hspace{0.01in}x_{0}.\infty}\text{d}_{q}x\,x^{-n}%
\hspace{0.01in}\delta_{q}(x)  &  =\frac{(-1)^{n}q^{n(n\hspace{0.01in}+1)/2}%
}{[[\hspace{0.01in}n]]_{q}!}\int_{-x_{0}.\infty}^{\hspace{0.01in}x_{0}.\infty
}\text{d}_{q}x\,D_{q}^{n}\delta_{q}(x)\nonumber\\
&  =\frac{1}{[[\hspace{0.01in}n]]_{q}!}\int_{-x_{0}.\infty}^{\hspace
{0.01in}x_{0}.\infty}\text{d}_{q}x\,\delta_{q}(q^{n}x)\cdot D_{q}^{n}1=0.
\end{align}

The result of Eq.~(\ref{JacAblQDelN}) allows one to rewrite the $q$-delta
functions of Eq.~(\ref{DefVerDelFktQ1Dim}):%
\begin{equation}
\delta_{q}((\bar{\ominus}\,qa)\,\bar{\oplus}\,x)=\sum_{k\hspace{0.01in}%
=\hspace{0.01in}0}^{\infty}\frac{q^{k(k\hspace{0.01in}+1)/2}}{[[k]]_{q}%
!}(-a)^{k}\hspace{0.01in}D_{q}^{k}\delta_{q}(x)=\sum_{k\hspace{0.01in}%
=\hspace{0.01in}0}^{\infty}a^{k}\delta_{q}(x)\,x^{-k},
\label{AltDar1DimTraDelFkt0}%
\end{equation}
and similarly%
\begin{equation}
\delta_{q}(x\,\bar{\oplus}\,(\bar{\ominus}\,qa))=\sum_{k\hspace{0.01in}%
=\hspace{0.01in}0}^{\infty}x^{-k}\delta_{q}(x)\hspace{0.01in}a^{k}.
\label{AltDar1DimTraDelFkt}%
\end{equation}
Using these expressions together with Eq.~(\ref{ErwChaIdeEindQDelt}), one
obtains for $n\in\mathbb{N}$:%
\begin{gather}
\int_{-x_{0}.\infty}^{\hspace{0.01in}x_{0}.\infty}\text{d}_{q}x\,\delta
_{q}((\bar{\ominus}\,qa)\,\bar{\oplus}\,x)\,x^{-n}=\int_{-x_{0}.\infty
}^{\hspace{0.01in}x_{0}.\infty}\text{d}_{q}x\sum_{k\hspace{0.01in}%
=\hspace{0.01in}-\infty}^{\infty}a^{k}\hspace{0.01in}\delta_{q}(x)\,x^{-k}%
\hspace{0.01in}x^{-n}\nonumber\\
=\sum_{k\hspace{0.01in}=\hspace{0.01in}-\infty}^{\infty}a^{k}\int
_{-x_{0}.\infty}^{\hspace{0.01in}x_{0}.\infty}\text{d}_{q}x\,\delta
_{q}(x)\,x^{-(k\hspace{0.01in}+\hspace{0.01in}n)}=\sum_{k\hspace
{0.01in}=\hspace{0.01in}-\infty}^{\infty}a^{k}\hspace{0.01in}\delta
_{k,-n}=a^{-n}. \label{CharIdeEinDimQDel5}%
\end{gather}

The $q$-delta functions are related to the basis functions $\phi_{m}^{\pm}(x)$
defined in Eq.~(\ref{BasVekFktBraLin}). For $a\in\mathbb{G}_{q,\hspace
{0.01in}x_{0}}$ and $0<q<1$, the identities in Eq.~(\ref{HerChaIdeDel1Dim})
imply%
\begin{equation}
\left.  \delta_{q}((\bar{\ominus}\,qa)\,\bar{\oplus}\,x)\right\vert
_{a\hspace{0.01in}=\hspace{0.01in}\pm\hspace{0.01in}x_{0}q^{m}}=\left.
\delta_{q}(x\,\bar{\oplus}\,(\bar{\ominus}\,qa))\right\vert _{a\hspace
{0.01in}=\hspace{0.01in}\pm\hspace{0.01in}x_{0}q^{m}}=\frac{\phi_{m}^{\pm}%
(x)}{(1-q)\hspace{0.01in}x_{0}\hspace{0.01in}q^{m}}. \label{DarQDelKroDel1Dim}%
\end{equation}
The functions $\phi_{m}^{\pm}(x)$ can be represented by smooth functions.
Define%
\begin{equation}
\psi(x,a)=\left\{
\begin{array}
[c]{ll}%
0 & \text{if\ }x\leq-a,\\
\operatorname*{e}\nolimits^{-\operatorname*{e}\nolimits^{x/(x^{2}%
-\hspace{0.01in}a^{2})}}a^{2}\operatorname*{e} & \text{if\ }-a<x<a,\\
a^{2}\operatorname*{e} & \text{if\ }a\leq x,
\end{array}
\right.
\end{equation}
with partial derivative%
\begin{equation}
\frac{\partial}{\partial x}\psi(x,a)=\left\{
\begin{array}
[c]{ll}%
0 & \text{if }x\leq-a,\\
\operatorname*{e}\nolimits^{x/(x^{2}-\hspace{0.01in}a^{2})-\operatorname*{e}%
\nolimits^{x/(x^{2}-\hspace{0.01in}a^{2})}}\frac{x^{2}+\hspace{0.01in}a^{2}%
}{(x^{2}-\hspace{0.01in}a^{2})^{2}}\,a^{2}\operatorname*{e} & \text{if
}-a<x<a,\\
0 & \text{if }a\leq x.
\end{array}
\right.
\end{equation}
This function is infinitely differentiable in $x$, symmetric about $x=0$, and
bell-shaped with maximum at the origin, where $\partial_{x}\psi(0,a)=1$
\cite{Grossmann:1988ah}. We define the auxiliary function $b_{q}%
(\hspace{0.01in}y)$, which assigns to $y$ half the distance from $y$ to the
nearest lattice point $q\hspace{0.01in}y$ or $q^{-1}y$:%
\begin{equation}
b_{q}(\hspace{0.01in}y)=\frac{1}{2}\min\left\{  |(1-q^{-1})\hspace
{0.01in}y|,|(1-q)\hspace{0.01in}y|\right\}  .
\end{equation}
Using $\partial_{x}\psi(x,a)$ and $b_{q}(\hspace{0.01in}y)$, we define%
\begin{equation}
\phi_{q}(x,y)=\frac{\partial}{\partial x}\psi(x-y,b_{q}(\hspace{0.01in}y)).
\label{DefKonDelFkt1Dim}%
\end{equation}
This function satisfies%
\begin{equation}
\phi_{q}(x,x)=1,\qquad\forall x\in\mathbb{R}, \label{ChaIdePhi1}%
\end{equation}
and%
\begin{equation}
\phi_{q}(x,y)=0,\qquad x\hspace{0.02in}\mathbb{\notin}\left[  \hspace
{0.01in}y-b_{q}(\hspace{0.01in}y),y+b_{q}(\hspace{0.01in}y)\right]  .
\label{ChaIdePhi2}%
\end{equation}
Thus, the functions $\phi_{m}^{\hspace{0.01in}\varepsilon}(x)$ on the
one-di\-men\-sion\-al $q$-lat\-tice $\mathbb{G}_{q,\hspace{0.01in}x_{0}}$ can
be represented by smooth functions:%
\begin{equation}
\phi_{m}^{\hspace{0.01in}\varepsilon}(x)=\phi_{q}(x,\varepsilon x_{0}q^{m}).
\label{DefPhimEps}%
\end{equation}
This result leads to the following representations [cf.
Eq.~(\ref{DarQDelKroDel1Dim})]:%
\begin{equation}
\delta_{q}((\bar{\ominus}\,qa)\,\bar{\oplus}\,x)=\delta_{q}(x\,\bar{\oplus
}\,(\bar{\ominus}\,qa))=\frac{\phi_{q}(x,a)}{(1-q)\hspace{-0.01in}\left\vert
a\right\vert }=\frac{\phi_{q}(x,a)}{(1-q)\hspace{-0.01in}\left\vert
x\right\vert }. \label{DarQDelKroDelAlg1Dim}%
\end{equation}

Finally, we introduce a scaling operator $\Lambda$, defined by%
\begin{equation}
\Lambda\hspace{0.01in}f(x)=f(qx), \label{ActScaOpe}%
\end{equation}
under which the $q$-delta function transforms according to \cite{Wachter:2024}%
:%
\begin{equation}
\Lambda\hspace{-0.01in}^{n}\hspace{0.01in}\delta_{q}(x)=\delta_{q}%
(q^{n}x)=q^{-n}\delta_{q}(x). \label{SkaDelFktEinQ}%
\end{equation}
We verify these identities as follows:%
\begin{gather}
\int_{-\infty}^{\hspace{0.01in}\infty}\text{d}_{q}x\left[  \hspace
{0.01in}\Lambda\hspace{-0.01in}^{n}\delta_{q}(x)\right]  \hspace
{0.01in}g(x)=\int_{-\infty}^{\hspace{0.01in}\infty}\text{d}_{q}\hspace
{0.01in}x\,\delta_{q}(q^{n}x)\,g(x)\nonumber\\
=q^{-n}\int_{-\infty}^{\hspace{0.01in}\infty}\text{d}_{q}x\,\delta
_{q}(x)\,g(q^{-n}x)=q^{-n}g(0)=\int_{-\infty}^{\hspace{0.01in}\infty}%
\text{d}_{q}p\,q^{-n}\delta_{q}(x)\,g(x).
\end{gather}
Since this result holds for all test functions $g\in\mathcal{S}_{q}$, the
identities in Eq.~(\ref{SkaDelFktEinQ}) are confirmed.

\subsection{$q$-Fourier Transforms\label{KapEinDimQFouTraN}}

The $q$-de\-formed Fourier transforms establish maps between two function
spaces $\mathcal{M}_{q,x}$ and $\mathcal{M}_{q,p}$, where $x$ and $p$ denote
position and momentum variables, respectively.

Using the $q$-de\-formed exponentials introduced in Eq.~(\ref{AusQExpEin}) of
Chap.~\ref{KapqAnaExpTriFkt}, we define $q$\textit{-ana\-logues of
one-di\-men\-sion\-al plane waves} \cite{Kempf:1994yd}:%
\begin{equation}
\exp_{q}(x|\text{i}p)=\sum_{k\hspace{0.01in}=\hspace{0.01in}0}^{\infty}%
\frac{x^{k}(\text{i}p)^{k}}{[[k]]_{q}!},\qquad\exp_{q^{-1}}(-\text{i}%
p|\hspace{0.01in}x)=\sum_{k\hspace{0.01in}=\hspace{0.01in}0}^{\infty}%
\frac{(-\text{i}p)^{k}\hspace{0.01in}x^{k}}{[[k]]_{q^{-1}}!}%
.\label{DefEinQEbeWel}%
\end{equation}
The momentum operator $\hat{P}=\,$i$^{-1}D_{q}$ acts on these $q$-de\-formed
plane waves as follows \cite{Wachter:2024}:%
\begin{align}
\text{i}^{-1}D_{q,x}\triangleright\exp_{q}(x|\text{i}p) &  =\exp
_{q}(x|\text{i}p)\hspace{0.01in}p,\nonumber\\
\text{i}^{-1}D_{q,x}\triangleright\exp_{q^{-1}}(-\text{i}p|x) &
=-\exp_{q^{-1}}(-\text{i}p|qx)\hspace{0.01in}p.\label{EigFktImp1Dim}%
\end{align}

To construct an improper $q$-in\-te\-gral over the lattice $\mathbb{G}%
_{q^{1/2},\,x_{0}}$ [cf. Eq.~(\ref{DefQLat}) of the previous chapter], we
combine an improper $q$-in\-te\-gral over the lattice $\mathbb{G}%
_{q,\hspace{0.01in}x_{0}}$ with one over the lattice $\mathbb{G}%
_{q,\hspace{0.01in}q^{1/2}x_{0}}$ (for $0<q<1$)
\cite{Cerchiai:1999,Hinterding:2000ph,Wess:math-ph9910013}:%
\begin{align}
&  \frac{1}{1+q^{1/2}}\left(  \int_{-x_{0}.\infty}^{\hspace{0.01in}%
x_{0}.\infty}\text{d}_{q}x\,f(x)+\int_{-x_{0}q^{1/2}.\infty}^{\hspace
{0.01in}x_{0}q^{1/2}.\infty}\text{d}_{q}x\,f(x)\right)  =\nonumber\\
&  \qquad\qquad=(1-q^{1/2})\sum_{m\hspace{0.01in}=\hspace{0.01in}-\infty
}^{\infty}x_{0}\hspace{0.01in}q^{m/2}\left[  f(x_{0}\hspace{0.01in}q^{m/2})+f(-x_{0}%
\hspace{0.01in}q^{m/2})\right] \nonumber\\
&  \qquad\qquad=\int_{-x_{0}.\infty}^{\hspace{0.01in}x_{0}.\infty}%
\text{d}_{q^{1/2}}x\,f(x). \label{VerJacInt1Dim}%
\end{align}
The $q^{1/2}$-in\-te\-gral in Eq.~(\ref{VerJacInt1Dim}) is composed of
$q$-in\-te\-grals that are invariant under $q$-trans\-la\-tions [cf.
Eqs.~(\ref{qTraInvEinJacIntUnk}) and (\ref{qTraInvEinJacIntUnk2}) from
Chap.~\ref{KapQIntTrig}]. Consequently, the $q^{1/2}$-in\-te\-gral is also
invariant under $q$-trans\-la\-tions:%
\begin{equation}
\int_{-x.\infty}^{\hspace{0.01in}x.\infty}\text{d}_{q^{1/2}}z\,f(z\,\bar
{\oplus}\,a)=\int_{-x.\infty}^{\hspace{0.01in}x.\infty}\text{d}_{q^{1/2}%
}z\,f(a\,\bar{\oplus}\,z)=\int_{-x.\infty}^{\hspace{0.01in}x.\infty}%
\text{d}_{q^{1/2}}z\,f(z).
\end{equation}
The in\-te\-gration-by-parts formula likewise holds for the improper $q^{1/2}%
$-in\-te\-gral, since it inherits this property from the $q$-in\-te\-grals it
comprises [cf. Eq.~(\ref{RegParIntBraLinK}) in Chap.~\ref{KapQIntTrig})]:%
\begin{equation}
\int\text{d}_{q^{1/2}}x\,f(x)\,D_{q}^{k}g(x)=(-1)^{k}q^{-k(k\hspace
{0.01in}-1)/2}\int\text{d}_{q^{1/2}}x\left[  D_{q}^{k}\,f(x)\right]
g(q^{k}x). \label{ParQHalInt}%
\end{equation}

Using the $q^{1/2}$-in\-te\-gral from Eq.~(\ref{VerJacInt1Dim}), we define the
$q$\textit{-Fourier transforms} \cite{Wachter:2024}:%
\begin{align}
\mathcal{F}_{q^{1/2}}(\phi)(\hspace{0.01in}p)  &  =\operatorname*{vol}%
\nolimits_{q}^{-1/2}\int_{-x_{0}.\infty}^{\hspace{0.01in}x_{0}.\infty}%
\text{d}_{q^{1/2}}x\,\phi(x)\hspace{0.01in}\exp_{q}(x|\text{i}%
p),\nonumber\\[0.04in]
\mathcal{F}_{q^{1/2}}^{\hspace{0.01in}-1}(\psi)(x)  &  =\operatorname*{vol}%
\nolimits_{q}^{-1/2}\int_{-p_{0}.\infty}^{\hspace{0.01in}p_{0}.\infty}%
\text{d}_{q^{1/2}}p\,\psi(\hspace{0.01in}p)\hspace{0.01in}\exp_{q^{-1}%
}(-\text{i}p|qx), \label{DefQFouEinDimQHal}%
\end{align}
where [cf. Eq.~(\ref{DefTheZ}) of Chap.~\ref{KapQIntTrig}]%
\begin{equation}
\operatorname*{vol}\nolimits_{q}=\frac{8\left[  \Theta_{q}(x_{0}%
\hspace{0.01in}p_{0})+\Theta_{q}(q^{1/2}x_{0}\hspace{0.01in}p_{0})\right]
}{(1+q^{1/2})^{2}}. \label{qDefVolEleEinDim}%
\end{equation}
The $q$-de\-formed volume element has the following classical limit [cf.
Eq.~(\ref{TheQFktKlaGre}) of Chap.~\ref{KapQIntTrig}]:%
\begin{equation}
\lim_{q\hspace{0.01in}\rightarrow1^{-}}\operatorname*{vol}\nolimits_{q}%
=\frac{8\left(  \frac{\pi}{2}+\frac{\pi}{2}\right)  }{4}=2\pi.
\label{KlaGreVolEle1Dim}%
\end{equation}

We summarize several fundamental properties of the $q$-Fourier transforms
\cite{olshanetsky1998q.alg}. To this end, we reconsider the scaling operator
introduced in Eq.~(\ref{ActScaOpe}) of Chap.~\ref{KapQDisEin}:%
\begin{equation}
\Lambda\hspace{0.01in}f(x)=f(qx),\qquad\Lambda\hspace{0.01in}g(\hspace
{0.01in}p)=g(q^{-1}p).
\end{equation}
The scaling operator $\Lambda$ commutes with the $q$-Fourier transforms in the
following manner:%
\begin{equation}
\mathcal{F}_{q^{1/2}}\hspace{0.01in}\Lambda=q^{-1}\Lambda\hspace
{0.01in}\mathcal{F}_{q^{1/2}},\qquad\mathcal{F}_{q^{1/2}}^{\hspace{0.01in}%
-1}\Lambda=q\hspace{0.01in}\Lambda\hspace{0.01in}\mathcal{F}_{q^{1/2}%
}^{\hspace{0.01in}-1}.
\end{equation}
Moreover, the $q$-Fourier transforms interchange derivative operators and
multiplication operators:%
\begin{equation}
\mathcal{F}_{q^{1/2}}(D_{q,\hspace{0.01in}x}\hspace{0.01in}\phi)=-\text{i}%
\Lambda\hspace{0.01in}p\hspace{0.01in}\mathcal{F}_{q^{1/2}}(\phi
),\qquad\mathcal{F}_{q^{1/2}}(x\phi)=-\text{i}D_{q,\hspace{0.01in}p}%
\hspace{0.01in}\mathcal{F}_{q^{1/2}}(\phi), \label{VerFouQ1DimAblOrt1}%
\end{equation}
and%
\begin{equation}
\mathcal{F}_{q^{1/2}}^{\hspace{0.01in}-1}(D_{q,\hspace{0.01in}p}%
\hspace{0.01in}\psi)=\text{i}\hspace{0.01in}\mathcal{F}_{q^{1/2}}%
^{\hspace{0.01in}-1}(\psi)\,x,\qquad\mathcal{F}_{q^{1/2}}^{\hspace{0.01in}%
-1}(\hspace{0.01in}p\hspace{0.01in}\psi)=\text{i}\hspace{0.01in}q^{-1}%
\Lambda^{-1}D_{q,\hspace{0.01in}x}\hspace{0.01in}\mathcal{F}_{q}%
^{\hspace{0.01in}-1}(\psi). \label{VerFouQ1DimAblOrt2}%
\end{equation}

The $q$-Fourier transforms are unitary maps on the Hilbert space $L_{q}^{2}$,
equipped with the scalar product defined in Eq.~(\ref{SkaProqDefEinBraLin}) of
the previous\ chapter. However, the $q$-Fourier transforms act slightly
differently on the two arguments in the $q$-scalar product. Explicitly, it
holds \cite{Wachter:2024}%
\begin{equation}
\left\langle \hspace{0.01in}g\hspace{0.01in},f\right\rangle _{q^{1/2}%
,\hspace{0.01in}x}=\left\langle \psi\hspace{0.01in},\phi\right\rangle
_{q^{1/2},\hspace{0.01in}p}, \label{ParIdeFour1dimQ}%
\end{equation}
where%
\begin{align}
\phi(\hspace{0.01in}p)  &  =\mathcal{F}_{q^{1/2}}(f)(\hspace{0.01in}%
p)=\operatorname*{vol}\nolimits_{q}^{-1/2}\int_{-x_{0}.\infty}^{\hspace
{0.01in}x_{0}.\infty}\text{d}_{q^{1/2}}x\,\exp_{q}(\text{i}p|x)\hspace
{0.01in}f(x),\nonumber\\[0.04in]
\psi(\hspace{0.01in}p)  &  =\mathcal{\tilde{F}}_{q^{1/2}}(\hspace
{0.01in}g)(\hspace{0.01in}p)=\operatorname*{vol}\nolimits_{q}^{-1/2}%
\int_{-x_{0}.\infty}^{\hspace{0.01in}x_{0}.\infty}\text{d}_{q^{1/2}}%
x\,\exp_{q^{-1}}(\text{i}p|qx)\,g(x),
\end{align}
and%
\begin{align}
f(x)  &  =\mathcal{F}_{q^{1/2}}^{\hspace{0.01in}-1}(\phi
)(x)=\operatorname*{vol}\nolimits_{q}^{-1/2}\int_{-p_{0}.\infty}%
^{\hspace{0.01in}p_{0}.\infty}\text{d}_{q^{1/2}}p\,\phi(\hspace{0.01in}%
p)\exp_{q^{-1}}(-\text{i}p|qx),\nonumber\\[0.04in]
g(x)  &  =\mathcal{\tilde{F}}_{q^{1/2}}^{\hspace{0.01in}-1}(\psi
)(x)=\operatorname*{vol}\nolimits_{q}^{-1/2}\int_{-p_{0}.\infty}%
^{\hspace{0.01in}p_{0}.\infty}\text{d}_{q^{1/2}}p\,\exp_{q}(x|\hspace
{-0.02in}-\hspace{-0.02in}\text{i}p)\,\psi(\hspace{0.01in}p).
\label{InvFourTraInd}%
\end{align}

Let $\delta_{q^{1/2}}$ denote the $q$-delta distribution with respect to the
$q^{1/2}$-in\-te\-gral:%
\begin{equation}
\int_{-x_{0}.\infty}^{\hspace{0.01in}x_{0}.\infty}\text{d}_{q^{1/2}}%
x\,\delta_{q^{1/2}}(x)\,f(x)=f(0). \label{ChaIdeQHalDelFktEinDim}%
\end{equation}
The identities in Eq.~(\ref{HerChaIdeDel1Dim}) of Chap.~\ref{KapQDisEin} can
be adapted so as to apply to the lattice $\mathbb{G}_{q^{1/2},\,x_{0}}$:%
\begin{align}
\int_{-x_{0}.\infty}^{\hspace{0.01in}x_{0}.\infty}\text{d}_{q^{1/2}}%
x\,\delta_{q^{1/2}}((\bar{\ominus}\,qa)\,\bar{\oplus}\,x)\,f(x)  &
=f(a),\nonumber\\[0.04in]
\int_{-x_{0}.\infty}^{\hspace{0.01in}x_{0}.\infty}\text{d}_{q^{1/2}}%
x\hspace{0.01in}f(x)\,\delta_{q^{1/2}}(x\,\bar{\oplus}\,(\bar{\ominus}\,qa))
&  =f(a), \label{ChaIdeQHalbDeltFkt}%
\end{align}
where, for $m\in\mathbb{Z}$ and $0<q<1$,%
\begin{gather}
\left.  \delta_{q^{1/2}}((\bar{\ominus}\,qa)\,\bar{\oplus}\,x)\right\vert
_{a\hspace{0.01in}=\hspace{0.01in}\pm x_{0}q^{m/2}}=\left.  \delta_{q^{1/2}%
}(x\,\bar{\oplus}\,(\bar{\ominus}\,qa))\right\vert _{a\hspace{0.01in}%
=\hspace{0.01in}\pm x_{0}q^{m/2}}\nonumber\\
=\frac{\phi_{m/2}^{\pm}(x)}{(1-q^{1/2})\hspace{0.01in}x_{0}\hspace
{0.01in}q^{m/2}}. \label{VerQDelqHalBasFkt}%
\end{gather}

The $q$-de\-formed plane waves [cf. Eq.~(\ref{DefEinQEbeWel})]%
\begin{equation}
u_{p}(x)=\operatorname*{vol}\nolimits_{q}^{-1/2}\exp_{q}(\text{i}x|p),\qquad
u_{p}^{\ast}(x)=\operatorname*{vol}\nolimits_{q}^{-1/2}\exp_{q^{-1}}%
(-\text{i}p|qx)
\end{equation}
satisfy \textit{completeness relations} and \textit{orthogonality conditions}
\cite{Wachter:2024} (for $n,m\in\mathbb{Z}$; $\varepsilon,\varepsilon
^{\hspace{0.01in}\prime}=\pm1$):%
\begin{align}
&  \int_{-p_{0}.\infty}^{\hspace{0.01in}p_{0}.\infty}\text{d}_{q^{1/2}%
}p\,u_{p}(\varepsilon\hspace{0.01in}x_{0}q^{n/2})\,u_{p}^{\ast}(\varepsilon
^{\hspace{0.01in}\prime}x_{0}q^{m/2})=\frac{\delta_{\varepsilon\varepsilon
^{\prime}}\delta_{nm}}{(1-q^{1/2})\hspace{0.01in}x_{0}\hspace{0.01in}q^{n/2}%
}\nonumber\\
&  \qquad\qquad=\delta_{q^{1/2}}(x\,\bar{\oplus}\,(\bar{\ominus}%
\,qy))|_{x\hspace{0.01in}=\hspace{0.01in}\varepsilon\hspace{0.01in}%
x_{0}q^{n/2},\,y\hspace{0.01in}=\hspace{0.01in}\varepsilon^{\prime}%
x_{0}q^{m/2}},\label{QVolRelNeuNorExp1Dim}%
\end{align}
and%
\begin{align}
&  \int_{-x_{0}.\infty}^{\,x_{0}.\infty}\text{d}_{q^{1/2}}x\,u_{\varepsilon
\hspace{0.01in}p_{0}q^{n/2}}^{\ast}(x)\,u_{\varepsilon^{\prime}p_{0}q^{m/2}%
}(x)=\frac{\delta_{\varepsilon\varepsilon^{\prime}}\delta_{nm}}{(1-q^{1/2}%
)\hspace{0.01in}p_{0}\hspace{0.01in}q^{n/2}}\nonumber\\
&  \qquad\qquad=\delta_{q^{1/2}}((\bar{\ominus}\,qp)\,\bar{\oplus}\,p^{\prime
})|_{p\hspace{0.01in}=\hspace{0.01in}\varepsilon\hspace{0.01in}p_{0}%
q^{n/2},\,p^{\prime}=\hspace{0.01in}\varepsilon^{\prime}p_{0}q^{m/2}%
}.\label{QOrtRelNeuNorExp1Dim}%
\end{align}
The last identity in Eq.~(\ref{QVolRelNeuNorExp1Dim}) or
Eq.~(\ref{QOrtRelNeuNorExp1Dim}) follows from Eq.~(\ref{VerQDelqHalBasFkt})
together with Eq.~(\ref{BasVekFktBraLin}) from the previous chapter.

By employing the completeness relation in Eq.~(\ref{QVolRelNeuNorExp1Dim}), a
function $f(x)$ in position space $\mathcal{M}_{q,x}$ can be expanded in terms
of the functions $u_{p}^{\ast}(x)$ \cite{olshanetsky1998q.alg,Wachter:2024}:%
\begin{equation}
f(x)=\int_{-p_{0}.\infty}^{\hspace{0.01in}p_{0}.\infty}\text{d}_{q^{1/2}%
}p\,a_{p}\hspace{0.01in}u_{p}^{\ast}(x),\quad\text{where}\quad a_{p}%
=\mathcal{F}_{q^{1/2}}(f)(\hspace{0.01in}p). \label{FouDarFkt1DimQ}%
\end{equation}
Similarly, a function $g(\hspace{0.01in}p)$ in momentum space $\mathcal{M}%
_{q,p}$ can be expanded in terms of the functions $u_{p}(x)$:%
\begin{equation}
g(\hspace{0.01in}p)=\int_{-x_{0}.\infty}^{\hspace{0.01in}x_{0}.\infty}%
\text{d}_{q^{1/2}}x\,b_{x}\hspace{0.01in}u_{p}(x),\quad\text{where}\quad
b_{x}=\mathcal{F}_{q^{1/2}}^{\hspace{0.01in}-1}(\hspace{0.01in}g)(x).
\end{equation}

Finally, we summarize several \textit{useful formulas for }$q$\textit{-Fourier
transforms} on the lattice $\mathbb{G}_{q^{1/2},\,x_{0}}$ \cite{Wachter:2024}:%
\begin{equation}
\mathcal{F}_{q^{1/2}}(1)(\hspace{0.01in}p)=\operatorname*{vol}\nolimits_{q}%
^{1/2}\delta_{q^{1/2}}(\hspace{0.01in}p),\qquad\mathcal{F}_{q^{1/2}}%
(\delta_{q^{1/2}})(\hspace{0.01in}p)=\operatorname*{vol}\nolimits_{q}^{-1/2},
\label{QHalFouTraIdeEinDim}%
\end{equation}
and%
\begin{align}
\mathcal{F}_{q^{1/2}}(x^{n})(\hspace{0.01in}p)  &  =\text{i}^{n}%
\operatorname*{vol}\nolimits_{q}^{1/2}q^{-n(n\hspace{0.01in}+1)/2}%
\hspace{0.01in}[[\hspace{0.01in}n]]_{q}!\,p^{-n}\hspace{0.01in}\delta
_{q^{1/2}}(\hspace{0.01in}p),\nonumber\\
\mathcal{F}_{q^{1/2}}(x^{-n\hspace{0.01in}-1})(\hspace{0.01in}p)  &
=\frac{\text{i}^{n\hspace{0.01in}+1}(1+q^{1/2})}{4\hspace{0.01in}%
[[\hspace{0.01in}n]]_{q}!}\hspace{0.01in}\operatorname*{vol}\nolimits_{q}%
^{1/2}p^{n}\operatorname*{sgn}(\hspace{0.01in}p).
\end{align}
Similarly, for the inverse $q$-Fourier transform we obtain:%
\begin{equation}
\mathcal{F}_{q^{1/2}}^{\hspace{0.01in}-1}(1)(x)=\operatorname*{vol}%
\nolimits_{q}^{1/2}\delta_{q^{1/2}}(x),\qquad\mathcal{F}_{q^{1/2}}^{-1}%
(\delta_{q^{1/2}})(\hspace{0.01in}p)=\operatorname*{vol}\nolimits_{q}^{-1/2},
\end{equation}
and%
\begin{align}
\mathcal{F}_{q^{1/2}}^{\hspace{0.01in}-1}(\hspace{0.01in}p^{n})(x)  &
=\text{i}^{-n}\operatorname*{vol}\nolimits_{q}^{1/2}\,[[\hspace{0.01in}%
n]]_{q}!\,x^{-n}\hspace{0.01in}\delta_{q^{1/2}}(x),\nonumber\\
\mathcal{F}_{q^{1/2}}^{-1}(\hspace{0.01in}p^{-n\hspace{0.01in}-1})(x)  &
=\frac{(-\text{i})^{n\hspace{0.01in}+1}\hspace{0.01in}(1+q^{1/2})}%
{4\hspace{0.01in}[[\hspace{0.01in}n]]_{q}!}\,q^{n(n\hspace{0.01in}%
+1)/2}\operatorname*{vol}\nolimits_{q}^{1/2}x^{n}\operatorname*{sgn}(x).
\end{align}

\section{Three-Dimensional $q$-Delta Function\label{KapDreiDimDel}}

\subsection{Definition\label{DefQDelFktKap}}

We introduce a $q$-de\-formed analog of the three-di\-men\-sion\-al delta
function \cite{Kempf:1994yd,Wachter:2019A}:%
\begin{equation}
\delta_{q}^{3}(\mathbf{x})=\int\nolimits_{-\infty}^{\infty}\text{d}_{q}%
^{3}\hspace{0.01in}p\hspace{0.01in}\exp_{q}(\mathbf{p}|\text{i}\mathbf{x}).
\label{DreDimDelLDef}%
\end{equation}
The $q$-ex\-po\-nen\-tial in Eq.~(\ref{DreDimDelLDef}) factorizes into a
product of three one-di\-men\-sion\-al $q$-ex\-po\-nen\-tials [cf.
Eq.~(\ref{ExpEukExp2}) in App.~\ref{KapExp} and Eq.~(\ref{AusQExpEin}) in
Chap.~\ref{KapqAnaExpTriFkt}]:%
\begin{equation}
\exp_{q}(\text{i}\mathbf{p}|\mathbf{x})=\exp_{q^{4}}(\hspace{0.01in}%
\text{i}\hspace{0.01in}p^{+}|\hspace{0.01in}x_{+})\exp_{q^{2}}(\hspace
{0.01in}\text{i}p^{3}|\hspace{0.01in}x_{3})\exp_{q^{4}}(\hspace{0.01in}%
\text{i}\hspace{0.01in}p^{-}|\hspace{0.01in}x_{-}). \label{DreExpQPro}%
\end{equation}
Similarly, the $q$-in\-te\-gral in Eq.~(\ref{DreDimDelLDef}) factorizes into
three one-di\-men\-sion\-al $q$-in\-te\-grals [cf. Eq.~(\ref{VerIntEntSpa}) of
Chap.~\ref{KapParDer}]:%
\begin{equation}
\int\text{d}_{q}^{3}\hspace{0.01in}p\,f(\mathbf{p})=\int\text{d}_{q^{2}}%
p^{+}\hspace{-0.01in}\int\text{d}_{q}\hspace{0.01in}p^{3}\hspace{-0.01in}%
\int\text{d}_{q^{2}}p^{-}\,f(\mathbf{p}).
\end{equation}
With these definitions at hand, Eq.~(\ref{DreDimDelLDef}) can be expressed as
(integration bounds are $-\infty$ to $\infty$ unless stated otherwise):%
\begin{align}
\delta_{q}^{3}(\mathbf{x})  &  =\int\text{d}_{q^{2}}p^{-}\exp_{q^{4}}%
(\hspace{0.01in}p^{-}|\hspace{0.01in}\text{i}\hspace{0.01in}x_{-})\int
\text{d}_{q}\hspace{0.01in}p^{3}\exp_{q^{2}}(\hspace{0.01in}p^{3}%
|\hspace{0.01in}\text{i}\hspace{0.01in}x_{3})\int\text{d}_{q^{2}}p^{+}%
\exp_{q^{4}}(\hspace{0.01in}p^{+}|\hspace{0.01in}\text{i}\hspace{0.01in}%
x_{+})\nonumber\\
&  =\operatorname*{vol}\nolimits_{q^{4}}\operatorname*{vol}\nolimits_{q^{2}%
}^{1/2}\mathcal{F}_{q^{2}}(1)(\hspace{0.01in}x_{-})\,\mathcal{F}%
_{q}(1)(\hspace{0.01in}x_{3})\,\mathcal{F}_{q^{2}}(1)(\hspace{0.01in}%
x_{+})\nonumber\\
&  =\operatorname*{vol}\nolimits_{q^{4}}^{2}\operatorname*{vol}%
\nolimits_{q^{2}}\delta_{q^{2}}(\hspace{0.01in}x_{-})\,\delta_{q}%
(\hspace{0.01in}x_{3})\,\delta_{q^{2}}(\hspace{0.01in}x_{+})\nonumber\\
&  =\operatorname*{vol}\nolimits_{q^{4}}^{2}\operatorname*{vol}%
\nolimits_{q^{2}}\delta_{q^{2}}(-\hspace{0.01in}q^{-1}x^{+})\,\delta
_{q}(\hspace{0.01in}x^{3})\,\delta_{q^{2}}(-\hspace{0.01in}q\hspace
{0.01in}x^{-}). \label{BerDreDimDelFkt}%
\end{align}
In this derivation, we have used the one-di\-men\-sion\-al $q$-de\-formed
Fourier transforms as defined in Eq.~(\ref{DefQFouEinDimQHal}) of
Chap.~\ref{KapEinDimQFouTraN}, together with Eq.~(\ref{QHalFouTraIdeEinDim})
from the same chapter. In the final step, coordinates with superscript indices
were employed as in Eq.~(\ref{HebSenInd}) of App.~\ref{KapQuaZeiEle}.
Integrating the delta function $\delta_{q}^{3}(\mathbf{x})$ over the entire
position space yields the corresponding volume element:%
\begin{align}
\operatorname*{vol}  &  =\int\hspace{-0.01in}\text{d}_{q}^{3}x\int\text{d}%
_{q}^{3}\hspace{0.01in}p\hspace{0.01in}\exp_{q}(\mathbf{p}|\text{i}%
\mathbf{x})=\int\hspace{-0.01in}\text{d}_{q}^{3}x\,\delta_{q}^{3}%
(\mathbf{x})\nonumber\\
&  =\operatorname*{vol}\nolimits_{q^{4}}^{2}\operatorname*{vol}%
\nolimits_{q^{2}}\hspace{-0.03in}\int\hspace{-0.01in}\text{d}_{q^{2}}%
x_{-}\,\delta_{q^{2}}(\hspace{0.01in}x_{-})\hspace{-0.01in}\int\hspace
{-0.01in}\text{d}_{q}x_{3}\,\delta_{q}(\hspace{0.01in}x_{3})\hspace
{-0.01in}\int\hspace{-0.01in}\text{d}_{q^{2}}x_{+}\,\delta_{q^{2}}%
(\hspace{0.01in}x_{+})\nonumber\\
&  =\operatorname*{vol}\nolimits_{q^{4}}^{2}\operatorname*{vol}%
\nolimits_{q^{2}}. \label{VolEleDef}%
\end{align}
Finally, by applying Eq.~(\ref{KlaGreVolEle1Dim}) of
Chap.~\ref{KapEinDimQFouTraN}, the classical limit of the volume element is
recovered as%
\begin{equation}
\lim_{q\hspace{0.01in}\rightarrow1}\,\operatorname*{vol}=\lim_{q\hspace
{0.01in}\rightarrow1}\,\operatorname*{vol}\nolimits_{q^{4}}^{2}%
\operatorname*{vol}\nolimits_{q^{2}}=\left(  2\pi\right)  ^{3}.
\end{equation}

\subsection{Approximate Expression}

Our aim is to derive an approximate expression for $\delta_{q}^{3}%
(\mathbf{x}\oplus(\ominus\hspace{0.01in}\kappa^{-1}\mathbf{y}))$. We begin by
applying the operator representation of $q$-trans\-la\-tions to the $q$-delta
function $\delta_{q}^{3}(\mathbf{x})$ [cf. Eq.~(\ref{ForQTra}) in
App.~\ref{KapExp}]. If $q$-de\-formed Euclidean space is to provide a
physically relevant description, the deformation parameter $q$ must be close
to unity.\ Accordingly, we neglect all contributions of order $\lambda
=q-q^{-1}$:%
\begin{align}
\delta_{q}^{3}(\mathbf{x}\oplus\mathbf{y})\approx &  \sum_{k_{+}%
=\hspace{0.01in}0}^{\infty}\sum_{k_{3}=\hspace{0.01in}0}^{\infty}\sum
_{k_{-}=\hspace{0.01in}0}^{\infty}\frac{\hat{U}_{x}^{-1}\triangleright
(x^{-})^{k_{-}}(x^{3})^{k_{3}}(x^{+})^{k_{+}}}{[[\hspace{0.01in}%
k_{-}]]_{q^{-4}}!\hspace{0.01in}[[\hspace{0.01in}k_{3}]]_{q^{-2}}%
!\hspace{0.01in}[[\hspace{0.01in}k_{+}]]_{q^{-4}}!}\nonumber\\
&  \times\left.  \hat{U}_{y}^{-1}D_{q^{-4},\hspace{0.01in}y^{-}}^{k_{-}%
}\hspace{0.01in}D_{q^{-2},\hspace{0.01in}y^{3}}^{k_{3}}\hspace{0.01in}%
D_{q^{-4},\hspace{0.01in}y^{+}}^{k_{+}}\hat{U}_{y}\triangleright\delta_{q}%
^{3}(\hspace{0.01in}\mathbf{y})\right\vert _{\substack{y_{-}\rightarrow
\hspace{0.01in}q^{-2k_{3}}y^{-}\\\,y^{3}\rightarrow\hspace{0.01in}q^{-2k_{+}%
}y^{3}}}. \label{TraQDelL3DimAnf}%
\end{align}
The operator representation of $q$-trans\-la\-tions in Eq.~(\ref{ForQTra}) of
App.~\ref{KapExp} is compatible with the normal ordering defined in
Eq.~(\ref{InvNorOrd}) of App.~\ref{KapQuaZeiEle}. By contrast, the expression
for $\delta_{q}^{3}(\mathbf{x})$ in Eq.~(\ref{BerDreDimDelFkt}) employs the
alternative normal ordering introduced in Eq.~(\ref{StePro0})
of\ App.~\ref{KapQuaZeiEle}. To reconcile these conventions in
Eq.~(\ref{TraQDelL3DimAnf}), we have inserted the reordering operators
$\hat{U}$ and $\hat{U}^{-1}$ [cf. Eqs.~(\ref{UmOrdOpeDreQuan}) and
(\ref{UmOrdInvOpeDreQuan}) in App.~\ref{KapQuaZeiEle}]. Neglecting terms of
order $\lambda$, the action of $\hat{U}$ on the $q$-delta function becomes%
\begin{gather}
\hat{U}_{x}\triangleright\delta_{q}^{3}(\mathbf{x})=q^{-2\hat{n}_{3}(\hat
{n}_{+}+\hspace{0.01in}\hat{n}_{-})}\delta_{q}^{3}(\mathbf{x})+\mathcal{O}%
(\lambda)\nonumber\\
\approx\int\text{d}_{q}\hspace{0.01in}p^{3}\sum_{n_{3}=\hspace{0.01in}%
0}^{\infty}\frac{(\hspace{0.01in}p^{3})^{n_{3}}(\text{i}\hspace{0.01in}%
x^{3})^{n_{3}}}{[[\hspace{0.01in}n_{3}]]_{q^{2}}!}\operatorname*{vol}%
\nolimits_{q^{4}}^{2}\delta_{q^{2}}(-\hspace{0.01in}q^{-2n_{3}+1}%
x^{-})\,\delta_{q^{2}}(-\hspace{0.01in}q^{-2n_{3}-1}x^{+}). \label{UmoDelFkt}%
\end{gather}
Here, we have employed the power series expansion of Eq.~(\ref{DefEinQEbeWel})
in Chap.~\ref{KapqAnaExpTriFkt}. The expression in Eq.~(\ref{UmoDelFkt}) can
be simplified using the scaling properties of one-di\-men\-sion\-al $q$-delta
functions and Jackson integrals [cf. Eq.~(\ref{SkaDelFktEinQ}) in
Chap.~\ref{KapQDisEin}]:%
\begin{align}
&  \int\text{d}_{q}\hspace{0.01in}p^{3}\sum_{n_{3}=\hspace{0.01in}0}^{\infty
}\frac{(\hspace{0.01in}p^{3})^{n_{3}}(\text{i}\hspace{0.01in}x^{3})^{n_{3}}%
}{[[\hspace{0.01in}n_{3}]]_{q^{2}}!}\operatorname*{vol}\nolimits_{q^{4}}%
^{2}\delta_{q^{2}}(-\hspace{0.01in}q^{-2n_{3}+1}x^{-})\,\delta_{q^{2}%
}(-\hspace{0.01in}q^{-2n_{3}-1}x^{+})=\nonumber\\
&  \qquad=\int\text{d}_{q}\hspace{0.01in}p^{3}\sum_{n_{3}=\hspace{0.01in}%
0}^{\infty}\frac{(\hspace{0.01in}p^{3})^{n_{3}}(\text{i}\hspace{0.01in}%
q^{4}x_{3})^{n_{3}}}{[[\hspace{0.01in}n_{3}]]_{q^{2}}!}\operatorname*{vol}%
\nolimits_{q^{4}}^{2}\delta_{q^{2}}(x_{+})\,\delta_{q^{2}}(x_{-})\nonumber\\
&  \qquad=\operatorname*{vol}\nolimits_{q^{2}}\delta_{q}(\hspace{0.01in}%
q^{4}x_{3})\,\operatorname*{vol}\nolimits_{q^{4}}^{2}\delta_{q^{2}}%
(x_{+})\,\delta_{q^{2}}(x_{-})\nonumber\\
&  \qquad=q^{-4}\operatorname*{vol}\delta_{q^{2}}(x_{+})\,\delta_{q}%
(\hspace{0.01in}x_{3})\,\delta_{q^{2}}(\hspace{0.01in}x_{-}).
\label{UmoDelFkt2}%
\end{align}
Combining Eqs.~(\ref{UmoDelFkt})\ and (\ref{UmoDelFkt2}) gives%
\begin{equation}
\hat{U}_{x}\triangleright\delta_{q}^{3}(\mathbf{x})=q^{-4}\operatorname*{vol}%
\delta_{q^{2}}(\hspace{0.01in}x_{-})\,\delta_{q}(\hspace{0.01in}x_{3}%
)\,\delta_{q^{2}}(x_{+})+\mathcal{O}(\lambda).
\end{equation}
Employing this result together with Eq.~(\ref{JacAblQDelN}) in
Chap.~\ref{KapQDisEin}, one obtains%
\begin{align}
&  \left.  D_{q^{-4},\hspace{0.01in}y^{-}}^{k_{-}}\hspace{0.01in}%
D_{q^{-2},\hspace{0.01in}y^{3}}^{k_{3}}\hspace{0.01in}D_{q^{-4},\hspace
{0.01in}y^{+}}^{k_{+}}\hat{U}_{y}\triangleright\delta_{q}^{3}(\hspace
{0.01in}\mathbf{y})\right\vert _{y^{-}\rightarrow\hspace{0.01in}q^{-2k_{3}%
}y^{-},\,y^{3}\rightarrow\hspace{0.01in}q^{-2k_{+}}y^{3}}=\nonumber\\
&  \qquad=\operatorname*{vol}\,(-1)^{k_{+}+\hspace{0.01in}k_{3}+k_{-}%
}q^{2k_{+}(k_{+}+\hspace{0.01in}1)+k_{3}(k_{3}+1)+2k_{-}(k_{-}+1)-\hspace
{0.01in}4}\nonumber\\
&  \qquad\hspace{0.17in}\times\lbrack\lbrack\hspace{0.01in}k_{+}]]_{q^{-4}%
}!\hspace{0.01in}[[\hspace{0.01in}k_{3}]]_{q^{-2}}!\hspace{0.01in}%
[[\hspace{0.01in}k_{-}]]_{q^{-4}}!\nonumber\\
&  \qquad\hspace{0.17in}\times\frac{\delta_{q^{2}}(-\hspace{0.01in}%
q^{-2k_{3}+1}y^{-})\,\delta_{q}(q^{-2k_{+}}y^{3})\,\delta_{q^{2}}%
(-\hspace{0.01in}q^{-1}y^{+})}{(q^{-2k_{3}}y^{-})^{k_{-}}(q^{-2k_{+}}%
y^{3})^{k_{3}}(\hspace{0.01in}y^{+})^{k_{+}}}+\mathcal{O}(\lambda).
\end{align}
Substituting this result into Eq.~(\ref{TraQDelL3DimAnf}) finally yields%
\begin{align}
\delta_{q}^{3}(\mathbf{x}\oplus\mathbf{y})\approx\,  &  \operatorname*{vol}%
\sum_{k_{+}=\hspace{0.01in}0}^{\infty}\sum_{k_{3}=\hspace{0.01in}0}^{\infty
}\sum_{k_{-}=\hspace{0.01in}0}^{\infty}q^{2k_{+}(k_{+}+\hspace{0.01in}%
1)+k_{3}(k_{3}+1)+2k_{-}(k_{-}+1)-4}\nonumber\\
&  \times\big (\hat{U}_{x}^{-1}\triangleright(x^{-})^{k_{-}}(x^{3})^{k_{3}%
}(x^{+})^{k_{+}}\big )\nonumber\\
&  \times\hat{U}_{y}^{-1}\triangleright\frac{\delta_{q^{2}}(-\hspace
{0.01in}q^{-2k_{3}+1}y^{-})\,\delta_{q}(q^{-2k_{+}}y^{3})\,\delta_{q^{2}%
}(-\hspace{0.01in}q^{-1}y^{+})}{(-\hspace{0.01in}q^{-2k_{3}}y^{-})^{k_{-}%
}(-\hspace{0.01in}q^{-2k_{+}}y^{3})^{k_{3}}(-\hspace{0.01in}y^{+})^{k_{+}}}.
\label{TraQDelL3DimZwi}%
\end{align}

To derive an approximate expression for $\delta_{q}^{3}((\ominus
\hspace{0.01in}\kappa^{-1}\mathbf{x})\oplus\mathbf{y})$, we first compute the
antipode of a monomial in the $x$-co\-or\-di\-nates. Neglecting terms
proportional to powers of $\lambda$, we find [cf. Eq.~(\ref{AntUnKonMon3dim})
in App.~\ref{KapExp}]:%
\begin{align}
\ominus((x^{-})^{k_{-}}(x^{3})^{k_{3}}(x^{+})^{k_{+}})=\, &  q^{-2k_{+}%
(k_{+}-1)-k_{3}(k_{3}-1)-2k_{-}(k_{-}-1)-2k_{3}(k_{+}+\hspace{0.01in}k_{-}%
)}\nonumber\\
&  \times(-x^{+})^{k_{+}}(-x^{3})^{k_{3}}(-x^{-})^{k_{-}}+\mathcal{O}%
(\lambda).\label{AntMonLam0NaeN}%
\end{align}
We omit the operator $\hat{U}_{x}^{-1}$, since the operator representation of the
antipode already corresponds to the normal ordering defined in
Eq.~(\ref{StePro0}) of App.~\ref{KapQuaZeiEle}. Making use of the scaling
properties of one-di\-men\-sion\-al $q$-delta functions [cf.
Eq.~(\ref{SkaDelFktEinQ}) in Chap.~\ref{KapQDisEin}], we find:%
\begin{gather}
\frac{\delta_{q^{2}}(-\hspace{0.01in}q^{-2k_{3}+1}y^{-})\,\delta
_{q}(q^{-2k_{+}}y^{3})\,\delta_{q^{2}}(-\hspace{0.01in}q^{-1}y^{+}%
))}{(q^{-2k_{3}}y^{-})^{k_{-}}(q^{-2k_{+}}y^{3})^{k_{3}}(\hspace{0.01in}%
y^{+})^{k_{+}}}=\nonumber\\
=q^{2k_{3}k_{-}+\hspace{0.01in}2k_{+}k_{3}+2k_{3}+2k_{+}}\,\frac{\delta
_{q^{2}}(-\hspace{0.01in}q\hspace{0.01in}y^{-})\,\delta_{q}(\hspace
{0.01in}y^{3})\,\delta_{q^{2}}(-\hspace{0.01in}q^{-1}y^{+})}{(\hspace
{0.01in}y^{-})^{k_{-}}(\hspace{0.01in}y^{3})^{k_{3}}(\hspace{0.01in}%
y^{+})^{k_{+}}}.\label{BruDelPot3Dim}%
\end{gather}
Proceeding in analogy to Eq.~(\ref{UmoDelFkt}), we obtain:%
\begin{align}
&  \hat{U}_{y}^{-1}\triangleright\frac{\delta_{q^{2}}(-\hspace{0.01in}%
q\hspace{0.01in}y^{-})\,\delta_{q}(\hspace{0.01in}y^{3})\,\delta_{q^{2}%
}(-\hspace{0.01in}q^{-1}y^{+})}{(\hspace{0.01in}y^{-})^{k_{-}}(\hspace
{0.01in}y^{3})^{k_{3}}(\hspace{0.01in}y^{+})^{k_{+}}}=\nonumber\\
&  \qquad=q^{2\hat{n}_{3}(\hat{n}_{+}+\hspace{0.01in}\hat{n}_{-})}%
\,\frac{\delta_{q^{2}}(-\hspace{0.01in}q^{-1}y^{+})\,\delta_{q}(\hspace
{0.01in}y^{3})\,\delta_{q^{2}}(-\hspace{0.01in}q\hspace{0.01in}y^{-}%
)}{(\hspace{0.01in}y^{+})^{k_{+}}(\hspace{0.01in}y^{3})^{k_{3}}(\hspace
{0.01in}y^{-})^{k_{-}}}+\mathcal{O}(\lambda)\nonumber\\
&  \qquad\approx\frac{1}{\operatorname*{vol}\nolimits_{q^{2}}}\int\text{d}%
_{q}\hspace{0.01in}p^{3}\sum_{n_{3}\hspace{0.01in}=\hspace{0.01in}0}^{\infty
}\frac{(\text{i}\hspace{0.01in}p^{3})^{n_{3}}(\hspace{0.01in}y^{3}%
)^{n_{3}-k_{3}}}{[[\hspace{0.01in}n_{3}]]_{q^{2}}!}\hspace{0.01in}\nonumber\\
&  \qquad\hspace{0.17in}\times\frac{\delta_{q^{2}}(-\hspace{0.01in}%
q^{2(n_{3}-\hspace{0.01in}k_{3})-1}y^{+})}{(\hspace{0.01in}q^{2(n_{3}-k_{3}%
)}y^{+})^{k_{+}}}\hspace{0.01in}\frac{\delta_{q^{2}}(-\hspace{0.01in}%
q^{2(n_{3}-\hspace{0.01in}k_{3})+1}y^{-})}{(\hspace{0.01in}q^{2(n_{3}-k_{3}%
)}y^{-})^{k_{-}}}.
\end{align}
Using the same method as in Eq.~(\ref{UmoDelFkt2}), this expression can be
rewritten as:%
\begin{align}
&  \frac{1}{\operatorname*{vol}\nolimits_{q^{2}}}\int\text{d}_{q}%
\hspace{0.01in}p^{3}\sum_{n_{3}\hspace{0.01in}=\hspace{0.01in}0}^{\infty}%
\frac{(\text{i}\hspace{0.01in}p^{3})^{n_{3}}(\hspace{0.01in}q^{-4-2k_{-}%
-2k_{+}}y^{3})^{n_{3}-k_{3}}}{[[\hspace{0.01in}n_{3}]]_{q^{2}}!}%
\hspace{0.01in}\frac{\delta_{q^{2}}(-\hspace{0.01in}q^{-1}y^{+})\,\delta
_{q^{2}}(-\hspace{0.01in}q\hspace{0.01in}y^{-})}{(\hspace{0.01in}y^{+}%
)^{k_{+}}\hspace{0.01in}(\hspace{0.01in}y^{-})^{k_{-}}}=\nonumber\\
&  \qquad\qquad=\frac{\delta_{q}(q^{-4-2k_{-}-2k_{+}}y^{3})}{(\hspace
{0.01in}q^{-4-2k_{-}-2k_{+}}y^{3})^{k_{3}}}\frac{\delta_{q^{2}}(-\hspace
{0.01in}q^{-1}y^{+})\hspace{0.01in}\delta_{q^{2}}(-\hspace{0.01in}%
q\hspace{0.01in}y^{-})}{(\hspace{0.01in}y^{+})^{k_{+}}\hspace{0.01in}%
(\hspace{0.01in}y^{-})^{k_{-}}}\nonumber\\
&  \qquad\qquad=q^{2(k_{+}+\hspace{0.01in}k_{-}+2)(k_{3}+1)}\hspace
{0.01in}\frac{\delta_{q^{2}}(-\hspace{0.01in}q^{-1}y^{+})\hspace{0.01in}%
\delta_{q}(y^{3})\hspace{0.01in}\delta_{q^{2}}(-\hspace{0.01in}q\hspace
{0.01in}y^{-})}{(\hspace{0.01in}y^{+})^{k_{+}}\hspace{0.01in}(\hspace
{0.01in}y^{3})^{k_{3}}\hspace{0.01in}(\hspace{0.01in}y^{-})^{k_{-}}%
}.\label{UmoProDelFin}%
\end{align}
Combining Eqs.~(\ref{TraQDelL3DimZwi})-(\ref{UmoProDelFin}), we finally arrive
at the approximate expression\footnote{The constant $\kappa$ takes the value
$q^{6}$ \cite{Wachter:2019A}.}%
\begin{align}
\delta_{q}^{3}((\ominus\hspace{0.01in}\kappa^{-1}\mathbf{x})\oplus
\mathbf{y})\approx &  \,\operatorname*{vol}\sum_{k_{+}=\hspace{0.01in}%
0}^{\infty}\sum_{k_{3}=\hspace{0.01in}0}^{\infty}\sum_{k_{-}=\hspace{0.01in}%
0}^{\infty}q^{2(k_{+}+\hspace{0.01in}k_{-})k_{3}}\hspace{0.01in}(q^{2}%
x^{+})^{k_{+}}(q^{2}x^{3})^{k_{3}}(x^{-})^{k_{-}}\nonumber\\
&  \qquad\qquad\qquad\times\frac{\delta_{q^{2}}(-\hspace{0.01in}q^{-1}%
y^{+})\,\delta_{q}(\hspace{0.01in}y^{3})\,\delta_{q^{2}}(-\hspace
{0.01in}q\hspace{0.01in}y^{-})}{(\hspace{0.01in}y^{+})^{k_{+}}(\hspace
{0.01in}y^{3})^{k_{3}}(\hspace{0.01in}y^{-})^{k_{-}}}%
.\label{DarDelDreiDimLam0Del1}%
\end{align}

\subsection{Consistency Checks\label{AppExp}}

We demonstrate that the approximate expression in
Eq.~(\ref{DarDelDreiDimLam0Del1}) is consistent with the star product on
three-di\-men\-sion\-al $q$-de\-formed Euclidean space [cf.
Eq.~(\ref{StaProForExp}) in App.~\ref{KapQuaZeiEle}] when all terms involving
powers of $\lambda$ are neglected. In particular, we verify the identities%
\begin{align}
f(\mathbf{y})  &  =\operatorname*{vol}\nolimits^{-1}\hspace{-0.02in}%
\int\text{d}_{q}^{3}x\,f(\mathbf{x})\circledast\delta_{q}^{3}((\ominus
\hspace{0.01in}\kappa^{-1}\mathbf{x})\oplus\mathbf{y})\nonumber\\
&  =\operatorname*{vol}\nolimits^{-1}\hspace{-0.02in}\int\text{d}_{q}%
^{3}x\,\delta_{q}^{3}((\ominus\hspace{0.01in}\kappa^{-1}\mathbf{y}%
)\oplus\mathbf{x})\circledast f(\mathbf{x}), \label{AlgChaIdeqDelFkt}%
\end{align}
where $f$ is assumed to be analytic around the origin and thus expandable in a
power series. Hence, it suffices to verify these identities for monomials in
the $x$-co\-or\-di\-nates.

Using Eqs.~(\ref{TraQDelL3DimZwi})-(\ref{BruDelPot3Dim}) together with the
one-di\-men\-sion\-al $q$-delta functions in Eqs.~(\ref{AltDar1DimTraDelFkt0})
and (\ref{DarQDelKroDelAlg1Dim}) of Chap.~\ref{KapQDisEin}, we obtain the
approximate expression%
\begin{gather}
\operatorname*{vol}\nolimits^{-1}\hspace{-0.01in}\delta_{q}^{3}((\ominus
\hspace{0.01in}\kappa^{-1}\mathbf{x})\oplus\mathbf{y})\approx\nonumber\\
\approx q^{-4}\sum_{k_{+}=\hspace{0.01in}0}^{\infty}\sum_{k_{3}=\hspace
{0.01in}0}^{\infty}\sum_{k_{-}=\hspace{0.01in}0}^{\infty}(-\hspace
{0.01in}q^{-1}x^{+})^{k_{+}}(q^{-2}x^{3})^{k_{3}}(-\hspace{0.01in}%
q\hspace{0.01in}q^{-2}x^{-})^{k_{-}}\nonumber\\
\times\,\hat{U}_{y}^{-1}\triangleright\frac{\delta_{q^{2}}(-\hspace
{0.01in}q\hspace{0.01in}y^{-})\,\delta_{q}(\hspace{0.01in}y^{3})\,\delta
_{q^{2}}(-\hspace{0.01in}q^{-1}y^{+})}{(-\hspace{0.01in}q\hspace{0.01in}%
y^{-})^{k_{-}}(\hspace{0.01in}y^{3})^{k_{3}}(-\hspace{0.01in}q^{-1}%
y^{+})^{k_{+}}}\nonumber\\
=q^{-4}\hat{U}_{y}^{-1}\triangleright\frac{\phi_{q^{2}}(x^{+}\hspace
{-0.01in},y^{+})}{\left\vert \hspace{0.01in}y^{+}(1-q^{2})\right\vert }%
\frac{\phi_{q}(q^{-2}x^{3}\hspace{-0.01in},y^{3})}{\left\vert \hspace
{0.01in}y^{3}(1-q)\right\vert }\frac{\phi_{q^{2}}(q^{-2}x^{-}\hspace
{-0.01in},y^{-})}{\left\vert \hspace{0.01in}y^{-}(1-q^{2})\right\vert }.
\label{AltDarDelDreiDimLam0}%
\end{gather}
Using this approximation, we evaluate:%
\begin{align}
&  \operatorname*{vol}\nolimits^{-1}\hspace{-0.02in}\int\text{d}_{q}%
^{3}x\,(x^{+})^{n_{+}}(x^{3})^{n_{3}}(x^{-})^{n_{-}}\hspace{-0.01in}%
\circledast\delta_{q}^{3}((\ominus\hspace{0.01in}\kappa^{-1}\mathbf{x}%
)\oplus\mathbf{y})\approx\nonumber\\
&  \qquad\approx q^{-4}\hspace{0.01in}\hat{U}_{y}^{-1}\triangleright
\int\text{d}_{q}^{3}x\,q^{2\hat{n}_{x^{3}}\hat{n}_{x^{\prime+}}+\hspace
{0.01in}2\hat{n}_{x^{-}}\hat{n}_{x^{\prime3}}}\triangleright(x^{+})^{n_{+}%
}(x^{3})^{n_{3}}(x^{-})^{n_{-}}\nonumber\\
&  \qquad\qquad\qquad\times\left.  \frac{\phi_{q^{2}}(x^{\hspace{0.01in}%
\prime+}\hspace{-0.01in},y^{+})}{\left\vert \hspace{0.01in}y^{+}%
(1-q^{2})\right\vert }\frac{\phi_{q}(q^{-2}x^{\hspace{0.01in}\prime
\hspace{0.01in}3}\hspace{-0.01in},y^{3})}{\left\vert \hspace{0.01in}%
y^{3}(1-q)\right\vert }\frac{\phi_{q^{2}}(q^{-2}x^{\hspace{0.01in}\prime
-}\hspace{-0.01in},y^{-})}{\left\vert \hspace{0.01in}y^{-}(1-q^{2})\right\vert
}\right\vert _{\mathbf{x}^{\prime}\rightarrow\hspace{0.01in}\mathbf{x}%
}\nonumber\\
&  \qquad=q^{-4}\hspace{0.01in}\hat{U}_{y}^{-1}\triangleright\int\text{d}%
_{q}^{3}x\,(-\hspace{0.01in}q\hspace{0.01in}x_{-})^{n_{+}}(x_{3})^{n_{3}%
}(-\hspace{0.01in}q^{-1}x_{+})^{n_{-}}\nonumber\\
&  \qquad\qquad\qquad\times\frac{\phi_{q^{2}}(q^{2n_{3}}x_{-},y_{-}%
)}{\left\vert \hspace{0.01in}y_{-}(1-q^{2})\right\vert }\frac{\phi
_{q}(q^{-2+2n_{-}}x_{3},y_{3})}{\left\vert \hspace{0.01in}y_{3}%
(1-q)\right\vert }\frac{\phi_{q^{2}}(q^{-2}x_{+},y_{+})}{\left\vert
\hspace{0.01in}y_{+}(1-q^{2})\right\vert }\nonumber\\[0.04in]
&  \qquad=q^{-4}\hspace{0.01in}\hat{U}_{y}^{-1}\triangleright q^{-2n_{3}%
}(-\hspace{0.01in}q\hspace{0.01in}q^{-2n_{3}}y_{-})^{n_{+}}q^{2-2n_{-}%
}(q^{2-2n_{-}}y_{3})^{n_{3}}q^{2}(-\hspace{0.01in}q^{-1}q^{2}y_{+})^{n_{-}%
}\nonumber\\[0.04in]
&  \qquad\approx q^{-2n_{3}(n_{+}+\hspace{0.01in}n_{-})}q^{2\hat{n}_{3}%
(\hat{n}_{+}+\hspace{0.01in}\hat{n}_{-})}\triangleright(\hspace{0.01in}%
y^{-})^{n_{-}}(\hspace{0.01in}y^{3})^{n_{3}}(\hspace{0.01in}y^{+})^{n_{+}%
}\nonumber\\
&  \qquad=(\hspace{0.01in}y^{+})^{n_{+}}(\hspace{0.01in}y^{3})^{n_{3}}%
(\hspace{0.01in}y^{-})^{n_{-}}. \label{CharIdeDelDrei1}%
\end{align}
In the first two steps, we applied the operator representation of the star
product [cf. Eq.~(\ref{StaProForExp}) in Chap.~\ref{KapQuaZeiEle}] while
neglecting all $\lambda$-de\-pen\-dent terms. In the third step, we used the
following identities valid for $y\in\mathbb{G}_{q,\hspace{0.01in}x_{0}}$ [cf.
Eqs.~(\ref{ChaIdePhi1}) and (\ref{ChaIdePhi2}) in Chap.~\ref{KapQDisEin}]:%
\begin{gather}
\int\text{d}_{q}x\,f(x)\frac{\phi_{q}(q^{\hspace{0.01in}n}x,y)}{\left\vert
\hspace{0.01in}y\hspace{0.01in}(1-q)\right\vert }=\sum_{\varepsilon
\hspace{0.01in}=\hspace{0.01in}\pm}^{\infty}\sum_{j\hspace{0.01in}%
=\hspace{0.01in}-\infty}^{\infty}\left\vert \hspace{0.01in}\varepsilon
x_{0}\hspace{0.01in}q^{\hspace{0.01in}j}(1-q)\right\vert \,f(\varepsilon
x_{0}\hspace{0.01in}q^{\hspace{0.01in}j})\frac{\phi_{q}(\varepsilon x_{0}\hspace{0.01in}q^{n+j}%
\hspace{-0.01in},y)}{\left\vert \hspace{0.01in}y\hspace{0.01in}%
(1-q)\right\vert }\nonumber\\
=\left\vert \hspace{0.01in}q^{-n}y\hspace{0.01in}(1-q)\right\vert
\,f(q^{-n}y)\frac{1}{\left\vert \hspace{0.01in}y\hspace{0.01in}%
(1-q)\right\vert }=q^{-n}f(q^{-n}y).
\end{gather}
In the final two steps of Eq.~(\ref{CharIdeDelDrei1}), the action of $\hat
{U}^{-1}$ was written explicitly [cf. Eq.~(\ref{UmOrdInvOpeDreQuan}) in
Chap.~\ref{KapQuaZeiEle}], again neglecting all $\lambda$-de\-pen\-dent contributions.

Using the approximation in Eq.~(\ref{DarDelDreiDimLam0Del1}) of the previous
subsection, we similarly evaluate:%
\begin{align}
&  \operatorname*{vol}\nolimits^{-1}\hspace{-0.02in}\int\text{d}_{q}%
^{3}y\,\delta_{q}^{3}((\ominus\hspace{0.01in}\kappa^{-1}\mathbf{x}%
)\oplus\mathbf{y})\circledast(\hspace{0.01in}y^{+})^{n_{+}}(\hspace
{0.01in}y^{3})^{n_{3}}(\hspace{0.01in}y^{-})^{n_{-}}\approx\nonumber\\
&  \qquad\approx\int\text{d}_{q}^{3}y\,\sum_{k_{+}=\hspace{0.01in}0}^{\infty
}\sum_{k_{3}=\hspace{0.01in}0}^{\infty}\sum_{k_{-}=\hspace{0.01in}0}^{\infty
}q^{2(k_{+}+\hspace{0.01in}k_{-})k_{3}}\hspace{0.01in}(q^{2}x^{+})^{k_{+}%
}(q^{2}x^{3})^{k_{3}}(x^{-})^{k_{-}}\nonumber\\
&  \qquad\qquad\qquad\times q^{2\hat{n}_{y^{3}}\hat{n}_{y^{\prime+}}%
+\hspace{0.01in}2\hat{n}_{y^{-}}\hat{n}_{y^{\prime3}}}\triangleright
\frac{\delta_{q^{2}}(-\hspace{0.01in}q^{-1}y^{+})\,\delta_{q}(\hspace
{0.01in}y^{3})\,\delta_{q^{2}}(-\hspace{0.01in}q\hspace{0.01in}y^{-}%
)}{(\hspace{0.01in}y^{+})^{k_{+}}(\hspace{0.01in}y^{3})^{k_{3}}(\hspace
{0.01in}y^{-})^{k_{-}}}\nonumber\\
&  \qquad\qquad\qquad\times\left.  (\hspace{0.01in}y^{\prime+})^{n_{+}%
}(\hspace{0.01in}y^{\prime3})^{n_{3}}(\hspace{0.01in}y^{\prime-})^{n_{-}%
}\right\vert _{\mathbf{y}^{\prime}\rightarrow\hspace{0.01in}\mathbf{y}%
}\nonumber\\
&  \qquad=\int\text{d}_{q}^{3}y\,\sum_{k_{+}=\hspace{0.01in}0}^{\infty}%
\sum_{k_{3}=\hspace{0.01in}0}^{\infty}\sum_{k_{-}=\hspace{0.01in}0}^{\infty
}q^{2(k_{+}+\hspace{0.01in}k_{-})k_{3}}\hspace{0.01in}(q^{2}x^{+})^{k_{+}%
}(q^{2}x^{3})^{k_{3}}(x^{-})^{k_{-}}\nonumber\\
&  \qquad\qquad\qquad\times\frac{\delta_{q^{2}}(-\hspace{0.01in}q^{-1}%
y^{+})\,\delta_{q}(\hspace{0.01in}q^{2n_{+}}y^{3})\,\delta_{q^{2}}%
(-\hspace{0.01in}q\hspace{0.01in}q^{2n_{3}}y^{-})}{(\hspace{0.01in}%
y^{+})^{k_{+}-\hspace{0.01in}n_{+}}q^{2n_{+}k_{3}}(\hspace{0.01in}%
y^{3})^{k_{3}-n_{3}}q^{2n_{3}k_{-}}(\hspace{0.01in}y^{-})^{k_{-}%
-\hspace{0.01in}n_{-}}}. \label{CharIdeDelFkt3}%
\end{align}
Using the scaling properties of the one-di\-men\-sion\-al $q$-delta functions,
the last expression in Eq.~(\ref{CharIdeDelFkt3}) can be rewritten as follows:%
\begin{align}
&  \int\text{d}_{q}^{3}y\,\sum_{k_{+}=\hspace{0.01in}0}^{\infty}\sum
_{k_{3}=\hspace{0.01in}0}^{\infty}\sum_{k_{-}=\hspace{0.01in}0}^{\infty
}q^{2(k_{+}+\hspace{0.01in}k_{-})k_{3}}\hspace{0.01in}(q^{2}x^{+})^{k_{+}%
}(q^{2}x^{3})^{k_{3}}(x^{-})^{k_{-}}\nonumber\\
&  \qquad\qquad\times\frac{\delta_{q^{2}}(\hspace{0.01in}y_{-})\,q^{-2n_{+}%
}\delta_{q}(\hspace{0.01in}y_{3})\,q^{-2n_{3}}\delta_{q^{2}}(\hspace
{0.01in}y_{+})}{q^{2n_{+}k_{3}+2n_{3}k_{-}}(-\hspace{0.01in}q)^{k_{+}%
-\hspace{0.01in}n_{+}-\hspace{0.01in}k_{-}+\hspace{0.01in}n_{-}}%
(\hspace{0.01in}y_{-})^{k_{+}-\hspace{0.01in}n_{+}}(\hspace{0.01in}%
y_{3})^{k_{3}-\hspace{0.01in}n_{3}}(\hspace{0.01in}y_{-})^{k_{-}%
-\hspace{0.01in}n_{-}}}\nonumber\\
&  \qquad=\sum_{k_{+}=\hspace{0.01in}0}^{\infty}\sum_{k_{3}=\hspace{0.01in}%
0}^{\infty}\sum_{k_{-}=\hspace{0.01in}0}^{\infty}q^{2(k_{+}+\hspace
{0.01in}k_{-})k_{3}-2n_{+}k_{3}-2n_{3}k_{-}}\hspace{0.01in}(q^{2}x^{+}%
)^{k_{+}}(q^{2}x^{3})^{k_{3}}(x^{-})^{k_{-}}\nonumber\\
&  \qquad\qquad\times(-\hspace{0.01in}q)^{-k_{+}+\hspace{0.01in}n_{+}%
+\hspace{0.01in}k_{-}-\hspace{0.01in}n_{-}}\hspace{0.01in}q^{-2n_{+}%
-\hspace{0.01in}2n_{3}}\hspace{0.01in}\delta_{k_{+},n_{+}}\hspace
{0.01in}\delta_{k_{3},n_{3}}\hspace{0.01in}\delta_{k_{-},n_{-}}\nonumber\\
&  \qquad=q^{2(n_{+}+\hspace{0.01in}n_{-})n_{3}-2n_{+}n_{3}-2n_{3}n_{-}%
}\hspace{0.01in}q^{-2n_{+}-\hspace{0.01in}2n_{3}}\hspace{0.01in}(q^{2}%
x^{+})^{n_{+}}(q^{2}x^{3})^{n_{3}}(x^{-})^{n_{-}}\nonumber\\
&  \qquad=(x^{+})^{n_{+}}(x^{3})^{n_{3}}(x^{-})^{n_{-}}.
\end{align}
Here, we changed to $y$-co\-or\-di\-nates with lower indices and used
Eq.~(\ref{CharIdeEinDimQDel5}) of Chap.~\ref{KapQDisEin}.

\section{Vacuum Energy\label{KapVakEne}}

\subsection{Classical Calculation of the Zero-Point Energy\label{ClasCal}}

The quantum harmonic oscillator possesses a non-van\-ish\-ing ground-state
energy, known as the zero-point energy. A quantum field can be regarded as an
infinite collection of such oscillators, and the sum of their zero-point
energies yields the vacuum energy associated with the quantum field. We focus
here on the vacuum energy of a free scalar field in Euclidean space.

For a Klein-Gordon field, the vacuum energy arises from virtual scalar
particles that momentarily emerge from the vacuum state $|0\rangle$. Using
creation and annihilation operators acting on the vacuum state, the vacuum
energy can be expressed as \cite{Weinberg:1989,Mukhanov:2007zz,Maggiore:2010}:%
\begin{align}
\langle H\rangle_{\operatorname{vac}}  &  =\frac{1}{(2\pi\hbar)^{3}}%
\int\text{d}^{3}x\int\text{d}^{3}p\,\frac{1}{2}E_{\mathbf{p}}\,\langle
0|\hat{a}_{\mathbf{p}}\operatorname{e}^{\frac{\text{i}}{\hbar}\mathbf{p}%
\cdot\mathbf{x}}\operatorname{e}^{-\frac{\text{i}}{\hbar}\mathbf{p}%
\cdot\mathbf{x}}\hat{a}_{\mathbf{p}}^{\dagger}|0\rangle\nonumber\\
&  =\frac{1}{(2\pi\hbar)^{3}}\int\text{d}^{3}x\int\text{d}^{3}%
p\,\operatorname{e}^{\mathbf{0}\cdot\mathbf{x}}\hspace{0.01in}\frac{1}%
{2}E_{\mathbf{p}}\nonumber\\
&  =\int\frac{\text{d}^{3}p}{(2\pi\hbar)^{3}}\,(2\pi)^{3}\hspace{0.01in}%
\delta^{3}(\mathbf{0})\hspace{0.01in}\frac{1}{2}E_{\mathbf{p}}.
\label{VacEneCla}%
\end{align}
Here, $E_{\mathbf{p}}$ is the energy of a free scalar particle of momentum
$\mathbf{p}$ and mass $m$:%
\begin{equation}
E_{\mathbf{p}}=\sqrt{c^{\hspace{0.01in}2}\hspace{-0.02in}\left\vert
\mathbf{p}\right\vert ^{2}\hspace{-0.02in}+m^{2}c^{\hspace{0.01in}4}}.
\end{equation}
The divergent factor $(2\pi)^{3}\delta^{3}(\mathbf{0})$
in\ Eq.~(\ref{VacEneCla}) originates from the integration over all of
Euclidean space:%
\begin{equation}
(2\pi)^{3}\hspace{0.01in}\delta^{3}(\mathbf{0})=\lim\limits_{V\rightarrow
\hspace{0.01in}\infty}\int_{V}\text{d}^{3}x\,\operatorname{e}^{\mathbf{0}%
\cdot\mathbf{x}}=\lim\limits_{V\rightarrow\hspace{0.01in}\infty}V.
\end{equation}
Accordingly, the momentum integral in the final expression of
Eq.~(\ref{VacEneCla}) may be interpreted as the vacuum energy density:%
\begin{equation}
\rho_{0}=\lim_{V\rightarrow\hspace{0.01in}\infty}\frac{\langle H\rangle
_{\operatorname{vac}}}{V}=\int\frac{\text{d}^{3}p}{(2\pi\hbar)^{3}}\,\frac
{1}{2}E_{\mathbf{p}}. \label{VacEneDic}%
\end{equation}

The integral in Eq.~(\ref{VacEneDic}) is divergent. To compute the vacuum
energy density, we confine the quantum field within a box of volume $V=L^{3}$
and introduce the Planck length $\ell_{P}$ as the smallest physically meaningful length scale:%
\begin{equation}
\ell_{P}=\sqrt{\frac{\hbar\hspace{0.01in}G}{c^{\hspace{0.01in}3}}}=1.616255\times
10^{-35}\text{m}. \label{WerPlaLae}%
\end{equation}
Since the virtual particles are restricted to the finite box volume, their
momentum components can assume only discrete values:%
\begin{equation}
p_{i}=\frac{2\pi\hbar}{L}\hspace{0.01in}n_{i},\qquad n_{i}=0,\pm1,\ldots
\end{equation}
Additionally, the wavelengths of the normal modes of the Klein-Gordon field
cannot be shorter than the Planck length. Hence, the maximum momentum of a
virtual particle is estimated as%
\begin{equation}
\left\vert \mathbf{p}\right\vert =\frac{2\pi\hbar}{\lambda}\leq\frac{2\pi
\hbar}{\ell_{P}}=p_{\max}.
\end{equation}
Using these assumptions, the vacuum energy density can be computed as
follows:\footnote{As pointed out in Ref.~\cite{akhmedov2002vacuum}, the
expression for the vacuum energy density in Eq.~(\ref{BerEneKlas}) violates
relativistic Lorentz invariance of the vacuum. Ref.~\cite{Maggiore_2011}
suggests that this issue may be resolved through renormalization. Here, we
focus on Eq.~(\ref{BerEneKlas}) because it provides a direct link to the
zero-point energy considerations in $q$-de\-formed Euclidean space.}%
\begin{align}
\rho_{0}  &  =\lim_{V\rightarrow\hspace{0.01in}\infty}\frac{\langle
H\rangle_{\operatorname{vac}}}{V}=\lim_{V\rightarrow\hspace{0.01in}\infty
}\frac{1}{V}\sum\nolimits_{n}\frac{1}{2}E_{\mathbf{p}_{n}}\nonumber\\
&  =\lim_{V\rightarrow\hspace{0.01in}\infty}\frac{1}{V}\hspace{-0.02in}%
\int\text{d}^{3}n\,\frac{1}{2}E_{\mathbf{p}}=\lim_{V\rightarrow\hspace
{0.01in}\infty}\frac{1}{V}\frac{V}{(2\pi\hbar)^{3}}\int\text{d}^{3}p\,\frac
{1}{2}E_{\mathbf{p}}\nonumber\\
&  =\frac{1}{(2\pi\hbar)^{3}}\,4\pi\hspace{-0.02in}\int_{0}^{p_{\max}%
}\text{d\hspace{-0.01in}}\left\vert \mathbf{p}\right\vert \,\frac{1}%
{2}\text{\hspace{-0.01in}}\left\vert \mathbf{p}\right\vert ^{2}\sqrt
{c^{\hspace{0.01in}2}\hspace{-0.01in}\left\vert \mathbf{p}\right\vert
^{2}+m^{2}c^{\hspace{0.01in}4}}\nonumber\\
&  \approx\frac{2\pi}{(2\pi\hbar)^{3}}\int_{0}^{p_{\max}}\text{d\hspace
{-0.01in}}\left\vert \mathbf{p}\right\vert \,c\hspace{-0.01in}\left\vert
\mathbf{p}\right\vert ^{3}=\frac{\pi c\hspace{0.01in}p_{\max}^{4}}{2(2\pi
\hbar)^{3}}=\pi^{2}\frac{\hbar\hspace{0.01in}c}{\ell_{P}^{\,4}}\approx
1.2\times10^{114}\,\text{J/m}^{3}. \label{BerEneKlas}%
\end{align}
This result is grossly incompatible with the observed vacuum energy density of
order 10$^{-9}$J$/$m$^{3}$, as inferred from cosmological data
\cite{Weinberg:1989}. Resolving this discrepancy is beyond our present scope.
However, as will be shown in the following chapters, in $q$-de\-formed
Euclidean space a small vacuum energy density at large scales can coexist with
a large vacuum energy density at small (Planck) scales.

In $q$-de\-formed Euclidean space, evaluating $\langle H\rangle
_{\operatorname{vac}}$ is considerably more involved than evaluating $\langle
H^{2}\rangle_{\operatorname{vac}}$. We demonstrate that the order of magnitude
of the vacuum energy density given in Eq.~(\ref{BerEneKlas}) can also be
estimated from $\langle H^{2}\rangle_{\operatorname{vac}}$. Following the same
reasoning as in Eq.~(\ref{BerEneKlas}) yields%
\begin{align}
\lim_{V\rightarrow\hspace{0.01in}\infty}\frac{\langle H^{2}\rangle
_{\operatorname{vac}}}{V}  &  =\lim_{V\rightarrow\hspace{0.01in}\infty}%
\frac{1}{V}\sum\nolimits_{n}\frac{1}{4}E_{\mathbf{p}_{n}}^{2}=\lim
_{V\rightarrow\hspace{0.01in}\infty}\frac{1}{V}\hspace{-0.02in}\int
\text{d}^{3}n\,\frac{1}{4}E_{\mathbf{p}}^{2}\nonumber\\
&  =\lim_{V\rightarrow\hspace{0.01in}\infty}\frac{1}{V}\frac{V}{(2\pi
\hbar)^{3}}\int\text{d}^{3}p\,\frac{1}{4}E_{\mathbf{p}}^{2}\nonumber\\
&  =\frac{1}{(2\pi\hbar)^{3}}\,4\pi\hspace{-0.02in}\int_{0}^{p_{\max}%
}\text{d\hspace{-0.01in}}\left\vert \mathbf{p}\right\vert \,\frac{1}%
{4}\text{\hspace{-0.01in}}\left\vert \mathbf{p}\right\vert ^{2}(c^{\hspace
{0.01in}2}\hspace{-0.01in}\left\vert \mathbf{p}\right\vert ^{2}+m^{2}%
c^{\hspace{0.01in}4})\nonumber\\
&  \approx\frac{\pi}{(2\pi\hbar)^{3}}\int_{0}^{p_{\max}}\text{d\hspace
{-0.01in}}\left\vert \mathbf{p}\right\vert \,c^{\hspace{0.01in}2}%
\hspace{-0.01in}\left\vert \mathbf{p}\right\vert ^{4}=\frac{\pi c^{\hspace
{0.01in}2}\hspace{0.01in}p_{\max}^{5}}{5(2\pi\hbar)^{3}}=\frac{(2\pi)^{3}}%
{10}\frac{\hbar^{2}\hspace{0.01in}c^{\hspace{0.01in}2}}{\ell_{P}^{\,5}}.
\label{DicEneQua}%
\end{align}
Interpreting $\ell_{P}^{\,3}$ as the minimal spatial volume $V_{\min}$, we obtain
from Eq.~(\ref{DicEneQua}) the following estimate for the vacuum energy
density:%
\begin{align}
\rho_{0}  &  \approx\frac{\sqrt{V_{\min}\cdot\lim_{V\rightarrow\hspace
{0.01in}\infty}\frac{\langle H^{2}\rangle_{\operatorname{vac}}}{V}}}{V_{\min}%
}=\frac{\sqrt{\ell_{P}^{\,3}\cdot\frac{(2\pi)^{3}}{10}\frac{\hbar^{2}%
\hspace{0.01in}c^{\hspace{0.01in}2}}{\ell_{P}^{\,5}}}}{\ell_{P}^{\,3}}\nonumber\\
&  =\frac{(2\pi)^{3/2}}{\sqrt{10}}\frac{\hbar\hspace{0.01in}c}{\ell_{P}^{\,4}%
}\approx2.3\times10^{114}\,\text{J/m}^{3}. \label{BerEneQuaDenCla}%
\end{align}
This result differs only slightly from Eq.~(\ref{BerEneKlas}), supporting the
use of this method in the $q$-de\-formed Euclidean setting (see
Chap.~\ref{VacEneQuaPoi}).

\subsection{Vacuum Energy for a $q$-De\-formed Scalar Field\label{VacEneQEuc}}

Our objective is to determine the vacuum energy of the $q$-de\-formed
Euclidean space resulting from a $q$-de\-formed scalar field governed by the
Hamiltonian operator $H$. For this purpose, we examine the matrix elements of
$H$:%
\begin{align}
\langle\mathbf{x}|H|\mathbf{x}^{\prime}\rangle &  =\int\text{d}_{q}%
^{3}p\,\langle\mathbf{x}|\mathbf{p}\rangle\overset{p}{\circledast
}E_{\mathbf{p}}\overset{p}{\circledast}\langle\mathbf{p}|\mathbf{x}^{\prime
}\rangle\nonumber\\
&  =\int\text{d}_{q}^{3}p\,u_{\mathbf{p}}(\mathbf{x})\overset{p}{\circledast
}E_{\mathbf{p}}\overset{p}{\circledast}(u^{\ast})_{\mathbf{p}}(\mathbf{x}%
^{\prime}). \label{VacErwHamImp}%
\end{align}
Here, $u_{\mathbf{p}}(\mathbf{x})$ and $(u^{\ast})_{\mathbf{p}}(\mathbf{x}%
^{\prime})$ denote the $q$-de\-formed momentum eigenfunctions, defined in
Eqs.~(\ref{ImpEigFktqDefN}) and (\ref{DefDuaImpEigFktWdh}) of
App.~\ref{KapVolRelImpOrt}, respectively. The quantity $E_{\mathbf{p}}$
denotes the energy of a $q$-de\-formed scalar particle with momentum
$\mathbf{p}$. For a massive scalar particle, we assume that $E_{\mathbf{p}}$
admits a formal pow\-er-se\-ries expansion of the general form [cf.
Eq.~(\ref{EneKleGorSer}) in App.~\ref{KapPlaWavSol}]:%
\begin{equation}
E_{\mathbf{p}}=\sum\nolimits_{n\hspace{0.01in}=\hspace{0.01in}0}^{\infty}%
a_{n}\cdot\mathbf{p}^{2n}. \label{AllEntEngN}%
\end{equation}
Inserting this expansion into Eq.~(\ref{VacErwHamImp}) allows us to express
the Hamiltonian matrix elements in terms of powers of the $q$-de\-formed
Laplace operator [cf. Eq.~(\ref{qLapOpe}) in App.~\ref{KapPlaWavSol}]:%
\begin{align}
\langle\mathbf{x}|H|\mathbf{x}^{\prime}\rangle &  =\sum_{n\hspace
{0.01in}=\hspace{0.01in}0}^{\infty}a_{n}\hspace{0.01in}\langle\mathbf{x}%
|\mathbf{p}^{2n}|\mathbf{x}^{\prime}\rangle=\sum_{n\hspace{0.01in}%
=\hspace{0.01in}0}^{\infty}a_{n}\int\text{d}_{q}^{3}p\,u_{\mathbf{p}%
}(\mathbf{x})\overset{p}{\circledast}\mathbf{p}^{2n}\hspace{-0.01in}%
\overset{p}{\circledast}(u^{\ast})_{\mathbf{p}}(\mathbf{x}^{\prime
})\nonumber\\
&  =\sum_{n\hspace{0.01in}=\hspace{0.01in}0}^{\infty}a_{n}\int\text{d}_{q}%
^{3}p\,\big(\text{i}^{-2}g_{AB}\hspace{0.01in}\partial_{x}^{\hspace{0.01in}%
A}\partial_{x}^{\hspace{0.01in}B}\big)^{n}\triangleright\,u_{\hspace
{0.01in}\mathbf{p}}(\mathbf{x})\circledast(u^{\ast})_{\mathbf{p}}%
(\mathbf{x}^{\prime})\nonumber\\
&  =\operatorname*{vol}\nolimits^{-1}\sum_{n\hspace{0.01in}=\hspace{0.01in}%
0}^{\infty}a_{n}\big(-\mathbf{\partial}_{x}\circ\mathbf{\partial}_{x}%
\big)^{n}\triangleright\delta_{q}^{3}(\mathbf{x}\oplus(\ominus\hspace
{0.01in}\kappa^{-1}\mathbf{x}^{\prime})). \label{EntWicHamAlg}%
\end{align}
In obtaining this result, we have used the eigenvalue equations for the
$q$-de\-formed momentum eigenfunctions [cf. Eq.~(\ref{EigGleImpOpeImpEigFkt0})
in App.~\ref{KapVolRelImpOrt}] and their completeness relation [cf.
Eq.~(\ref{VolRelZeiWelDreDim}) in App.~\ref{KapVolRelImpOrt}].

To evaluate the action of $(-\,\mathbf{\partial}\circ\mathbf{\partial})^{n}$
on the $q$-delta function, we employ the identity [cf. Eq.~(\ref{EntPotP}) in
App.~\ref{KapPlaWavSol}]%
\begin{equation}
(-\,\mathbf{\partial}\circ\mathbf{\partial})^{n}=\sum_{v\hspace{0.01in}%
=\hspace{0.01in}0}^{n}\frac{(q+q^{-1})^{n-\hspace{0.01in}v}}{(-q^{2})^{v}%
}\hspace{0.01in}%
\genfrac{[}{]}{0pt}{}{n}{v}%
_{q^{4}}\,(\partial_{-})^{n-\hspace{0.01in}v}\,(\partial_{\hspace{0.01in}%
3})^{2\hspace{0.01in}v}\,(\partial_{+})^{n-\hspace{0.01in}v}.
\label{DarPotAblQuaDreEnt}%
\end{equation}
From Eqs.~(\ref{EntWicHamAlg}) and (\ref{DarPotAblQuaDreEnt}) it follows that
one must determine how monomials of partial derivatives act on the
$q$-de\-formed delta function $\delta_{q}^{3}(\mathbf{x}\oplus(\ominus
\hspace{0.01in}\kappa^{-1}\mathbf{y}))$. Using Eq.~(\ref{MehAblDreDim}) of
App.~\ref{KapParDer} and Eq.~(\ref{AltDarDelDreiDimLam0}) of
Chap.~\ref{AppExp}, one obtains the following approximate expression:%
\begin{gather}
\operatorname*{vol}\nolimits^{-1}\hspace{0.01in}(\partial_{-})^{m_{-}%
}(\partial_{\hspace{0.01in}3})^{m_{3}}(\partial_{+})^{m_{+}}\triangleright
\delta_{q}^{3}(\mathbf{x}\oplus(\ominus\hspace{0.01in}\kappa^{-1}%
\mathbf{\tilde{y}}))\approx\nonumber\\[0.02in]
\approx q^{-2m_{3}(m_{+}+\hspace{0.02in}m_{-})-4}\,d_{m_{-}}^{(-)}%
(x^{-}\hspace{-0.01in},\tilde{y}^{-})\,d_{m_{3},\hspace{0.01in}m_{-}}%
^{(3)}(x^{3}\hspace{-0.01in},\tilde{y}^{3})\,d_{m_{+},\hspace{0.01in}m_{3}%
}^{(+)}(x^{+}\hspace{-0.01in},\tilde{y}^{+}), \label{WirImpPotDreQDel}%
\end{gather}
where%
\begin{align}
d_{m_{-}}^{(-)}(x^{-}\hspace{-0.01in},\tilde{y}^{-})  &  =\frac{D_{q^{4}%
,\hspace{0.01in}x^{-}}^{m_{-}}\phi_{q^{2}}(q^{-2}x^{-}\hspace{-0.01in}%
,\tilde{y}^{-})}{\left\vert (1-q^{2})\hspace{0.01in}\tilde{y}^{-}\right\vert
},\nonumber\\[0.1in]
d_{m_{3},\hspace{0.01in}m_{-}}^{(3)}(x^{3}\hspace{-0.01in},\tilde{y}^{3})  &
=\frac{D_{q^{2},\hspace{0.01in}x^{3}}^{m_{3}}\phi_{q}(q^{2(m_{-}-1)}%
x^{3}\hspace{-0.01in},\tilde{y}^{3})}{\left\vert (1-q)\hspace{0.01in}\tilde
{y}^{3}\right\vert },\nonumber\\[0.1in]
d_{m_{+},\hspace{0.01in}m_{3}}^{(+)}(x^{+}\hspace{-0.01in},\tilde{y}^{+})  &
=\frac{D_{q^{4},\hspace{0.01in}x^{+}}^{m_{+}}\phi_{q^{2}}(q^{2m_{3}}%
x^{+}\hspace{-0.01in},\tilde{y}^{+})}{\left\vert (1-q^{2})\hspace
{0.01in}\tilde{y}^{+}\right\vert }. \label{WirImpPotDreQDelEin}%
\end{align}
As the operator $\hat{U}_{y}^{-1}$ has been omitted, the $\tilde{y}%
$-co\-or\-di\-nates in the above expressions are to be understood with respect
to the normal ordering defined in Eq.~(\ref{InvNorOrd}) of
App.~\ref{KapQuaZeiEle}. By combining the results of\ Eqs.~(\ref{EntWicHamAlg}%
)-(\ref{WirImpPotDreQDel}), we obtain the following approximate expression for
the matrix elements of $H$:%
\begin{align}
\langle\mathbf{x}|H|\mathbf{\tilde{y}}\rangle\approx &  \operatorname*{vol}%
\nolimits^{-1}\sum_{n\hspace{0.01in}=\hspace{0.01in}0}^{\infty}a_{n}%
\sum_{v\hspace{0.01in}=\hspace{0.01in}0}^{n}\frac{(-1)^{v}\hspace
{0.01in}(q+q^{-1})^{n-v}}{q^{2v+8v(n-v)+4}}\hspace{0.01in}%
\genfrac{[}{]}{0pt}{}{n}{v}%
_{q^{4}}\,\nonumber\\
&  \qquad\qquad\times d_{n-v}^{(-)}(x^{-}\hspace{-0.01in},\tilde{y}%
^{-})\,d_{2v,\hspace{0.01in}n-v}^{(3)}(x^{3}\hspace{-0.01in},\tilde{y}%
^{3})\,d_{n-v,\hspace{0.01in}2v}^{(+)}(x^{+}\hspace{-0.01in},\tilde{y}^{+}).
\label{MatEleHExp}%
\end{align}

We assume that the coordinates of the $q$-de\-formed Euclidean space are
restricted to a $q$-lat\-tice, defined by%
\begin{equation}
(x^{+},x^{3},x^{-})\in\{(\pm\hspace{0.01in}\alpha_{+}\hspace{0.01in}q^{2k_{+}%
},\pm\hspace{0.01in}\alpha_{3}\hspace{0.01in}q^{k_{3}},\pm\hspace
{0.01in}\alpha_{-}\hspace{0.01in}q^{2k_{-}})|\hspace{0.01in}k_{+},k_{3}%
,k_{-}\in\mathbb{Z}\}, \label{xKooExpQGit}%
\end{equation}
where the lattice parameters satisfy%
\begin{equation}
\alpha_{\pm}\in\left(  \min(1,q^{2});\max(1,q^{2})\right)  \quad
\text{and}\quad\alpha_{3}\in\left(  \min(1,q);\max(1,q)\right)  .
\end{equation}
Under this assumption, the expressions in Eq.~(\ref{WirImpPotDreQDelEin}) can
be further simplified. Employing Eq.~(\ref{MehAnwJacAblFkt}) from
Chap.~\ref{KapqAnaExpTriFkt} to evaluate the high\-er-or\-der Jackson
derivatives in Eq.~(\ref{WirImpPotDreQDelEin}), we obtain:%
\begin{align}
&  \left.  d_{m_{3},\hspace{0.01in}m_{-}}^{(3)}(x^{3}\hspace{-0.01in}%
,\tilde{y}^{3})\right\vert _{x^{3}=\hspace{0.01in}\pm\hspace{0.01in}\alpha
_{3}\hspace{0.01in}q^{k_{3}},\hspace{0.02in}\tilde{y}^{3}=\hspace{0.01in}%
\pm\hspace{0.01in}\alpha_{3}\hspace{0.01in}q^{\hspace{0.01in}l_{3}}%
}=\nonumber\\[0.02in]
&  \qquad=\sum_{j_{3}=\hspace{0.01in}0}^{m_{3}}%
\genfrac{[}{]}{0pt}{}{m_{3}}{j_{3}}%
_{q^{-2}}\frac{(-1)^{j_{3}}q^{-j_{3}(j_{3}-1)}\,\phi_{q}(q^{2m_{-}%
-\hspace{0.01in}2+2j_{3}+\hspace{0.01in}k_{3}}\hspace{-0.01in},q^{\hspace
{0.01in}l_{3}})}{[(1-q^{2})(\pm\hspace{0.01in}\alpha_{3}\hspace{0.01in}%
q^{k_{3}})]^{m_{3}}\left\vert 1-q\right\vert \alpha_{3}\hspace{0.01in}%
q^{\hspace{0.01in}l_{3}}}. \label{JacAbl3DeltX3N}%
\end{align}
Using the identity [cf. Eqs.~(\ref{ChaIdePhi1}) and (\ref{ChaIdePhi2}) in
Chap.~\ref{KapQDisEin}]%
\begin{equation}
\phi_{q}(q^{m},q^{n})=\delta_{m,n}, \label{IdeQDel}%
\end{equation}
the sum in\ Eq.~(\ref{JacAbl3DeltX3N}) reduces to%
\begin{align}
&  \left.  d_{m_{3},\hspace{0.01in}m_{-}}^{(3)}(x^{3}\hspace{-0.01in}%
,\tilde{y}^{3})\right\vert _{x^{3}=\hspace{0.01in}\pm\hspace{0.01in}\alpha
_{3}\hspace{0.01in}q^{k_{3}},\hspace{0.02in}\tilde{y}^{3}=\hspace{0.01in}%
\pm\hspace{0.01in}\alpha_{3}\hspace{0.01in}q^{\hspace{0.01in}l_{3}}%
}=\nonumber\\[0.02in]
&  \qquad=\frac{(-1)^{(l_{3}-\hspace{0.01in}k_{3})/2-m_{-}+\hspace{0.01in}%
1}\hspace{0.01in}q^{-[(l_{3}-\hspace{0.01in}k_{3})/2-m_{-}+\hspace
{0.01in}1][(l_{3}-\hspace{0.01in}k_{3})/2-m_{-}]}}{[(1-q^{2})(\pm
\hspace{0.01in}\alpha_{3}\hspace{0.01in}q^{k_{3}})]^{m_{3}}\left\vert
1-q\right\vert \alpha_{3}\hspace{0.01in}q^{\hspace{0.01in}l_{3}}%
}\nonumber\\[0.02in]
&  \qquad\hspace{0.17in}\times%
\genfrac{[}{]}{0pt}{}{m_{3}}{(l_{3}-k_{3})/2-m_{-}+1}%
_{q^{-2}}\Xi(l_{3}-k_{3}). \label{JacAbl3DeltX3End}%
\end{align}
Note that the sum in\ Eq.~(\ref{JacAbl3DeltX3N}) vanishes unless $(l_{3}%
-k_{3})/2-m_{-}+1\in\mathbb{N}_{0}$. To capture this restriction, we
introduced a Kronecker delta based on the floor function\footnote{The floor
function $\lfloor x\rfloor$ gives the greatest integer less than or equal to
$x$.}:%
\begin{equation}
\Xi(x)=\delta_{\hspace{0.01in}2\lfloor x/2\rfloor,\hspace{0.01in}x}=\left\{
\begin{array}
[c]{ll}%
1 & \text{if\ }x\ \text{is an even integer,}\\
0 & \text{otherwise.}%
\end{array}
\right.
\end{equation}
Analogously, one obtains%
\begin{align}
&  \left.  d_{m_{+},\hspace{0.01in}m_{3}}^{(+)}(x^{+}\hspace{-0.01in}%
,\tilde{y}^{+})\right\vert _{x^{+}=\hspace{0.01in}\pm\hspace{0.01in}\alpha
_{+}\hspace{0.01in}q^{2k_{+}}\hspace{-0.01in},\hspace{0.02in}\tilde{y}%
^{+}=\hspace{0.01in}\pm\hspace{0.01in}\alpha_{+}\hspace{0.01in}q^{2l_{+}}%
}=\nonumber\\[0.02in]
&  \qquad=\frac{(-1)^{(l_{+}-\hspace{0.01in}k_{+}-\hspace{0.01in}m_{3}%
)/2}q^{-(l_{+}-\hspace{0.01in}k_{+}-\hspace{0.01in}m_{3})(l_{+}-\hspace
{0.01in}k_{+}-\hspace{0.01in}m_{3}-2)/2}}{[(1-q^{4})(\pm\hspace{0.01in}%
\alpha_{+}\hspace{0.01in}q^{2k_{+}})]^{m_{+}}\left\vert 1-q^{2}\right\vert
\alpha_{+}\hspace{0.01in}q^{2l_{+}}}\nonumber\\[0.02in]
&  \qquad\hspace{0.17in}\times%
\genfrac{[}{]}{0pt}{}{m_{+}}{(l_{+}\hspace{-0.01in}-k_{+}\hspace
{-0.01in}-m_{3})/2}%
_{q^{-4}}\Xi(l_{+}\hspace{-0.01in}-k_{+}\hspace{-0.01in}-m_{3}),
\label{JacMPlOrtEigM}%
\end{align}
and%
\begin{align}
&  \left.  d_{m_{-}}^{(-)}(x^{-}\hspace{-0.01in},\tilde{y}^{-})\right\vert
_{x^{-}=\hspace{0.01in}\pm\hspace{0.01in}\alpha_{-}\hspace{0.01in}q^{2k_{-}%
}\hspace{-0.01in},\hspace{0.02in}\tilde{y}^{-}=\hspace{0.01in}\pm
\hspace{0.01in}\alpha_{-}\hspace{0.01in}q^{2l_{-}}}=\nonumber\\[0.02in]
&  \qquad=\frac{(-1)^{(l_{-}-\hspace{0.01in}k_{-}+1)/2}\hspace{0.01in}%
q^{-(l_{-}-\hspace{0.01in}k_{-}+1)(l_{-}-\hspace{0.01in}k_{-}-1)/2}}%
{[(1-q^{4})(\pm\hspace{0.01in}\alpha_{-}\hspace{0.01in}q^{2k_{-}})]^{m_{-}%
}\left\vert 1-q^{2}\right\vert \alpha_{-}\hspace{0.01in}q^{2l_{-}}%
}\nonumber\\[0.02in]
&  \qquad\hspace{0.17in}\times%
\genfrac{[}{]}{0pt}{}{m_{-}}{(l_{-}\hspace{-0.01in}-k_{-}\hspace
{-0.01in}+1)/2}%
_{q^{-4}}\Xi(l_{-}\hspace{-0.01in}-k_{-}\hspace{-0.01in}+1).
\label{JacAbl3DeltXM}%
\end{align}

Finally, let $S$ be a region in the $q$-de\-formed Euclidean space
$\mathbb{R}_{q}^{3}$, associated with the state%
\begin{equation}
|S\rangle=\int\nolimits_{S}\text{d}_{q}^{3}x\,|\mathbf{x}\rangle.
\end{equation}
The corresponding expectation value%
\begin{equation}
\langle H\rangle_{S}=\frac{\langle S\hspace{0.01in}|H|S\rangle}{\langle
S\hspace{0.01in}|S\rangle}=\frac{\int\nolimits_{S}\text{d}_{q}^{3}%
x\int\nolimits_{S}\text{d}_{q}^{3}x^{\prime}\,\langle\mathbf{x}|H|\mathbf{x}%
^{\prime}\rangle}{\int\nolimits_{S}\text{d}_{q}^{3}x\int\nolimits_{S}%
\text{d}_{q}^{3}x^{\prime}\,\langle\mathbf{x}|\mathbf{x}^{\prime}\rangle}
\label{BerZuOrt}%
\end{equation}
represents the \textit{vacuum energy contained in }$S$\textit{ for the }%
$q$\textit{-de\-formed Klein-Gordon field}. In what follows, we evaluate
$\langle H\rangle_{S}$ in two cases: (i) when $S$ is a neighborhood of a
quasipoint, and (ii) when $S$ coincides with the entire $q$-de\-formed
Euclidean space.

\subsection{Vacuum Energy Density Around a Quasipoint\label{VacEneQuaPoi}}

The classical calculation presented in Eq.~(\ref{BerEneKlas}) of
Chap.~\ref{ClasCal} yields an extremely large value for the vacuum energy
density. Our aim is to investigate whether a similar result arises for a
Klein-Gordon field defined on $q$-de\-formed Euclidean space.

The evaluation of $\langle H\rangle_{S}$ is technically involved. In contrast,
the computation of $\langle H^{2}\rangle_{S}$ is considerably simpler,
especially in the case of a massless $q$-de\-formed Klein-Gordon field. In
this situation, it holds [cf. Eq.~(\ref{BerZuOrt}) of the previous chapter]:%
\begin{equation}
\langle H^{2}\rangle_{S}=\frac{\langle S\hspace{0.01in}|c^{\hspace{0.01in}%
2}\mathbf{p}^{2}|S\rangle}{\langle S\hspace{0.01in}|S\rangle}=\frac
{\int\nolimits_{S}\text{d}_{q}^{3}x\int\nolimits_{S}\text{d}_{q}^{3}x^{\prime
}\,\langle\mathbf{x}|c^{\hspace{0.01in}2}\mathbf{p}^{2}|\mathbf{x}^{\prime
}\rangle}{\int\nolimits_{S}\text{d}_{q}^{3}x\int\nolimits_{S}\text{d}_{q}%
^{3}x^{\prime}\,\langle\mathbf{x}|\mathbf{x}^{\prime}\rangle}.
\end{equation}
We therefore focus on the matrix elements of $\mathbf{p}^{2}$ [cf.
Eq.~(\ref{EntWicHamAlg})]:%
\begin{equation}
\langle\mathbf{x}|\mathbf{p}^{2}|\mathbf{\tilde{y}}\rangle
=-\operatorname*{vol}\nolimits^{-1}\hspace{-0.01in}\mathbf{\partial}%
_{x}\hspace{-0.01in}\circ\mathbf{\partial}_{x}\triangleright\delta_{q}%
^{3}(\mathbf{x}\oplus(\ominus\hspace{0.01in}\kappa^{-1}\mathbf{\tilde{y}})).
\label{MatEleP2Ort}%
\end{equation}
By combining the results of\ Eqs.~(\ref{DarPotAblQuaDreEnt}%
)-(\ref{WirImpPotDreQDel}) with those of Eqs.~(\ref{JacAbl3DeltX3End}%
)-(\ref{JacAbl3DeltXM}), we obtain the following approximate expression:%
\begin{gather}
\left.  \langle\mathbf{x}|\mathbf{p}^{2}|\mathbf{\tilde{y}}\rangle\right\vert
_{x^{A}=\hspace{0.01in}\pm\alpha_{A}q^{(2-\delta_{A3})k_{A}}\hspace
{-0.01in},\hspace{0.02in}\tilde{y}^{A}=\hspace{0.01in}\pm\alpha_{A}%
q^{(2-\delta_{A3})l_{A}}}\approx\nonumber\\
\approx\sum_{v\hspace{0.01in}=\hspace{0.01in}0}^{1}\frac{(-1)^{v}%
\hspace{0.01in}(q+q^{-1})^{1-v}}{q^{-8v^{2}+10v+4}}\hspace{0.01in}%
\genfrac{[}{]}{0pt}{}{1}{v}%
_{q^{4}}\,b_{\hspace{0.01in}l_{3},\hspace{0.01in}k_{3},\hspace{0.01in}v}%
^{(3)}\hspace{0.01in}b_{\hspace{0.01in}l_{+},\hspace{0.01in}k_{+}%
,\hspace{0.01in}v}^{(+)}\hspace{0.01in}b_{\hspace{0.01in}l_{-},\hspace
{0.01in}k_{-},\hspace{0.01in}v}^{(-)}\nonumber\\[0.04in]
\times%
\genfrac{[}{]}{0pt}{}{2v}{(l_{3}-k_{3})/2+v}%
_{q^{-2}}%
\genfrac{[}{]}{0pt}{}{1-v}{(l_{+}\hspace{-0.01in}-k_{+})/2\hspace{-0.01in}-v}%
_{q^{-4}}\,%
\genfrac{[}{]}{0pt}{}{1-v}{(l_{-}\hspace{-0.01in}-k_{-}+1)/2}%
_{q^{-4}}\nonumber\\[0.04in]
\times\,\Xi(l_{3}-k_{3})\,\Xi(l_{+}\hspace{-0.01in}-k_{+})\,\Xi(l_{-}%
\hspace{-0.01in}-k_{-}\hspace{-0.01in}+1), \label{AusVakKom}%
\end{gather}
with%
\begin{align}
b_{\hspace{0.01in}l_{3},\hspace{0.01in}k_{3},\hspace{0.01in}v}^{(3)}  &
=\frac{(-1)^{(l_{3}-k_{3})/2}\hspace{0.01in}q^{-[(l_{3}-k_{3})/2+v][(l_{3}%
-k_{3})/2v-1]}}{[(1-q^{2})(\pm\hspace{0.01in}\alpha_{3}\hspace{0.01in}%
q^{k_{3}})]^{2v}\left\vert 1-q\right\vert \alpha_{3}\hspace{0.01in}%
q^{\hspace{0.01in}l_{3}}},\nonumber\\
b_{\hspace{0.01in}l_{+},\hspace{0.01in}k_{+},\hspace{0.01in}v}^{(+)}  &
=\frac{(-1)^{(l_{+}-\hspace{0.01in}k_{+})/2}\hspace{0.01in}q^{-(l_{+}%
-\hspace{0.01in}k_{+}-\hspace{0.01in}2v)(l_{+}-\hspace{0.01in}k_{+}%
-\hspace{0.01in}2v-2)/2}}{[(1-q^{4})(\pm\hspace{0.01in}\alpha_{+}%
\hspace{0.01in}q^{2k_{+}})]^{1-v}\left\vert 1-q^{2}\right\vert \alpha
_{+}\hspace{0.01in}q^{2l_{+}}},\nonumber\\
b_{\hspace{0.01in}l_{-},\hspace{0.01in}k_{-},\hspace{0.01in}v}^{(-)}  &
=\frac{(-1)^{(l_{-}-\hspace{0.01in}k_{-}+1)/2}\hspace{0.01in}q^{-(l_{-}%
-\hspace{0.01in}k_{-}+1)(l_{-}-\hspace{0.01in}k_{-}-1)/2}}{[(1-q^{4}%
)(\pm\hspace{0.01in}\alpha_{-}\hspace{0.01in}q^{2k_{-}})]^{1-v}\left\vert
1-q^{2}\right\vert \alpha_{-}\hspace{0.01in}q^{2l_{-}}}.
\end{align}

For the matrix element in Eq.~(\ref{AusVakKom}) to be non-zero, the variables
$l_{A}$ and $k_{A}$ $(A\in\{+,3,-\})$ must satisfy specific constraints. This
follows from the property of the $q$-bi\-no\-mi\-al coefficients,%
\begin{equation}%
\genfrac{[}{]}{0pt}{}{n}{k}%
_{q^{m}}\hspace{-0.02in}=0\quad\text{if}\quad k<0\quad\text{or}\quad k>n.
\label{ProQBin}%
\end{equation}
Accordingly, the summand on the right-hand side of Eq.~(\ref{AusVakKom})
corresponding to a given value of $v$ vanishes unless all of the following
inequalities hold:%
\begin{align}
0  &  \leq(l_{+}\hspace{-0.01in}-k_{+})/2\hspace{-0.01in}-v\leq1-v,\nonumber\\
0  &  \leq(l_{-}\hspace{-0.01in}-k_{-}+1)/2\leq1-v,\nonumber\\
0  &  \leq(l_{3}-k_{3})/2+v\leq2v.
\end{align}
These inequalities can be equivalently written as%
\begin{align}
2v  &  \leq l_{+}\hspace{-0.01in}-k_{+}\leq2,\nonumber\\
-1  &  \leq l_{-}\hspace{-0.01in}-k_{-}\leq1-2v,\nonumber\\
-2v  &  \leq l_{3}-k_{3}\leq2v.
\end{align}
Since\ the summation index $v$ in Eq.~(\ref{AusVakKom}) takes only the integer values
$0$ and\ $1$, the matrix element in Eq.~(\ref{AusVakKom}) can be 
non-van\-ish\-ing only if the values of $l_{A}$ and $k_{A}$ satisfy the\ above
inequalities for at least one of these values. We therefore restrict attention to the following, less stringent system of inequalities:%
\begin{align}
0  &  \leq l_{+}\hspace{-0.01in}-k_{+}\leq2,\nonumber\\
-1  &  \leq l_{-}\hspace{-0.01in}-k_{-}\leq1,\nonumber\\
-2  &  \leq l_{3}-k_{3}\leq2. \label{BedMatElem}%
\end{align}
This\ system of inequalities\ characterizes\ the\ domain\ of quasipoints%
\begin{equation}
(\pm\hspace{0.01in}\alpha_{+}\hspace{0.01in}q^{2l_{+}}\hspace{-0.01in}%
,\pm\hspace{0.01in}\alpha_{3}\hspace{0.01in}q^{l_{3}},\pm\hspace{0.01in}%
\alpha_{-}\hspace{0.01in}q^{2l_{-}})
\end{equation}
that can yield non-van\-ish\-ing matrix elements of $\mathbf{p}^{2}$ with a
fixed quasipoint
\begin{equation}
(\pm\hspace{0.01in}\alpha_{+}\hspace{0.01in}q^{2k_{+}}\hspace{-0.01in}%
,\pm\hspace{0.01in}\alpha_{3}\hspace{0.01in}q^{k_{3}},\pm\hspace{0.01in}%
\alpha_{-}\hspace{0.01in}q^{2k_{-}}).
\end{equation}

Motivated by this observation, we consider the vacuum expectation value of
$\mathbf{p}^{2}$ in the superposition of quasipoint states [cf.
Eq.~\ref{BerZuOrt} in Chap.~\ref{VacEneQEuc} and Eq.~(\ref{QDefVolDreWdh})]%
\begin{equation}
|S_{z}\rangle=\sum_{s_{\pm}\hspace{0.01in}=\hspace{0.01in}-2}^{2}\,\sum
_{s_{3}\hspace{0.01in}=\hspace{0.01in}-2}^{2}\left.  |\mathbf{x}%
\rangle\,\text{d}_{q}^{3}\hspace{0.01in}x\right\vert _{_{x^{A}=\hspace
{0.01in}\pm\alpha_{A}\hspace{0.01in}q^{(2-\delta_{A3})(k_{A}+s_{A})}}},
\end{equation}
where the volume of the quasipoint $\mathbf{x}$ is given by:%
\begin{equation}
\text{d}_{q}^{3}\hspace{0.01in}x=\text{d}_{q^{2}}x^{+}\hspace{0.01in}%
\text{d}_{q}\hspace{0.01in}x^{3}\hspace{0.01in}\text{d}_{q^{2}}x^{-}%
=\left\vert (1-q)(1-q^{2})^{2}\hspace{0.01in}x^{+}x^{3}x^{-}\right\vert
.\label{QDefVolDreWdh}%
\end{equation}
The relevant expectation value is then%
\begin{gather}
\frac{\langle S_{z}|\mathbf{p}^{2}|S_{z}\rangle}{V_{S_{z}}}=\nonumber\\
=\frac{1}{V_{S_{z}}}\sum_{s_{\pm},\hspace{0.01in}t_{\pm}=\hspace{0.01in}%
-2}^{2}\,\sum_{s_{3},\hspace{0.01in}t_{3}=\hspace{0.01in}-2}^{2}\left.
\text{d}_{q}^{3}\hspace{0.01in}x\,\langle\mathbf{x}|\mathbf{p}^{2}%
|\mathbf{\tilde{y}}\rangle\,\text{d}_{q}^{3}\hspace{0.01in}\tilde
{y}\right\vert _{_{\substack{x^{A}=\hspace{0.01in}\pm\alpha_{A}\hspace
{0.01in}q^{(2-\delta_{A3})(k_{A}+s_{A})}\\\tilde{y}^{B}=\hspace{0.01in}%
\pm\alpha_{B}\hspace{0.01in}q^{(2-\delta_{B3})(k_{B}+t_{B})}}}}%
,\label{ErwPQua}%
\end{gather}
where the volume of the region $S_{z}$ is given by:%
\begin{equation}
V_{S_{z}}=\langle S_{z}|S_{z}\rangle=\operatorname*{vol}\nolimits^{-1}%
\hspace{-0.02in}\int\nolimits_{S_{z}}\text{d}_{q}^{3}x\int\nolimits_{S_{z}%
}\text{d}_{q}^{3}x^{\prime}\,\delta_{q}^{3}(\mathbf{x}\oplus(\ominus
\hspace{0.01in}\kappa^{-1}\mathbf{x}^{\prime})).\label{VolReg}%
\end{equation}
Inserting Eq.~(\ref{AusVakKom}) and  (\ref{QDefVolDreWdh}) into Eq.~(\ref{ErwPQua}), we obtain the
approximate expression%
\begin{align}
\langle S_{z}|\mathbf{p}^{2}|S_{z}\rangle &  \approx\sum_{v\hspace
{0.01in}=\hspace{0.01in}0}^{1}(-1)^{v}\frac{(q+q^{-1})^{1-\hspace{0.01in}v}%
}{q^{10v-8v^{2}+\hspace{0.01in}4}}\hspace{0.01in}\sum_{s_{\pm},\hspace
{0.01in}t_{\pm}=\hspace{0.01in}-2}^{2}\,\sum_{s_{3},\hspace{0.01in}%
t_{3}=\hspace{0.01in}-2}^{2}\hspace{0.01in}%
\genfrac{[}{]}{0pt}{}{2v}{(t_{3}-s_{3})/2+v}%
_{q^{-2}}\nonumber\\
&  \qquad\qquad\times\Theta(t_{+}\hspace{-0.01in}-s_{+}\hspace{-0.01in}%
-2v)\,\Theta(s_{+}\hspace{-0.01in}-t_{+}\hspace{-0.01in}+2)\nonumber\\
&  \qquad\qquad\times\Theta(t_{-}\hspace{-0.01in}-s_{-}\hspace{-0.01in}%
+1)\,\Theta(s_{-}\hspace{-0.01in}-t_{-}\hspace{-0.01in}-2v+1)\nonumber\\
&  \qquad\qquad\times\Xi(t_{+}-\hspace{0.01in}s_{+})\,\Xi(t_{3}-s_{3}%
)\,\Xi(t_{-}-\hspace{0.01in}s_{-}+1)\nonumber\\
&  \qquad\qquad\times c_{\hspace{0.01in}t_{3},\hspace{0.01in}s_{3}%
,\hspace{0.01in}v}(z^{3})\,c_{\hspace{0.01in}t_{+},\hspace{0.01in}%
s_{+},\hspace{0.01in}v}(z^{+})\,c_{\hspace{0.01in}t_{-},\hspace{0.01in}%
s_{-},\hspace{0.01in}v}(z^{-}),\label{VakErwP2}%
\end{align}
where%
\begin{align}
c_{\hspace{0.01in}t_{3},\hspace{0.01in}s_{3},\hspace{0.01in}v}^{(3)}(z^{3}) &
=\frac{(-1)^{(t_{3}-s_{3})/2}\hspace{0.01in}\left\vert (1-q)\hspace
{0.01in}q^{s_{3}}z^{3}\right\vert }{q^{[(t_{3}-s_{3})/2+\hspace{0.01in}%
v][(t_{3}-s_{3})/2+\hspace{0.01in}v\hspace{0.01in}-1]}\,[(1-q^{2}%
)\hspace{0.01in}q^{s_{3}}z^{3}]^{2v}},\nonumber\\
c_{\hspace{0.01in}t_{+},\hspace{0.01in}s_{+},\hspace{0.01in}v}^{(+)}(z^{+}) &
=\frac{(-1)^{(t_{+}-\hspace{0.01in}s_{+})/2}\hspace{0.01in}\left\vert
(1-q^{2})\hspace{0.01in}q^{2s_{+}}z^{+}\right\vert }{q^{(t_{+}-\hspace
{0.01in}s_{+}-\hspace{0.01in}2v)(t_{+}-\hspace{0.01in}s_{+}-\hspace
{0.01in}2v-2)/2}\,[(1-q^{4})\hspace{0.01in}q^{2s_{+}}z^{+}]^{1-\hspace
{0.01in}v}},\nonumber\\
c_{\hspace{0.01in}t_{-},\hspace{0.01in}s_{-},\hspace{0.01in}v}^{(-)}(z^{-}) &
=\frac{(-1)^{(t_{-}-s_{-}+\hspace{0.01in}1)/2}\hspace{0.01in}\left\vert
(1-q^{2})\hspace{0.01in}q^{2s_{-}}z^{-}\right\vert }{q^{(t_{-}-\hspace
{0.01in}s_{-}+1)(t_{-}-\hspace{0.01in}s_{-}-1)/2}\,[(1-q^{4})\hspace
{0.01in}q^{2s_{-}}z^{-}]^{1-\hspace{0.01in}v}}.
\end{align}
The coordinates $z^{+}$, $z^{3}$, and $z^{-}$ are defined as:%
\begin{equation}
z^{+}=\pm\hspace{0.01in}\alpha_{+}\hspace{0.01in}q^{2k_{+}},\hspace
{0.06in}z^{3}=\pm\hspace{0.01in}\alpha_{3}\hspace{0.01in}q^{k_{3}}%
,\hspace{0.06in}z^{-}=\pm\hspace{0.01in}\alpha_{-}\hspace{0.01in}q^{2k_{-}%
}.\label{DefZKoor}%
\end{equation}
To obtain the approximate expression in Eq.~(\ref{VakErwP2}), we replaced
$l_{A}$ and $k_{A}$ in Eq.~(\ref{AusVakKom}) by $k_{A}+t_{A}$ and $k_{A}%
+s_{A}$, respectively, where $t_{A}$ and $s_{A}$ denote the summation indices.
Since $v\in\{0,1\}$, we employed the identities%
\begin{equation}%
\genfrac{[}{]}{0pt}{}{1}{v}%
_{q^{4}}\hspace{-0.02in}=1,
\end{equation}
and%
\begin{align}%
\genfrac{[}{]}{0pt}{}{1-v}{(t_{+}\hspace{-0.01in}-s_{+}\hspace{-0.01in}-2v)/2}%
_{q^{-4}}\hspace{-0.02in} &  =\Theta(t_{+}\hspace{-0.01in}-s_{+}%
\hspace{-0.01in}-2v)\hspace{0.01in}\Theta(s_{+}\hspace{-0.01in}-t_{+}%
\hspace{-0.01in}+2),\nonumber\\[0.1in]%
\genfrac{[}{]}{0pt}{}{1-v}{(t_{-}\hspace{-0.01in}-s_{-}+1)/2}%
_{q^{-4}}\hspace{-0.02in} &  =\Theta(t_{-}\hspace{-0.01in}-s_{-}%
\hspace{-0.01in}+1)\hspace{0.01in}\Theta(s_{-}\hspace{-0.01in}-t_{-}%
\hspace{-0.01in}-2v+1),\label{VerBinKoe}%
\end{align}
where the Heaviside step functions 
arise from the property of the $q$-bi\-no\-mi\-al
coefficients given in Eq.~(\ref{ProQBin}).

In Eqs.~(\ref{AusVakKom}) and (\ref{VakErwP2}), the $x$- and $y$%
-co\-or\-di\-nates are not expressed in the same normal ordering, since the
operator $\hat{U}_{y}^{-1}$ was omitted in Eq.~(\ref{WirImpPotDreQDel}) for
the sake of simplicity. This omission, however, does not affect the validity
of the subsequent results, as the action of $\hat{U}_{y}^{-1}$\textbf{
}generates only corrections of order $h$, and our analysis is primarily
concerned with the limit $h\rightarrow0$.\footnote{The~deformation parameter
$q$ is related to $h$ by~$q=\operatorname{e}^{h}$. In the limit $h\rightarrow
0$, one has $q\rightarrow1$, corresponding to the disappearance of the
deformation. The parameter $h$ should not be confused with Planck's constant.}

It remains to evaluate the volume of the region $S_{z}$. Starting from%
\begin{equation}
V_{S_{z}}=\sum_{s_{\pm},\hspace{0.01in}t_{\pm}=\hspace{0.01in}-2}^{2}%
\,\sum_{s_{3},\hspace{0.01in}t_{3}=\hspace{0.01in}-2}^{2}\left.  \text{d}%
_{q}^{3}\hspace{0.01in}x\,\langle\mathbf{x}|\mathbf{\tilde{y}}\rangle
\,\text{d}_{q}^{3}\hspace{0.01in}\tilde{y}\right\vert _{_{\substack{x^{A}%
=\hspace{0.01in}\pm\alpha_{A}\hspace{0.01in}q^{(2-\delta_{A3})(k_{A}%
+\hspace{0.01in}s_{A})}\\\tilde{y}^{B}=\hspace{0.01in}\pm\alpha_{B}%
\hspace{0.01in}q^{(2-\delta_{B3})(k_{B}+t\hspace{0.01in}_{B})}}}}%
\end{equation}
and using%
\begin{equation}
\langle\mathbf{x}|\hspace{0.01in}\mathbf{\tilde{y}}\rangle=\operatorname*{vol}%
\nolimits^{-1}\delta_{q}^{3}(\mathbf{x}\oplus(\ominus\hspace{0.01in}%
\kappa^{-1}\mathbf{\tilde{y}})),\label{VolRelOrtZus}%
\end{equation}
we obtain%
\begin{align}
V_{S_{z}} &  \approx\sum_{s_{\pm},\hspace{0.01in}t_{\pm}=\hspace{0.01in}%
-2}^{2}\,\sum_{s_{3},\hspace{0.01in}t_{3}=\hspace{0.01in}-2}^{2}%
q^{-4}q^{2(s_{+}+\hspace{0.01in}s_{-})+\hspace{0.01in}s_{3}}\text{d}_{q}%
^{3}\hspace{0.01in}z\,q^{2(t_{+}+\hspace{0.01in}t_{-})+\hspace{0.01in}t_{3}%
}\text{d}_{q}^{3}\hspace{0.01in}z\nonumber\\
&  \hspace{0.17in}\times\frac{\phi_{q^{2}}(q^{2s_{+}}z^{+},q^{2t_{+}}%
z^{+})\,\phi_{q}(q^{s_{3}-\hspace{0.01in}2}z^{3},q^{t_{3}}z^{3})\,\phi_{q^{2}%
}(q^{2s_{-}-\hspace{0.01in}2}z^{-},q^{2t_{-}}z^{-})}{\left\vert q^{2(t_{+}%
+\hspace{0.01in}t_{-})+t_{3}}(1-q)(1-q^{2})^{2}\hspace{0.01in}z^{+}z^{3}%
z^{-}\right\vert }\nonumber\\
&  =\sum_{s_{\pm}=\hspace{0.01in}-2}^{2}\sum_{s_{3}=\hspace{0.01in}-2}%
^{2}q^{2(s_{+}+\hspace{0.01in}s_{-})+\hspace{0.01in}s_{3}-4}\hspace
{0.01in}\text{d}_{q}^{3}\hspace{0.01in}z\sum_{t_{\pm}=\hspace{0.01in}-2}%
^{2}\,\sum_{t_{3}=\hspace{0.01in}-2}^{2}\delta_{s_{+},t_{+}}\hspace
{0.01in}\delta_{s_{3}-\hspace{0.01in}2,t_{3}}\hspace{0.01in}\delta
_{s_{-}-\hspace{0.01in}1,t_{-}}\nonumber\\
&  =\sum_{s_{+}=\hspace{0.01in}-2}^{2}\sum_{s_{-}=\hspace{0.01in}-1}^{2}%
\sum_{s_{3}=\hspace{0.01in}0}^{2}q^{2(s_{+}+\hspace{0.01in}s_{-}%
)+\hspace{0.01in}s_{3}-4}\hspace{0.01in}\text{d}_{q}^{3}\hspace{0.01in}%
z=q^{-10}[[5]]_{q^{2}}[[4]]_{q^{2}}[[3]]_{q}\hspace{0.01in}\text{d}_{q}%
^{3}\hspace{0.01in}z.\label{NaeVolEleS}%
\end{align}
In the first step of the this derivation, we applied the approximate expression
from Eq.~(\ref{AltDarDelDreiDimLam0}) in Chap.~\ref{AppExp}. Furthermore, we
employed the identities [cf. Eqs.~(\ref{QDefVolDreWdh}) and (\ref{DefZKoor})]:%
\begin{align}
\left.  \text{d}_{q}^{3}\hspace{0.01in}x\right\vert _{_{x^{A}=\hspace
{0.01in}\pm\alpha_{A}q^{(2-\delta_{A3})(k_{A}+\hspace{0.01in}s_{A})}}} &
=q^{2(s_{+}+\hspace{0.01in}s_{-})+\hspace{0.01in}s_{3}}\text{d}_{q}^{3}%
\hspace{0.01in}z,\nonumber\\
\left.  \text{d}_{q}^{3}\hspace{0.01in}\tilde{y}\right\vert _{_{\tilde{y}%
^{A}=\hspace{0.01in}\pm\alpha_{A}q^{(2-\delta_{A3})(k_{A}+\hspace{0.01in}%
t_{A})}}} &  =q^{2(t_{+}+\hspace{0.01in}t_{-})+\hspace{0.01in}t_{3}}%
\text{d}_{q}^{3}\hspace{0.01in}z.
\end{align}
The simplification in the second step of Eq.~(\ref{NaeVolEleS}) follows from
Eqs.~(\ref{IdeQDel}) and (\ref{QDefVolDreWdh}). In the final step, we used the
identity [cf. Eq.~(\ref{DefAntSymQZah2}) in Chap.~\ref{KapqAnaExpTriFkt}]:%
\begin{equation}
\sum_{k\hspace{0.01in}=\hspace{0.01in}0}^{n}q^{k}=[[\hspace{0.01in}n+1]]_{q}.
\end{equation}

We next derive an approximate expression for the vacuum expectation value of
$\mathbf{p}^{2}$ in the limit $h\rightarrow0$. To this end, we employ the
expansions%
\begin{align}%
\genfrac{[}{]}{0pt}{}{m}{k}%
_{q^{\alpha}}\hspace{-0.02in} &  =\binom{m}{k}+\mathcal{O}(h), & q+q^{-1} &
=2+\mathcal{O}(h),\nonumber\\[0.03in]
1-q^{\alpha} &  =-\hspace{0.01in}\alpha\hspace{0.01in}h+\mathcal{O}(h^{2}), &
q^{\alpha} &  =1+\mathcal{O}(h),\label{NaehAusKla}%
\end{align}
which follow directly from $q=\operatorname{e}^{h}$. These relations imply the
following approximation for the volume of a quasipoint:%
\begin{equation}
\text{d}_{q}^{3}\hspace{0.01in}z=\left\vert 4\hspace{0.01in}h^{3}z^{+}%
z^{3}z^{-}\right\vert +\mathcal{O}(h^{4}).\label{NaeVolEle}%
\end{equation}
Substituting Eqs.~(\ref{NaehAusKla}) and (\ref{NaeVolEle}) into
Eq.~(\ref{NaeVolEleS}), we obtain:%
\begin{equation}
V_{S_{z}}=60\hspace{0.01in}\left\vert 4\hspace{0.01in}h^{3}z^{+}z^{3}%
z^{-}\right\vert +\mathcal{O}(h^{4}).\label{NaeVolS}%
\end{equation}
Finally, inserting the expansions from Eq.~(\ref{NaehAusKla}) into
Eq.~(\ref{VakErwP2}) and using the result from Eq.~(\ref{NaeVolS}), we arrive
at%
\begin{align}
\frac{\langle S_{z}|\mathbf{p}^{2}|S_{z}\rangle}{V_{S_{z}}}\approx &
\sum_{v\hspace{0.01in}=\hspace{0.01in}0}^{1}(-1)^{v}\frac{2^{1-\hspace
{0.01in}v}}{60}\hspace{0.01in}\sum_{s_{\pm},\hspace{0.01in}t_{\pm}%
=\hspace{0.01in}-2}^{2}\,\sum_{s_{3},\hspace{0.01in}t_{3}=\hspace{0.01in}%
-2}^{2}\,\binom{2v}{(t_{3}-s_{3})/2+v}\nonumber\\[0.02in]
&  \times\Theta(t_{3}-s_{3}+2v)\,\Theta(s_{3}-t_{3}+2v)\nonumber\\[0.02in]
&  \times\Theta(t_{+}\hspace{-0.01in}-s_{+}\hspace{-0.01in}-2v)\,\Theta
(s_{+}\hspace{-0.01in}-t_{+}\hspace{-0.01in}+2)\nonumber\\
&  \times\Theta(t_{-}\hspace{-0.01in}-s_{-}\hspace{-0.01in}+1)\,\Theta
(s_{-}\hspace{-0.01in}-t_{-}\hspace{-0.01in}-2v+1)\nonumber\\
&  \times\Xi(t_{+}-\hspace{0.01in}s_{+})\,\Xi(t_{3}-s_{3})\,\Xi(t_{-}%
-\hspace{0.01in}s_{-}+1)\nonumber\\
&  \times\frac{(-1)^{(t_{+}+\hspace{0.01in}t_{3}+\hspace{0.01in}t_{-}%
-\hspace{0.01in}s_{+}-\hspace{0.01in}s_{3}-\hspace{0.01in}s_{-}+1)/2}%
}{16\hspace{0.01in}h^{2}z^{+}z^{-}}\left(  \frac{4\hspace{0.01in}z^{+}z^{-}%
}{z^{3}z^{3}}\right)  ^{v}.\label{NaeAusVakEneN2}%
\end{align}
We can explicitly evaluate the sums over the indices $s_{A}$ and
$t_{A}$ on the right-hand side of Eq.~(\ref{NaeAusVakEneN2}). For $v=0$, one
finds%
\begin{align}
&  \sum_{s_{+},\hspace{0.01in}t_{+}=\hspace{0.01in}-2}^{2}(-1)^{(t_{+}%
-\hspace{0.01in}s_{+})/2}\,\Theta(t_{+}\hspace{-0.01in}-s_{+})\,\Theta
(s_{+}\hspace{-0.01in}-t_{+}\hspace{-0.01in}+2)\,\Xi(t_{+}-\hspace
{0.01in}s_{+})=2,\nonumber\\
&  \sum_{s_{-},\hspace{0.01in}t_{-}=\hspace{0.01in}-2}^{2}(-1)^{(t_{-}%
-\hspace{0.01in}s_{-}+1)/2}\,\Theta(t_{-}\hspace{-0.01in}-s_{-}\hspace
{-0.01in}+1)\,\Theta(s_{-}\hspace{-0.01in}-t_{-}+1)\,\Xi(t_{-}-\hspace
{0.01in}s_{-}+1)=0,\nonumber\\
&  \sum_{s_{3},\hspace{0.01in}t_{3}=\hspace{0.01in}-2}^{2}(-1)^{(t_{3}%
-s_{3})/2}\,\Theta(t_{3}-s_{3})\,\Theta(s_{3}-t_{3})\,\Xi(t_{3}-s_{3})=5.
\end{align}
For $v=1$, the evaluation yields%
\begin{align}
&  \sum_{s_{+},\hspace{0.01in}t_{+}=\hspace{0.01in}-2}^{2}(-1)^{(t_{+}%
-\hspace{0.01in}s_{+})/2}\,\Theta(t_{+}\hspace{-0.01in}-s_{+}\hspace
{-0.01in}-2)\,\Theta(s_{+}\hspace{-0.01in}-t_{+}\hspace{-0.01in}%
+2)\,\,\Xi(t_{+}-\hspace{0.01in}s_{+})=-3,\nonumber\\
&  \sum_{s_{-},\hspace{0.01in}t_{-}=\hspace{0.01in}-2}^{2}(-1)^{(t_{-}%
-\hspace{0.01in}s_{-}+1)/2}\,\Theta(t_{-}\hspace{-0.01in}-s_{-}\hspace
{-0.01in}+1)\,\Theta(s_{-}\hspace{-0.01in}-t_{-}\hspace{-0.01in}-1)\,\Xi
(t_{-}-\hspace{0.01in}s_{-}+1)=4,\nonumber\\
&  \sum_{s_{3},\hspace{0.01in}t_{3}=\hspace{0.01in}-2}^{2}(-1)^{(t_{3}%
-s_{3})/2}\binom{2}{(t_{3}-s_{3})/2+1}\,\Theta(t_{3}-s_{3}+2)\,\Theta
(s_{3}-t_{3}+2)\nonumber\\
&  \qquad\qquad\qquad\times\Xi(t_{3}-s_{3})=4.
\end{align}
Substituting these results into Eq.~(\ref{NaeAusVakEneN2}) leads to the
approximation%
\begin{equation}
\frac{\langle S_{z}|\mathbf{p}^{2}|S_{z}\rangle}{V_{S_{z}}}=\frac{1}%
{5\hspace{0.01in}hz^{3}hz^{3}}+\mathcal{O}(h^{-1}).\label{VakDicP2Nae}%
\end{equation}
\textbf{ }

We interpret $hz^{A}$ as the distance between adjacent quasipoints along the
$x^{A}$-di\-rec\-tion, representing the smallest measurable length in space.
Consequently, $hz^{A}$ cannot be smaller than the Planck length $\ell_{P}$. To
refine the result in Eq.~(\ref{VakDicP2Nae}), we incorporate the square of the
reduced Planck constant $\hbar=1.054572\times10^{-34}~$J$\cdot$s, which is
directly related to the square of the momentum density. By multiplying the
expression in Eq.~(\ref{VakDicP2Nae}) by the square of the speed of light $c$
and substituting the Planck length $\ell_{P}$ for $hz^{A}$ [cf.
Eq.~(\ref{WerPlaLae}) from Chap.~\ref{ClasCal}], we obtain the following
approximate expression for the expectation value of $c^{\hspace{0.01in}%
2}\mathbf{p}^{2}$ in the region $S_{z}$:%
\begin{equation}
\langle c^{\hspace{0.01in}2}\mathbf{p}^{2}\rangle_{S_{z}}=\frac{c^{\hspace
{0.01in}2}\langle S_{z}|\mathbf{p}^{2}|S_{z}\rangle}{V_{S_{z}}}\approx
\frac{\hbar^{2}c^{\hspace{0.01in}2}}{5\hspace{0.01in}\ell_{P}^{\,2}}.
\label{DicP2DreiDim}%
\end{equation}

By analogy with the computation in Eq.~(\ref{DicEneQua}) of
Chap.~\ref{ClasCal}, and using the relations%
\begin{equation}
(\Delta E)^{2}=\langle E^{2}\rangle-\langle E\rangle^{2},\qquad\langle
E\rangle^{2}\geq(\Delta E)^{2},
\end{equation}
where $\Delta E$ denotes the energy fluctuation, we conclude:%
\begin{equation}
\langle E\rangle^{2}=\langle E^{2}\rangle-(\Delta E)^{2}\leq\langle
E^{2}\rangle=\langle E\rangle^{2}+(\Delta E)^{2}\leq2\langle E\rangle^{2},
\end{equation}
i.e.%
\begin{equation}
\left\vert \langle E\rangle\right\vert \leq\sqrt{\langle E^{2}\rangle}%
\leq\sqrt{2}\left\vert \langle E\rangle\right\vert .
\end{equation}
It is therefore natural to employ the approximation [also cf.
Eq.~(\ref{BerEneQuaDenCla}) in Chap.~\ref{ClasCal}]%
\begin{equation}
\langle H\rangle_{S_{z}}\approx\sqrt{\langle H^{2}\rangle_{S_{z}}}.
\end{equation}
Accordingly, the vacuum energy density $\rho_{0}$ at a quasipoint, as
contributed by a massless $q$-de\-formed Klein-Gordon field, is estimated to
be:%
\begin{equation}
\rho_{0}\approx\frac{\sqrt{\langle c^{\hspace{0.01in}2}\mathbf{p}^{2}%
\rangle_{S_{z}}}}{\mathcal{V}_{S_{z}}}\approx\frac{\sqrt{\frac{\hbar
^{2}c^{\hspace{0.01in}2}}{5\hspace{0.01in}\ell_{P}^{\,2}}}}{240\hspace
{0.01in}\ell_{P}^{\,3}}=0.0019\,\frac{\hbar c}{\ell_{P}^{\,4}}=0.9\times
10^{111}\,\text{J/m}^{3}. \label{DicVacEneAbs}%
\end{equation}
This result does not differ significantly from the value presented in
Eq.~(\ref{BerEneKlas}) of Chap.~\ref{ClasCal}.\footnote{The minor discrepancy
between Eq.(\ref{DicVacEneAbs}) and Eq.~(\ref{BerEneKlas}) of
Chap.~\ref{ClasCal} arises mainly because the volume of the region $S_{z}$
contains approximately $60$ quasipoints [cf. Eq.\ (\ref{NaeVolEleS})].}

\subsection{Vacuum Energy of the Entire $q$-De\-formed Euclidean
Space\label{KapVacEneTot}}

The vacuum energy of the entire $q$-de\-formed Euclidean space $\mathbb{R}%
_{q}^{3}$ is defined through the limiting process%
\begin{equation}
\langle H\rangle_{\mathbb{R}_{q}^{3}}=\lim_{S\rightarrow\mathbb{R}_{q}^{3}%
}\frac{\langle S\hspace{0.01in}|H|S\rangle}{\langle S\hspace{0.01in}|S\rangle
}=\lim_{S\rightarrow\mathbb{R}_{q}^{3}}\frac{\int\nolimits_{S}\text{d}_{q}%
^{3}x\int\nolimits_{S}\text{d}_{q}^{3}x^{\prime}\,\langle\mathbf{x}%
|H|\mathbf{x}^{\prime}\rangle}{\int\nolimits_{S}\text{d}_{q}^{3}%
x\int\nolimits_{S}\text{d}_{q}^{3}x^{\prime}\,\langle\mathbf{x}|\mathbf{x}%
^{\prime}\rangle},\label{VakEneDef}%
\end{equation}
where $H$ denotes the Hamiltonian operator of a $q$-de\-formed scalar field.
To evaluate this expression, we introduce a sequence of spatial regions
$S_{m}$ satisfying%
\begin{equation}
\lim_{m\rightarrow\infty}S_{m}=\mathbb{R}_{q}^{3}.
\end{equation}
Each $S_{m}$ is chosen as a $q$-ana\-logue of a cuboid centered at the origin,
with edge lengths $2\hspace{0.01in}\alpha_{A}\hspace{0.01in}q^{(2-\delta
_{A3})\hspace{0.01in}m}$, where $A\in\left\{  +,3,-\right\}  $ and $\alpha_{A}\in\left(  1,q^{2-\delta_{A3}%
}\right)$. On the
$q$-lat\-tice, $S_{m}$ thus consists of all points of the form%
\begin{equation}
(z^{+}|\hspace{0.01in}z^{3}|\hspace{0.01in}z^{-})=(\varepsilon_{+}%
\hspace{0.01in}\alpha_{+}\hspace{0.01in}q^{2k_{+}}|\hspace{0.01in}%
\varepsilon_{3}\hspace{0.01in}\alpha_{3}\hspace{0.01in}q^{k_{3}}%
|\hspace{0.01in}\varepsilon_{-}\hspace{0.01in}\alpha_{-}\hspace{0.01in}%
q^{2k_{-}}),
\end{equation}
where, for $A\in\{+,3,-\}$ and $q>1$,\footnote{In this section, we assume
$q>1$. The arguments, however, remain valid for $0<q<1$ with minor
modifications.}%
\begin{equation}
k_{A}\in\left\{  z\in\mathbb{Z}|\hspace{0.01in}z\leq m\right\}  =\left\{
\ldots,-2,-1,0,1,2,\ldots m-1,m\right\}  ,
\end{equation}
and%
\begin{equation}
\varepsilon_{A}\in\{+,-\}  .
\end{equation}
Accordingly, the states appearing in Eq.~(\ref{VakEneDef}) take the form%
\begin{equation}
|S_{m}\rangle=\sum_{\varepsilon_{A}\hspace{0.01in}=\hspace{0.01in}\pm}%
\sum_{k_{A}\hspace{0.01in}=\hspace{0.01in}-\infty}^{m}\left.  |\mathbf{x}%
\rangle\,\text{d}_{q}^{3}\hspace{0.01in}x\right\vert _{_{x^{A}=\hspace
{0.01in}\pm\alpha_{A}\hspace{0.01in}q^{(2-\delta_{A3})\hspace{0.01in}k_{A}}}},
\end{equation}
with the summation shorthand%
\begin{equation}
\sum_{\varepsilon_{A}\hspace{0.01in}=\hspace{0.01in}\pm}=\sum_{\varepsilon
_{+},\hspace{0.01in}\varepsilon_{3},\hspace{0.01in}\varepsilon_{-}%
\hspace{0.01in}=\hspace{0.01in}\pm},\qquad\sum_{k_{A}\hspace{0.01in}=-\infty
}^{m}=\sum_{k_{+},\hspace{0.01in}k_{3},\hspace{0.01in}k_{-}\hspace
{0.01in}=\hspace{0.01in}-\infty}^{m}.
\end{equation}
Hence, the vacuum energy of $\mathbb{R}_{q}^{3}$ can be written as%
\begin{equation}
\langle H\rangle_{\mathbb{R}_{q}^{3}}=\lim_{m\rightarrow\infty}\frac{\langle
S_{m}\hspace{0.01in}|H|S_{m}\rangle}{\langle S_{m}\hspace{0.01in}|S_{m}%
\rangle}.\label{DefGesEne}%
\end{equation}

As in the previous chapter, we focus on the operator $H^{2}$, because
calculating its expectation values is technically simpler than working
directly with $H$. In the limit $q\rightarrow1$, the expectation value of $H$
can be inferred from that of $H^{2}$, so this choice is made without loss of generality.
For a massless $q$-de\-formed scalar field, we have%
\begin{equation}
\langle H^{2}\rangle_{S_{m}}=\frac{\langle S_{m}\hspace{0.01in}|H^{2}%
|S_{m}\rangle}{\langle S_{m}\hspace{0.01in}|S_{m}\rangle}=\frac{\langle
S_{m}\hspace{0.01in}|c^{\hspace{0.01in}2}\mathbf{p}^{2}|S_{m}\rangle}{\langle
S_{m}\hspace{0.01in}|S_{m}\rangle}. \label{ErwHQua}%
\end{equation}
The numerator involves the matrix elements of $\mathbf{p}^{2}$,%
\begin{equation}
\langle S_{m}\hspace{0.01in}|\mathbf{p}^{2}|S_{m}\rangle=\sum_{\varepsilon
_{A},\hspace{0.01in}\varepsilon_{B}\hspace{0.01in}=\hspace{0.01in}\pm}%
\,\sum_{k_{A},\hspace{0.01in}l_{B}\hspace{0.01in}=\hspace{0.01in}-\infty}%
^{m}\left.  \text{d}_{q}^{3}\hspace{0.01in}x\,\langle\mathbf{x}|\mathbf{p}%
^{2}|\mathbf{\tilde{y}}\rangle\,\text{d}_{q}^{3}\hspace{0.01in}\tilde
{y}\right\vert _{_{_{\substack{x^{A}=\hspace{0.01in}\varepsilon_{A}%
\hspace{0.01in}\alpha_{A}\hspace{0.01in}q^{(2-\delta_{A3})\hspace{0.01in}%
k_{A}}\\\tilde{y}^{B}=\hspace{0.01in}\varepsilon_{B}^{\prime}\hspace
{0.01in}\alpha_{B}\hspace{0.01in}q^{(2-\delta_{B3})\hspace{0.01in}l_{B}}}}}},
\label{MatSmP2}%
\end{equation}
while the denominator simply gives the $q$-vol\-ume of $S_{m}$:%
\begin{equation}
\langle S_{m}\hspace{0.01in}|S_{m}\rangle=\sum_{\varepsilon_{A},\varepsilon
_{B}^{\prime}\hspace{0.01in}=\hspace{0.01in}\pm}\,\sum_{k_{A},\hspace
{0.01in}l_{B}\hspace{0.01in}=\hspace{0.01in}-\infty}^{m}\left.  \text{d}%
_{q}^{3}\hspace{0.01in}x\,\langle\mathbf{x}|\mathbf{\tilde{y}}\rangle
\,\text{d}_{q}^{3}\hspace{0.01in}\tilde{y}\right\vert _{_{\substack{x^{A}%
=\hspace{0.01in}\varepsilon_{A}\hspace{0.01in}\alpha_{A}\hspace{0.01in}%
q^{(2-\delta_{A3})\hspace{0.01in}k_{A}}\\\tilde{y}^{B}=\hspace{0.01in}%
\varepsilon_{B}^{\prime}\hspace{0.01in}\alpha_{B}\hspace{0.01in}%
q^{(2-\delta_{B3})\hspace{0.01in}l_{B}}}}}. \label{VolSm}%
\end{equation}

To evaluate Eqs.~(\ref{MatSmP2}) and (\ref{VolSm}), we compute $\langle
\mathbf{x}|\mathbf{p}^{2}|\mathbf{\tilde{y}}\rangle$ and $\langle
\mathbf{x}|\mathbf{\tilde{y}}\rangle$. The identity in Eq.~(\ref{MatEleP2Ort})
of the previous chapter, combined with Eq.~(\ref{DarPotAblQuaDreEnt}) in
Chap.~\ref{VacEneQEuc} for $n=1$, yields%
\begin{align}
\langle\mathbf{x}|\mathbf{p}^{2}|\mathbf{\tilde{y}}\rangle=  &
-\operatorname*{vol}\nolimits^{-1}\hspace{-0.01in}\mathbf{\partial}_{x}%
\hspace{-0.01in}\circ\mathbf{\partial}_{x}\triangleright\delta_{q}%
^{3}(\mathbf{x}\oplus(\ominus\hspace{0.01in}\kappa^{-1}\mathbf{\tilde{y}%
}))\nonumber\\
=  &  \operatorname*{vol}\nolimits^{-1}(q+q^{-1})\,\partial_{-}\partial
_{+}\triangleright\delta_{q}^{3}(\mathbf{x}\oplus(\ominus\hspace{0.01in}%
\kappa^{-1}\mathbf{\tilde{y}}))\nonumber\\
&  -\operatorname*{vol}\nolimits^{-1}q^{-2}\,\partial_{3}\partial
_{3}\triangleright\delta_{q}^{3}(\mathbf{x}\oplus(\ominus\hspace{0.01in}%
\kappa^{-1}\mathbf{\tilde{y}})).
\end{align}
Substituting this result into Eq.~(\ref{MatSmP2}) and applying
Eq.~(\ref{WirImpPotDreQDel}) in Chap.~\ref{VacEneQEuc} gives%
\begin{align}
\langle S_{m}\hspace{0.01in}|\mathbf{p}^{2}|S_{m}\rangle\approx &
\sum_{\varepsilon_{A},\hspace{0.01in}\varepsilon_{B}^{\prime}\hspace
{0.01in}=\hspace{0.01in}\pm}\,\sum_{k_{A},\hspace{0.01in}l_{B}\hspace
{0.01in}=\hspace{0.01in}-\infty}^{m}\text{d}_{q}^{3}\hspace{0.01in}%
x\,q^{-4}(q+q^{-1})\,d_{1}^{(-)}(x^{-}\hspace{-0.01in},\tilde{y}%
^{-})\nonumber\\
&  \qquad\times\left.  d_{0,1}^{(3)}(x^{3}\hspace{-0.01in},\tilde{y}%
^{3})\,d_{1,0}^{(+)}(x^{+}\hspace{-0.01in},\tilde{y}^{+})\,\text{d}_{q}%
^{3}\hspace{0.01in}\tilde{y}\right\vert _{_{\substack{x^{A}=\hspace
{0.01in}\varepsilon_{A}\hspace{0.01in}\alpha_{A}\hspace{0.01in}q^{(2-\delta
_{A3})\hspace{0.01in}k_{A}}\\\tilde{y}^{B}=\hspace{0.01in}\varepsilon
_{B}^{\prime}\hspace{0.01in}\alpha_{B}\hspace{0.01in}q^{(2-\delta_{B3}%
)\hspace{0.01in}l_{B}}}}}\nonumber\\
&  -\sum_{\varepsilon_{A},\varepsilon_{B}^{\prime}\hspace{0.01in}%
=\hspace{0.01in}\pm}\,\sum_{k_{A},\hspace{0.01in}l_{B}\hspace{0.01in}%
=\hspace{0.01in}-\infty}^{m}\text{d}_{q}^{3}\hspace{0.01in}x\,q^{-6}%
\,d_{0}^{(-)}(x^{-}\hspace{-0.01in},\tilde{y}^{-})\nonumber\\
&  \qquad\times\left.  d_{2,0}^{(3)}(x^{3}\hspace{-0.01in},\tilde{y}%
^{3})\,d_{0,2}^{(+)}(x^{+}\hspace{-0.01in},\tilde{y}^{+})\,\text{d}_{q}%
^{3}\hspace{0.01in}\tilde{y}\right\vert _{_{\substack{x^{A}=\hspace
{0.01in}\varepsilon_{A}\hspace{0.01in}\alpha_{A}\hspace{0.01in}q^{(2-\delta
_{A3})\hspace{0.01in}k_{A}}\\\tilde{y}^{B}=\hspace{0.01in}\varepsilon
_{B}^{\prime}\hspace{0.01in}\alpha_{B}\hspace{0.01in}q^{(2-\delta_{B3}%
)\hspace{0.01in}l_{B}}}}}. \label{ErwSmP2}%
\end{align}
Similarly, Eq.~(\ref{VolRelOrtZus}) in the previous chapter, combined with
Eqs.~(\ref{DarPotAblQuaDreEnt}) and (\ref{WirImpPotDreQDel}) in
Chap.~\ref{VacEneQEuc} for $n=0$, yields%
\begin{align}
\langle\mathbf{x}|\mathbf{\tilde{y}}\rangle &  =\operatorname*{vol}%
\nolimits^{-1}\hspace{-0.01in}\delta_{q}^{3}(\mathbf{x}\oplus(\ominus
\hspace{0.01in}\kappa^{-1}\mathbf{\tilde{y}}))\nonumber\\
&  \approx q^{-4}\,d_{0}^{(-)}(x^{-}\hspace{-0.01in},\tilde{y}^{-}%
)\,d_{0,0}^{(3)}(x^{3}\hspace{-0.01in},\tilde{y}^{3})\,d_{0,0}^{(+)}%
(x^{+}\hspace{-0.01in},\tilde{y}^{+}).
\end{align}
Inserting this result into Eq.~(\ref{VolSm}) finally gives:%
\begin{align}
\langle S_{m}\hspace{0.01in}|S_{m}\rangle\approx &  \sum_{\varepsilon
_{A},\hspace{0.01in}\varepsilon_{B}^{\prime}\hspace{0.01in}=\hspace{0.01in}%
\pm}\sum_{k_{A},\hspace{0.01in}l_{B}=-\infty}^{m}\text{d}_{q}^{3}%
\hspace{0.01in}x\,q^{-4}\,d_{0}^{(-)}(x^{-}\hspace{-0.01in},\tilde{y}%
^{-})\,\nonumber\\
&  \qquad\times\left.  d_{0,0}^{(3)}(x^{3}\hspace{-0.01in},\tilde{y}%
^{3})\,d_{0,0}^{(+)}(x^{+}\hspace{-0.01in},\tilde{y}^{+})\,\text{d}_{q}%
^{3}\hspace{0.01in}\tilde{y}\right\vert _{_{\substack{x^{A}=\hspace
{0.01in}\varepsilon_{A}\hspace{0.01in}\alpha_{A}\hspace{0.01in}q^{(2-\delta
_{A3})\hspace{0.01in}k_{A}}\\\tilde{y}^{B}=\hspace{0.01in}\varepsilon
_{B}^{\prime}\hspace{0.01in}\alpha_{B}\hspace{0.01in}q^{(2-\delta_{B3}%
)\hspace{0.01in}l_{B}}}}}. \label{SkaSmWdh}%
\end{align}

The first term on the right-hand side of Eq.~(\ref{ErwSmP2}) factorizes into
three independent series, each associated with one spatial coordinate. These
series are given explicitly by [cf. Eq.~(\ref{QDefVolDreWdh}) of
Chap.~\ref{VacEneQuaPoi}]%
\begin{equation}
\sum_{\varepsilon_{3},\hspace{0.01in}\varepsilon_{3}^{\prime}=\hspace
{0.01in}\pm}\,\sum_{k_{3},\hspace{0.01in}l_{3}\hspace{0.01in}=\hspace
{0.01in}-\infty}^{m}\left.  \left\vert (1-q)^{2}\hspace{0.01in}x^{3}%
\hspace{0.01in}\tilde{y}^{3}\right\vert \,d_{0,1}^{(3)}(x^{3}\hspace
{-0.01in},\tilde{y}^{3})\right\vert _{_{\substack{x^{3}=\hspace{0.01in}%
\varepsilon_{3}\hspace{0.01in}\alpha_{3}\hspace{0.01in}q^{k_{3}}\\\tilde
{y}^{3}=\hspace{0.01in}\varepsilon_{3}^{\prime}\hspace{0.01in}\alpha
_{3}\hspace{0.01in}q^{l_{3}}}}}=2\hspace{0.01in}q\hspace{0.01in}\alpha
_{3}\hspace{0.01in}q^{m}, \label{MehSum01}%
\end{equation}
and%
\begin{align}
\sum_{\varepsilon_{-},\hspace{0.01in}\varepsilon_{-}^{\prime}=\hspace
{0.01in}\pm}\,\sum_{k_{-},l_{-}\hspace{0.01in}=\hspace{0.01in}-\infty}%
^{m}\left.  \left\vert (1-q^{2})^{2}\hspace{0.01in}x^{-}\hspace{0.01in}%
\tilde{y}^{-}\right\vert \,d_{1}^{(-)}(x^{-}\hspace{-0.01in},\tilde{y}%
^{-})\right\vert _{_{\substack{x^{-}=\hspace{0.01in}\varepsilon_{-}%
\hspace{0.01in}\alpha_{-}\hspace{0.01in}q^{2k_{-}}\\\tilde{y}^{-}%
=\hspace{0.01in}\varepsilon_{-}^{\prime}\hspace{0.01in}\alpha_{-}%
\hspace{0.01in}q^{2l_{-}}}}}  &  =0,\nonumber\\
\sum_{\varepsilon_{+},\hspace{0.01in}\varepsilon_{+}^{\prime}\hspace
{0.01in}=\hspace{0.01in}\pm}\,\sum_{k_{+},\hspace{0.01in}l_{+}\hspace
{0.01in}=\hspace{0.01in}-\infty}^{m}\left.  \left\vert (1-q^{2})^{2}%
\hspace{0.01in}x^{+}\hspace{0.01in}\tilde{y}^{+}\right\vert \hspace
{0.01in}d_{1,0}^{(+)}(x^{+}\hspace{-0.01in},\tilde{y}^{+})\right\vert
_{_{\substack{x^{+}=\hspace{0.01in}\varepsilon_{+}\hspace{0.01in}\alpha
_{+}\hspace{0.01in}q^{2k_{+}}\\\tilde{y}^{+}=\hspace{0.01in}\varepsilon
_{+}^{\prime}\hspace{0.01in}\alpha_{+}\hspace{0.01in}q^{2l_{+}}}}}  &  =0.
\label{MehSum1}%
\end{align}
These identities follow directly from Eqs.~(\ref{MehSumPhi}) and
(\ref{MehSumAblPhi}) in App.~\ref{AnhHerFor}, after incorporating the
relations in Eq.~(\ref{WirImpPotDreQDelEin}) of Chap.~\ref{VacEneQEuc}. The
second term on the right-hand side of Eq.~(\ref{ErwSmP2}) likewise factorizes
into three contributions, given by%
\begin{gather}
\sum_{\varepsilon_{3},\hspace{0.01in}\varepsilon_{3}^{\prime}\hspace
{0.01in}=\hspace{0.01in}\pm}\,\sum_{k_{3},\hspace{0.01in}l_{3}\hspace
{0.01in}=\hspace{0.01in}-\infty}^{m}\left.  \left\vert (1-q)^{2}%
\hspace{0.01in}x^{3}\hspace{0.01in}\tilde{y}^{3}\right\vert \,d_{2,0}%
^{(3)}(x^{3}\hspace{-0.01in},\tilde{y}^{3})\right\vert _{_{\substack{x^{3}%
=\hspace{0.01in}\varepsilon_{3}\hspace{0.01in}\alpha_{3}\hspace{0.01in}%
q^{k_{3}}\\\tilde{y}^{3}=\hspace{0.01in}\varepsilon_{3}^{\prime}%
\hspace{0.01in}\alpha_{3}\hspace{0.01in}q^{l_{3}}}}}=\nonumber\\
=\frac{2\hspace{0.01in}q^{-2}}{(1-q^{2})\hspace{0.01in}\alpha_{3}%
\hspace{0.01in}q^{m}}, \label{MehSum20}%
\end{gather}
and%
\begin{gather}
\sum_{\varepsilon_{-},\hspace{0.01in}\varepsilon_{-}^{\prime}\hspace
{0.01in}=\hspace{0.01in}\pm}\,\sum_{k_{-},\hspace{0.01in}l_{-}\hspace
{0.01in}=\hspace{0.01in}-\infty}^{m}\left.  \left\vert (1-q^{2})^{2}%
\hspace{0.01in}x^{-}\hspace{0.01in}\tilde{y}^{-}\right\vert \,d_{0}%
^{(-)}(x^{-}\hspace{-0.01in},\tilde{y}^{-})\right\vert _{_{\substack{x^{-}%
=\hspace{0.01in}\varepsilon_{-}\hspace{0.01in}\alpha_{-}\hspace{0.01in}%
q^{2k_{-}}\\\tilde{y}^{-}=\hspace{0.01in}\varepsilon_{-}^{\prime}%
\hspace{0.01in}\alpha-\hspace{0.01in}q^{2l_{-}}}}}=\nonumber\\
=2\hspace{0.01in}q^{2}\alpha_{-}\hspace{0.01in}q^{2m},\nonumber\\[0.07in]
\sum_{\varepsilon_{+},\hspace{0.01in}\varepsilon_{+}^{\prime}\hspace
{0.01in}=\hspace{0.01in}\pm}\,\sum_{k_{+},\hspace{0.01in}l_{+}\hspace
{0.01in}=\hspace{0.01in}-\infty}^{m}\left.  \left\vert (1-q^{2})^{2}%
\hspace{0.01in}x^{+}\hspace{0.01in}\tilde{y}^{+}\right\vert \,d_{0,2}%
^{(+)}(x^{+}\hspace{-0.01in},\tilde{y}^{+})\right\vert _{_{\substack{x^{+}%
=\hspace{0.01in}\varepsilon_{+}\hspace{0.01in}\alpha_{+}\hspace{0.01in}%
q^{2k_{+}}\\\tilde{y}^{+}=\hspace{0.01in}\varepsilon_{+}^{\prime}%
\hspace{0.01in}\alpha_{+}\hspace{0.01in}q^{2l_{+}}}}}=\nonumber\\
=2\hspace{0.01in}\alpha_{+}\hspace{0.01in}q^{2m}. \label{MehSum2}%
\end{gather}
These results follow from Eqs.~(\ref{MehSumPhi}) and (\ref{MehSumAbl2Phi})
in\ App.~\ref{AnhHerFor}, again using Eq.~(\ref{WirImpPotDreQDelEin}) in
Chap.~\ref{VacEneQEuc}.\ Substituting Eqs.~(\ref{MehSum01})-(\ref{MehSum2})
into Eq.~(\ref{ErwSmP2}) yields%
\begin{equation}
\langle S_{m}\hspace{0.01in}|\mathbf{p}^{2}|S_{m}\rangle\approx\frac
{8\hspace{0.01in}q^{-6}\hspace{0.01in}\alpha_{+}\hspace{0.01in}q^{2m}%
\hspace{0.01in}\alpha_{-}\hspace{0.01in}q^{2m}}{(q^{2}-1)\hspace{0.01in}%
\alpha_{3}\hspace{0.01in}q^{m}}. \label{AusSmp2Sm}%
\end{equation}
A similar argument, employing in addition the relations%
\begin{gather}
\sum_{\varepsilon_{3},\hspace{0.01in}\varepsilon_{3}^{\prime}=\hspace
{0.01in}\pm}\,\sum_{k_{3},\hspace{0.01in}l_{3}\hspace{0.01in}=\hspace
{0.01in}-\infty}^{m}\left.  \left\vert (1-q)^{2}\hspace{0.01in}x^{3}%
\hspace{0.01in}\tilde{y}^{3}\right\vert \,d_{0,0}^{(3)}(x^{3},\tilde{y}%
^{3})\right\vert _{_{\substack{x^{3}=\hspace{0.01in}\varepsilon_{3}%
\hspace{0.01in}\alpha_{3}\hspace{0.01in}q^{k_{3}}\\\tilde{y}^{3}%
=\hspace{0.01in}\varepsilon_{3}^{\prime}\hspace{0.01in}\alpha_{3}%
\hspace{0.01in}q^{l_{3}}}}}=\nonumber\\
=2\hspace{0.01in}q\hspace{0.01in}\alpha_{3}\hspace{0.01in}q^{m},
\label{MehSum30}%
\end{gather}
and%
\begin{gather}
\sum_{\varepsilon_{-},\hspace{0.01in}\varepsilon_{-}^{\prime}\hspace
{0.01in}=\hspace{0.01in}\pm}\,\sum_{k_{-},\hspace{0.01in}l_{-}\hspace
{0.01in}=\hspace{0.01in}-\infty}^{m}\left.  \left\vert (1-q^{2})^{2}%
\hspace{0.01in}x^{-}\hspace{0.01in}\tilde{y}^{-}\right\vert \,d_{0}%
^{(-)}(x^{-},\tilde{y}^{-})\right\vert _{_{\substack{x^{-}=\hspace
{0.01in}\varepsilon_{-}\hspace{0.01in}\alpha_{-}\hspace{0.01in}q^{2k_{-}%
}\\\tilde{y}^{-}=\hspace{0.01in}\varepsilon_{-}^{\prime}\hspace{0.01in}%
\alpha_{-}\hspace{0.01in}q^{2l_{-}}}}}=\nonumber\\
=2\hspace{0.01in}q^{2}\alpha_{-}\hspace{0.01in}q^{2m},\nonumber\\[0.07in]
\sum_{\varepsilon_{+},\hspace{0.01in}\varepsilon_{+}^{\prime}\hspace
{0.01in}=\hspace{0.01in}\pm}\,\sum_{k_{+},\hspace{0.01in}l_{+}\hspace
{0.01in}=\hspace{0.01in}-\infty}^{m}\left.  \left\vert (1-q^{2})^{2}%
\hspace{0.01in}x^{+}\hspace{0.01in}\tilde{y}^{+}\right\vert \hspace
{0.01in}d_{0,0}^{(+)}(x^{+},\tilde{y}^{+})\right\vert _{_{\substack{x^{+}%
=\hspace{0.01in}\varepsilon_{+}\hspace{0.01in}\alpha_{+}\hspace{0.01in}%
q^{2k_{+}}\\\tilde{y}^{+}=\hspace{0.01in}\varepsilon_{+}^{\prime}%
\hspace{0.01in}\alpha_{+}\hspace{0.01in}q^{2l_{+}}}}}=\nonumber\\
=2\hspace{0.01in}q^{2}\alpha_{+}\hspace{0.01in}q^{2m}, \label{MehSum3}%
\end{gather}
which follow from Eq.~(\ref{MehSumPhi}) in App.~\ref{AnhHerFor}, leads from
Eq.~(\ref{SkaSmWdh}) to%
\begin{equation}
\langle S_{m}\hspace{0.01in}|S_{m}\rangle\approx8\hspace{0.01in}%
q\hspace{0.01in}\alpha_{+}\hspace{0.01in}q^{2m}\hspace{0.01in}\alpha
_{-}\hspace{0.01in}q^{2m}\hspace{0.01in}\alpha_{3}\hspace{0.01in}q^{m}.
\label{AusSmSm}%
\end{equation}

Substituting Eqs.~(\ref{AusSmp2Sm}) and (\ref{AusSmSm}) into
Eq.~(\ref{ErwHQua}) gives the following approximate expression for the
expectation value $\langle H^{2}\rangle_{S_{m}}$, now explicitly including
$\hbar$:%
\begin{equation}
\langle H^{2}\rangle_{S_{m}}=\frac{\langle S_{m}\hspace{0.01in}|c^{\hspace
{0.01in}2}\mathbf{p}^{2}|S_{m}\rangle}{\langle S_{m}\hspace{0.01in}%
|S_{m}\rangle}\approx\frac{q^{-4}}{(q-q^{-1})}\frac{c^{\hspace{0.01in}2}%
\hbar^{2}}{(\alpha_{3}\hspace{0.01in}q^{m})^{2}}\approx\frac{c^{\hspace
{0.01in}2}\hbar^{2}}{2\hspace{0.01in}h\hspace{0.01in}z_{3}^{2}}.
\label{ErwEneQuaEpx}%
\end{equation}
In the last step of Eq.~(\ref{ErwEneQuaEpx}), we identified $\alpha_{3}q^{m}$
with the coordinate $z_{3}$, where $2\hspace{0.01in}z^{3}$ denotes the spatial
extent of the $q$-de\-formed cuboid, and applied the approximations from
Eq.~(\ref{NaehAusKla}) of the previous chapter.

From Eq.~(\ref{ErwEneQuaEpx}), it follows that $\langle H^{2}\rangle_{S_{m}}$
decreases monotonically as the region $S_{m}$ centered at the origin expands.
Using the inequality $\langle E\rangle^{2}\leq\langle E^{2}\rangle$ and the
definition in Eq.~(\ref{DefGesEne}),\ we conclude that, under assumptions
underlying Eq.~(\ref{ErwEneQuaEpx}), the vacuum energy of the complete
$q$-de\-formed Euclidean space vanishes:%
\begin{equation}
\langle H\rangle_{\mathbb{R}_{q}^{3}}=\lim_{m\rightarrow\infty}\langle
H\rangle_{S_{m}}\leq\lim_{m\rightarrow\infty}\sqrt{\langle H^{2}\rangle
_{S_{m}}}\approx\lim_{z_{3}\rightarrow\infty}\frac{c\hbar}{\sqrt
{2\hspace{0.01in}h}\hspace{0.01in}z_{3}}=0. \label{VacEneTotEuc}%
\end{equation}

The vanishing contribution of a massless Klein-Gordon field to the total
vacuum energy of the $q$-de\-formed Euclidean space can also be derived
directly from Eq.~(\ref{VakEneDef}). We begin by considering the matrix
elements of the Hamiltonian operator for a massive Klein-Gordon field in the
$q$-de\-formed Euclidean space [cf. Eqs.~(\ref{EntWicHamAlg}) in
Chap.~\ref{VacEneQEuc}, as well as Eq.~(\ref{EneKleGorSer}) in
App.~\ref{KapPlaWavSol}]:%
\begin{equation}
\langle\mathbf{x}|H|\mathbf{x}^{\prime}\rangle=%
c\sum_{n\hspace{0.01in}=\hspace{0.01in}0}^{\infty}\binom{1/2}%
{n}(m\hspace{0.01in}c)^{1-2n}\,\langle\mathbf{x}|\mathbf{p}^{2n}%
|\mathbf{x}^{\prime}\rangle. \label{ReiEntHamMat}%
\end{equation}
We integrate the matrix elements of $\mathbf{p}^{2n}$ over the spatial
coordinates of the initial and the final state:%
\begin{gather}
\int\text{d}_{q}^{3}x^{\prime}\int
\text{d}_{q}^{3}x\,\langle\mathbf{x}|\mathbf{p}^{2n}|\mathbf{x}^{\prime
}\rangle=\nonumber\\
=\int\text{d}_{q}^{3}x^{\prime}\int
\text{d}_{q}^{3}x\big(-\mathbf{\partial}_{x}\circ\mathbf{\partial}%
_{x}\big)^{n}\triangleright\delta_{q}^{3}(\mathbf{x}\oplus(\ominus
\hspace{0.01in}\kappa^{-1}\mathbf{x}^{\prime})). \label{Rq3Hrq3}%
\end{gather}
Using the identities in Eq.~(\ref{AlgChaIdeqDelFkt}) of Chap.~\ref{AppExp}, it
follows for $n>0$ that%
\begin{gather}
\operatorname*{vol}\nolimits^{-1}\hspace{-0.02in}\int_{\mathbb{R}_{q}^{3}%
}\text{d}_{q}^{3}x\int_{\mathbb{R}_{q}^{3}}\text{d}_{q}^{3}x^{\prime}%
\hspace{-0.01in}\big(-\mathbf{\partial}_{x}\hspace{-0.01in}\circ
\mathbf{\partial}_{x}\big)^{n}\triangleright\delta_{q}^{3}(\mathbf{x}%
\oplus(\ominus\hspace{0.01in}\kappa^{-1}\mathbf{x}^{\prime}))\nonumber\\
=\int_{\mathbb{R}_{q}^{3}}\text{d}_{q}^{3}x\hspace{0.01in}%
\big(-\mathbf{\partial}_{x}\hspace{-0.01in}\circ\mathbf{\partial}_{x}%
\big)^{n}\triangleright1=0. \label{TotErwP2k}%
\end{gather}
Substituting Eq.~(\ref{ReiEntHamMat}) into Eq.~(\ref{VakEneDef}) and applying
Eq.~(\ref{TotErwP2k}), we obtain%
\begin{align}
\langle H\rangle_{\mathbb{R}_{q}^{3}}  &  =c\sum_{n\hspace{0.01in}%
=\hspace{0.01in}0}^{\infty}\binom{1/2}{n}\lim_{S\rightarrow\mathbb{R}_{q}^{3}%
}\frac{\operatorname*{vol}\nolimits^{-1}\hspace{-0.02in}\int\nolimits_{S}%
\text{d}_{q}^{3}x\int\nolimits_{S}\text{d}_{q}^{3}x^{\prime}\hspace
{-0.01in}\langle\mathbf{x}|\mathbf{p}^{2n}|\mathbf{x}^{\prime}\rangle
}{(m\hspace{0.01in}c)^{2n\hspace{0.01in}-1}\operatorname*{vol}\nolimits^{-1}%
\hspace{-0.02in}\int\nolimits_{S}\text{d}_{q}^{3}x\int\nolimits_{S}%
\text{d}_{q}^{3}x^{\prime}\,\langle\mathbf{x}|\mathbf{x}^{\prime}\rangle
}\nonumber\\
&  =\lim_{S\rightarrow\mathbb{R}_{q}^{3}}\frac{m\hspace{0.01in}c^{\hspace
{0.01in}2}\operatorname*{vol}\nolimits^{-1}\hspace{-0.02in}\int\nolimits_{S}%
\text{d}_{q}^{3}x\int\nolimits_{S}\text{d}_{q}^{3}x^{\prime}\hspace
{-0.01in}\langle\mathbf{x}|\mathbf{x}^{\prime}\rangle}{\operatorname*{vol}%
\nolimits^{-1}\hspace{-0.02in}\int\nolimits_{S}\text{d}_{q}^{3}x\int
\nolimits_{S}\text{d}_{q}^{3}x^{\prime}\langle\mathbf{x}|\mathbf{x}^{\prime
}\rangle}=m\hspace{0.01in}c^{\hspace{0.01in}2}. \label{HerVacEne}%
\end{align}
Thus, the vacuum energy of a $q$-de\-formed Klein-Gordon field equals the rest
energy of a single scalar particle. Consequently, for a massless
$q$-de\-formed Klein-Gordon field the vacuum energy vanishes [cf.
Eq.~(\ref{VacEneTotEuc})].

The computation in Eq.~(\ref{HerVacEne}) involves an interchange of limit
operations, a potentially delicate step. To avoid this issue, we instead
evaluate the expectation value of the squared Hamiltonian:%
\begin{align}
\langle H^{2}\rangle_{\mathbb{R}_{q}^{3}}  &  =\lim_{S\rightarrow
\mathbb{R}_{q}^{3}}\frac{\int\nolimits_{S}\text{d}_{q}^{3}x\int\nolimits_{S}%
\text{d}_{q}^{3}x^{\prime}\hspace{-0.01in}\int\text{d}_{q}^{3}p\,u_{\hspace
{0.01in}\mathbf{p}}(\mathbf{x})\circledast E_{\mathbf{p}}^{\hspace{0.01in}%
2}\circledast(u^{\ast})_{\mathbf{p}}(\mathbf{x}^{\prime})}{\operatorname*{vol}%
\nolimits^{-1}\hspace{-0.02in}\int\nolimits_{S}\text{d}_{q}^{3}x\int
\nolimits_{S}\text{d}_{q}^{3}x^{\prime}\,\delta_{q}^{3}(\mathbf{x}%
\oplus(\ominus\hspace{0.01in}\kappa^{-1}\mathbf{x}^{\prime}))}\nonumber\\
&  =\lim_{S\rightarrow\mathbb{R}_{q}^{3}}\frac{\int\nolimits_{S}\text{d}%
_{q}^{3}x\int\nolimits_{S}\text{d}_{q}^{3}x^{\prime}\hspace{-0.01in}%
\int\text{d}_{q}^{3}p\,u_{\hspace{0.01in}\mathbf{p}}(\mathbf{x})\circledast
c^{\hspace{0.01in}2}(\mathbf{p}^{2}+(m\hspace{0.01in}c)^{2})\circledast
(u^{\ast})_{\mathbf{p}}(\mathbf{x}^{\prime})}{\operatorname*{vol}%
\nolimits^{-1}\hspace{-0.02in}\int\nolimits_{S}\text{d}_{q}^{3}x\int
\nolimits_{S}\text{d}_{q}^{3}x^{\prime}\,\delta_{q}^{3}(\mathbf{x}%
\oplus(\ominus\hspace{0.01in}\kappa^{-1}\mathbf{x}^{\prime}))}\nonumber\\
&  =\lim_{S\rightarrow\mathbb{R}_{q}^{3}}\frac{\int\nolimits_{S}\text{d}%
_{q}^{3}x\int\nolimits_{S}\text{d}_{q}^{3}x^{\prime}\hspace{-0.01in}\left(
\langle\mathbf{x}|c^{\hspace{0.01in}2}\mathbf{p}^{2}|\mathbf{x}^{\prime
}\rangle+(m\hspace{0.01in}c^{\hspace{0.01in}2})^{2}\langle\mathbf{x}%
|\mathbf{x}^{\prime}\rangle\right)  }{\int\nolimits_{S}\text{d}_{q}^{3}%
x\int\nolimits_{S}\text{d}_{q}^{3}x^{\prime}\,\langle\mathbf{x}|\mathbf{x}%
^{\prime}\rangle}\nonumber\\
&  =(m\hspace{0.01in}c^{\hspace{0.01in}2})^{2}.
\end{align}
Thus, the vacuum expectation value of $H^{2}$ equals the square of the rest
energy of a single scalar particle. In the massless case, this quantity
vanishes, $\langle H^{2}\rangle_{\mathbb{R}_{q}^{3}}=0$. Since the inequality
$\langle E\rangle^{2}\leq\langle E^{2}\rangle$ holds in general, it follows
that the vacuum energy for a massless Klein-Gordon field also vanishes:
$\langle H\rangle_{\mathbb{R}_{q}^{3}}=0$.

Finally, we show that the result in Eq.~(\ref{TotErwP2k}) is consistent with
the approximate expressions in Eqs.~(\ref{WirImpPotDreQDel}) and
(\ref{WirImpPotDreQDelEin})\ of Chap.~\ref{VacEneQEuc}. To this end, for fixed
values of the spatial coordinates in the initial state, we
integrate\ Eq.~(\ref{WirImpPotDreQDel}) of Chap.~\ref{VacEneQEuc} over all
spatial coordinates of the final state:%
\begin{gather}
\int_{\mathbb{R}_{q}^{3}}\text{d}_{q}^{3}x\left.  (\partial_{-})^{m_{-}%
}(\partial_{\hspace{0.01in}3})^{m_{3}}(\partial_{+})^{m_{+}}\triangleright
\delta_{q}^{3}(\mathbf{x}\oplus(\ominus\hspace{0.01in}\kappa^{-1}%
\mathbf{\tilde{y}}))\right\vert _{\tilde{y}^{A}=\hspace{0.01in}\pm
\hspace{0.01in}\alpha_{A}\hspace{0.01in}q^{(2-\delta_{A3})\hspace{0.01in}%
l_{A}}}\sim\nonumber\\[0.02in]
\sim\int_{\mathbb{R}_{q}^{3}}\text{d}_{q}^{3}x\,d_{m_{+},\hspace{0.01in}m_{3}%
}^{(+)}(x^{+}\hspace{-0.01in},\tilde{y}^{+})\,d_{m_{3},\hspace{0.01in}m_{-}%
}^{(3)}(x^{3}\hspace{-0.01in},\tilde{y}^{3})\,d_{m_{-}}^{(-)}(x^{-}%
\hspace{-0.01in},\tilde{y}^{-}).\label{SumEneFinZus}%
\end{gather}
The integrals on the right-hand side can be evaluated using the identities%
\begin{align}
&  \left.  \int\text{d}_{q}x^{3}\,d_{m_{3},\hspace{0.01in}m_{-}}^{(3)}%
(x^{3}\hspace{-0.01in},\tilde{y}^{3})\right\vert _{\tilde{y}^{3}%
=\hspace{0.01in}\pm\hspace{0.01in}\alpha_{3}\hspace{0.01in}q^{\hspace
{0.01in}l_{3}}}=\nonumber\\
&  \qquad=\frac{(q^{2(m_{3}-1)};q^{-2})_{m_{3}}}{(\pm\hspace{0.01in}\alpha
_{3}(1-q^{2}))^{m_{3}}\hspace{0.01in}q^{\hspace{0.01in}(l_{3}-2m_{-}%
+\hspace{0.01in}2)(m_{3}-1)}\hspace{0.01in}q^{\hspace{0.01in}l_{3}}%
},\label{IntX3AbQlDelt}%
\end{align}
and%
\begin{align}
&  \left.  \int\text{d}_{q^{2}}x^{+}\hspace{0.01in}d_{m_{+},\hspace
{0.01in}m_{3}}^{(+)}(x^{+}\hspace{-0.01in},\tilde{y}^{+})\right\vert
_{\tilde{y}^{+}=\hspace{0.01in}\pm\hspace{0.01in}\alpha_{+}\hspace
{0.01in}q^{2l_{+}}}=\nonumber\\
&  \qquad=\frac{(q^{4(m_{+}-1)};q^{-4})_{m_{+}}}{(\pm\hspace{0.01in}\alpha
_{+}(1-q^{4}))^{m_{+}}\hspace{0.01in}q^{2(l_{+}-\hspace{0.01in}m_{3}%
)(m_{+}-1)}\hspace{0.01in}q^{2l_{+}}},\nonumber\\[0.08in]
&  \left.  \int\text{d}_{q^{2}}x^{-}\hspace{0.01in}d_{m_{-}}^{(-)}%
(x^{-}\hspace{-0.01in},\tilde{y}^{-})\right\vert _{\tilde{y}^{-}%
=\hspace{0.01in}\pm\hspace{0.01in}\alpha_{-}\hspace{0.01in}q^{2l_{-}}%
}=\nonumber\\
&  \qquad=\frac{(q^{4(m_{-}-1)};q^{-4})_{m_{-}}}{(\pm\hspace{0.01in}\alpha
_{-}(1-q^{4}))^{m_{-}}\hspace{0.01in}q^{2(l_{-}+1)(m_{-}-1)}\hspace
{0.01in}q^{2l_{-}}}.\label{IntXMinAbQlDelt}%
\end{align}
As an example, we derive the first of these identities explicitly [cf.
Eq.~(\ref{WirImpPotDreQDelEin}) in Chap.~\ref{VacEneQEuc}]:%
\begin{align}
&  \left.  \int\text{d}_{q}x^{3}\,\frac{D_{q^{2},\hspace{0.01in}x^{3}}^{m_{3}%
}\phi_{q}(q^{2(m_{-}-1)}x^{3}\hspace{-0.01in},\tilde{y}^{3})}{\left\vert
(1-q)\hspace{0.01in}\tilde{y}^{3}\right\vert }\right\vert _{\tilde{y}%
^{3}=\hspace{0.01in}\pm\hspace{0.01in}\alpha_{3}\hspace{0.01in}q^{\hspace
{0.01in}l_{3}}}=\\
&  \quad=\sum_{k_{3}=\hspace{0.01in}-\infty}^{\infty}\left\vert 1-q\right\vert
\hspace{0.01in}\alpha_{3}\hspace{0.01in}q^{k_{3}}\sum_{j_{3}=\hspace{0.01in}%
0}^{m_{3}}\frac{(-1)^{j_{3}}q^{-j_{3}(j_{3}-1)}}{(\pm\hspace{0.01in}\alpha
_{3}\hspace{0.01in}q^{k_{3}}(1-q^{2}))^{m_{3}}}\frac{\delta_{-2+2m_{-}%
+\hspace{0.01in}2j_{3}+k_{3},\hspace{0.01in}l_{3}}}{\alpha_{3}\hspace
{0.01in}q^{\hspace{0.01in}l_{3}}|1-q|}%
\genfrac{[}{]}{0pt}{}{m_{3}}{j_{3}}%
_{q^{-2}}\nonumber\\
&  \quad=\sum_{j_{3}=\hspace{0.01in}0}^{m_{3}}\frac{(-1)^{j_{3}}%
q^{-j_{3}(j_{3}-1)}}{(\pm\hspace{0.01in}\alpha_{3}(1-q^{2}))^{m_{3}}%
q^{(l_{3}-\hspace{0.01in}2m_{-}-\hspace{0.01in}2j_{3}+2)(m_{3}-1)}%
q^{\hspace{0.01in}l_{3}}}%
\genfrac{[}{]}{0pt}{}{m_{3}}{j_{3}}%
_{q^{-2}}\nonumber\\
&  \quad=\frac{(q^{2(m_{3}-1)};q^{-2})_{m_{3}}}{(\pm\hspace{0.01in}\alpha
_{3}(1-q^{2}))^{m_{3}}\hspace{0.01in}q^{(l_{3}-\hspace{0.01in}2m_{-}%
+\hspace{0.01in}2)(m_{3}-1)}\hspace{0.01in}q^{\hspace{0.01in}l_{3}}}.
\end{align}
In this derivation, we started by expressing the Jackson integral explicitly
[cf. Eq.~(\ref{UneJackIntAll}) in Chap.~\ref{KapQIntTrig}] and noted that
$x^{3}$ and $\tilde{y}^{3}$ must share the same sign, as required by the
function $\phi_{q}$ [cf. Eqs.~(\ref{ChaIdePhi1}) and (\ref{ChaIdePhi2}) in
Chap.~\ref{KapQDisEin}]. Using Eq.~(\ref{MehAnwJacAblFkt}) from
Chap.~\ref{KapqAnaExpTriFkt}, we evaluated the higher-order Jackson
derivatives and applied the identity%
\begin{equation}
\phi_{q}(q^{-2+2m_{-}}q^{2j_{3}}\alpha_{3}\hspace{0.01in}q^{k_{3}},\alpha
_{3}\hspace{0.01in}q^{\hspace{0.01in}l_{3}})=\delta_{-2+2m_{-}+\hspace
{0.01in}2j_{3}+\hspace{0.01in}k_{3},\hspace{0.01in}l_{3}}.
\end{equation}
We concluded by rewriting the sum in terms of $q$-Pochhammer symbols
\cite{Klimyk:1997eb}:%
\begin{equation}
(z\hspace{0.01in};q)_{m}=(1-z)(1-z\hspace{0.01in}q)\ldots(1-z\hspace
{0.01in}q^{m-1})=\sum_{k\hspace{0.01in}=\hspace{0.01in}0}^{m}%
\genfrac{[}{]}{0pt}{}{m}{k}%
_{q}q^{k(k\hspace{0.01in}-1)/2}(-z)^{k}.\label{QBinThePocHam}%
\end{equation}
If $m_{+}$, $m_{-}$, or $m_{3}$ is a natural number, then%
\begin{gather}
\int_{\mathbb{R}_{q}^{3}}\text{d}_{q}^{3}x\left.  (\partial_{-})^{m_{-}%
}(\partial_{\hspace{0.01in}3})^{m_{3}}(\partial_{+})^{m_{+}}\triangleright
\delta_{q}^{3}(\mathbf{x}\oplus(\ominus\hspace{0.01in}\kappa^{-1}%
\mathbf{y}))\right\vert _{\tilde{y}^{A}=\hspace{0.01in}\pm\hspace
{0.01in}\alpha_{A}\hspace{0.01in}q^{\hspace{0.01in}l_{A}}}=\nonumber\\
=0+\mathcal{O}(h),\label{ZwiRecTotVac}%
\end{gather}
since at least one of the terms in Eqs.~(\ref{IntX3AbQlDelt}) and
(\ref{IntXMinAbQlDelt}) vanishes due to the property%
\begin{equation}
(q^{-m};q)_{n}=0\quad\text{for\quad}n=m+1,m+2,\ldots\qquad(m\in\mathbb{N}%
_{0}).\label{AbbBedPocSym}%
\end{equation}
From Eq.~(\ref{DarPotAblQuaDreEnt}) of Chap.~\ref{VacEneQEuc} and
Eq.~(\ref{ZwiRecTotVac}), we arrive at [cf. Eq.~(\ref{TotErwP2k})]%
\begin{equation}
\int_{\mathbb{R}_{q}^{3}}\text{d}_{q}^{3}x\hspace{0.02in}%
\big(-\mathbf{\partial}_{x}\circ\mathbf{\partial}_{x}\big)^{n}\triangleright
\delta_{q}^{3}(\mathbf{x}\oplus(\ominus\hspace{0.01in}\kappa^{-1}%
\mathbf{x}^{\prime}))=0+\mathcal{O}(h).\label{IntIniSta}%
\end{equation}

\section{Conclusion}

Our investigation focused on analyzing the ground-state energy of a
Klein-Gordon field in $q$-de\-formed Euclidean space. This zero-point energy
constitutes the primary contribution to the vacuum energy in such a space. On
a global scale, the vacuum energy of the $q$-de\-formed Euclidean space is
found to be exceedingly small. For a massless $q$-de\-formed Klein-Gordon
field, it even appears to vanish entirely.

A more detailed analysis of the energy density within extremely small regions
- such as quasipoints - reveals markedly higher values. Neighboring
quasipoints seem to exchange substantial amounts of energy with one another.
When the ground-state energy is evaluated over a restricted set of
quasipoints, the resulting vacuum energy density is correspondingly large.

In contrast, when the total ground-state energy of the $q$-de\-formed
Klein-Gordon field is computed over the entire $q$-de\-formed Euclidean space,
the energy exchanges between different quasipoints effectively cancel.
Consequently, the overall vacuum energy associated with the ground-state
energy is extremely small. In the massless case, it may in fact be exactly zero.

\appendix

\section{Star Product on $q$-De\-formed Eu\-clidean Space\label{KapQuaZeiEle}}

The \textit{three-di\-men\-sion\-al }$q$\textit{-de\-formed Euclidean space}
$\mathbb{R}_{q}^{3}$ is generated by $X^{+}$, $X^{3}$, and $X^{-}$, subject to
the commutation relations \cite{Lorek:1997eh}%
\begin{align}
X^{3}X^{+} &  =q^{2}X^{+}X^{3},\nonumber\\
X^{3}X^{-} &  =q^{-2}X^{-}X^{3},\nonumber\\
X^{-}X^{+} &  =X^{+}X^{-}+(q-q^{-1})\hspace{0.01in}X^{3}X^{3}%
.\label{RelQuaEukDre}%
\end{align}
A $q$-\textit{de\-formed analogue of the three-di\-men\-sion\-al Euclidean
metric} is given by \cite{Lorek:1997eh} (with rows and columns ordered as
$+,3,-$):%
\begin{equation}
g_{AB}=g^{AB}=\left(
\begin{array}
[c]{ccc}%
0 & 0 & -\hspace{0.01in}q\\
0 & 1 & 0\\
-\hspace{0.01in}q^{-1} & 0 & 0
\end{array}
\right)  .\label{MetDreiDim}%
\end{equation}
With this metric, indices can be raised and lowered according to%
\begin{equation}
X_{A}=g_{AB}\hspace{0.01in}X^{B},\qquad X^{A}=g^{AB}X_{B}.\label{HebSenInd}%
\end{equation}

We extend the algebra $\mathbb{R}_{q}^{3}$ by introducing a time coordinate
$X^{0}$, which commutes with the spatial generators $X^{+}$, $X^{3}$, and
$X^{-}$ \cite{Wachter:2020A}:%
\begin{equation}
X^{0}X^{A}=X^{A}X^{0},\text{\qquad}A\in\{+,3,-\}. \label{ZusRelExtDreEukQUa}%
\end{equation}
The extended algebra $\mathbb{R}_{q}^{3,1}$ is generated by the four
coordinates $X^{i}$ with $i\in\{0,+,3,-\}$. Each element $F\in$ $\mathbb{R}%
_{q}^{3,1}$ admits an expansion in terms of nor\-mal-or\-dered
monomials\textbf{\ }(reflecting the \textit{Poincar\'{e}-Birkhoff-Witt
property}):%
\begin{equation}
F=\sum\limits_{n_{+},\ldots,\hspace{0.01in}n_{0}}a_{\hspace{0.01in}n_{+}%
\ldots\hspace{0.01in}n_{0}}\,(X^{+})^{n_{+}}(X^{3})^{n_{3}}(X^{-})^{n_{-}%
}(X^{0})^{n_{0}},\quad\quad a_{\hspace{0.01in}n_{+}\ldots\hspace{0.01in}n_{0}%
}\in\mathbb{C}.
\end{equation}
There is a vector space isomorphism%
\begin{equation}
\mathcal{W}:\mathbb{C}[\hspace{0.01in}x^{+},x^{3},x^{-},t\hspace
{0.01in}]\rightarrow\mathbb{R}_{q}^{3,1} \label{VecRauIsoInv}%
\end{equation}
defined by%
\begin{equation}
\mathcal{W}\left(  (x^{+})^{n_{+}}(x^{3})^{n_{3}}(x^{-})^{n_{-}}%
t^{\hspace{0.01in}n_{0}}\right)  =(X^{+})^{n_{+}}(X^{3})^{n_{3}}(X^{-}%
)^{n_{-}}(X^{0})^{n_{0}}. \label{StePro0}%
\end{equation}
In general, one has%
\begin{equation}
\mathbb{C}[\hspace{0.01in}x^{+},x^{3},x^{-},t\hspace{0.01in}]\ni f\mapsto
F\in\mathbb{R}_{q}^{3,1},
\end{equation}
with%
\begin{align}
f  &  =\sum\limits_{n_{+},\ldots,\hspace{0.01in}n_{0}}a_{\hspace{0.01in}%
n_{+}\ldots\hspace{0.01in}n_{0}}\,(x^{+})^{n_{+}}(x^{3})^{n_{3}}(x^{-}%
)^{n_{-}}t^{\hspace{0.01in}n_{0}},\nonumber\\
F  &  =\sum\limits_{n_{+},\ldots,\hspace{0.01in}n_{0}}a_{\hspace{0.01in}%
n_{+}\ldots\hspace{0.01in}n_{0}}\,(X^{+})^{n_{+}}(X^{3})^{n_{3}}(X^{-}%
)^{n_{-}}(X^{0})^{n_{0}}. \label{AusFfNorOrd}%
\end{align}
The isomorphism $\mathcal{W}$ acts as the \textit{Moyal-Weyl map}, associating
an operator $F$ to a com\-plex-val\-ued function $f$
\cite{Bayen:1977ha,1997q.alg.....9040K,Madore:2000en,Moyal:1949sk}.

To extend this vector space isomorphism to an algebra isomorphism, we
introduce the \textit{star product},\textit{\ }defined through the
homomorphism property%
\begin{equation}
\mathcal{W}\left(  f\circledast g\right)  =\mathcal{W}\left(  f\right)
\cdot\mathcal{W}\left(  \hspace{0.01in}g\right)  . \label{HomBedWeyAbb}%
\end{equation}
Since the Mo\-yal-Weyl map $\mathcal{W}$ is invertible, the star product can
be expressed as:%
\begin{equation}
f\circledast g=\mathcal{W}^{\hspace{0.01in}-1}\big (\,\mathcal{W}\left(
f\right)  \cdot\mathcal{W}\left(  \hspace{0.01in}g\right)  \big ).
\label{ForStePro}%
\end{equation}

The product of two nor\-mal-or\-dered monomials can itself be expanded into a
series of nor\-mal-or\-dered monomials (see Ref.~\cite{Wachter:2002A} for
details):%
\begin{gather}
(X^{+})^{n_{+}}\ldots\hspace{0.01in}(X^{0})^{n_{0}}\cdot(X^{+})^{m_{+}}%
\ldots\hspace{0.01in}(X^{0})^{m_{0}}=\nonumber\\
=\sum_{\underline{k}\hspace{0.01in}=\hspace{0.01in}0}a_{\underline{k}%
}^{\hspace{0.01in}\underline{n},\underline{m}}\,(X^{+})^{k_{+}}\ldots
\hspace{0.01in}(X^{0})^{k_{0}}. \label{EntProMon}%
\end{gather}
This expansion leads to a general formula for the star product of two power
series in commutative space-time coordinates (with $\lambda=q-q^{-1}$):\
\begin{gather}
f(\mathbf{x},t)\circledast g(\mathbf{x},t)=\nonumber\\
\sum_{k\hspace{0.01in}=\hspace{0.01in}0}^{\infty}\lambda^{k}\hspace
{0.01in}\frac{(x^{3})^{2k}}{[[k]]_{q^{4}}!}\,q^{2(\hat{n}_{3}\hspace
{0.01in}\hat{n}_{+}^{\prime}+\,\hat{n}_{-}\hat{n}_{3}^{\prime})}%
D_{q^{4},\hspace{0.01in}x^{-}}^{k}f(\mathbf{x},t)\,D_{q^{4},\hspace
{0.01in}x^{\prime+}}^{k}g(\mathbf{x}^{\prime},t)\big|_{x^{\prime}%
\rightarrow\hspace{0.01in}x}. \label{StaProForExp}%
\end{gather}
This expression\ involves the operators%
\begin{equation}
\hat{n}_{A}=x^{A}\frac{\partial}{\partial x^{A}} \label{NOpeDef}%
\end{equation}
together with the Jackson derivatives defined in Eq.~(\ref{DefJacAbl}) of
Chap.~\ref{KapqAnaExpTriFkt}.

In certain contexts it is convenient to employ an alternative normal ordering,
different from that in Eq.~(\ref{StePro0}):%
\begin{equation}
\widetilde{\mathcal{W}}\left(  (x^{+})^{n_{+}}(x^{3})^{n_{3}}(x^{-})^{n_{-}%
}t^{\hspace{0.01in}n_{0}}\right)  =(X^{-})^{n_{-}}(X^{3})^{n_{3}}%
(X^{+})^{n_{+}}(X^{0})^{n_{0}}. \label{InvNorOrd}%
\end{equation}
Using the commutation relations in Eq.~(\ref{RelQuaEukDre}), we can change the
normal ordering:%
\begin{align}
&  (X^{+})^{n_{+}}(X^{3})^{n_{3}}(X^{-})^{n_{-}}(X^{0})^{n_{0}}=\nonumber\\
&  \qquad=\sum_{k\hspace{0.01in}=\hspace{0.01in}0}^{\infty}\left(
-\lambda\right)  ^{k}q^{-2n_{3}(n_{+}+\hspace{0.01in}n_{-}-\hspace{0.01in}%
k)}\,(B_{q^{-1}})_{k}^{n_{-},\hspace{0.01in}n_{+}}\nonumber\\
&  \qquad\qquad\times(X^{-})^{n_{-}-\hspace{0.01in}k}(X^{3})^{n_{3}%
+\hspace{0.01in}2k}(X^{+})^{n_{+}-\hspace{0.01in}k}(X^{0})^{n_{0}},
\label{UmrNorOrd}%
\end{align}
where%
\begin{equation}
(B_{q})_{k}^{i,j}=\frac{1}{[[k]]_{q^{4}}!}\frac{[[i]]_{q^{4}}!\,[[j]]_{q^{4}%
}!}{[[i-k]]_{q^{4}}!\,[[j-k]]_{q^{4}}!}. \label{BKoef}%
\end{equation}
The expansion in Eq. (\ref{UmrNorOrd}) leads to an operator that maps a
commutative power series referring to the ordering in Eq.~(\ref{StePro0}) into
one referring to the ordering in Eq.~(\ref{InvNorOrd}) (see
Ref.~\cite{Wachter:2004ph} for details):%
\begin{equation}
\hat{U}f=\sum_{k\hspace{0.01in}=\hspace{0.01in}0}^{\infty}\left(
-\lambda\right)  ^{k}\frac{(x^{3})^{2k}}{[[k]]_{q^{-4}}!}\,q^{-2\hat{n}%
_{3}(\hat{n}_{+}+\hspace{0.01in}\hat{n}_{-}+\hspace{0.01in}k)}D_{q^{-4}%
,\hspace{0.01in}x^{+}}^{k}D_{q^{-4},\hspace{0.01in}x^{-}}^{k}f.
\label{UmOrdOpeDreQuan}%
\end{equation}
The inverse operator $U^{-1}$ is given by:%
\begin{equation}
\hat{U}^{-1}f=\sum_{k\hspace{0.01in}=\hspace{0.01in}0}^{\infty}\lambda
^{k}\hspace{0.01in}\frac{(x^{3})^{2k}}{[[k]]_{q^{4}}!}\,q^{2\hat{n}_{3}%
(\hat{n}_{+}+\hspace{0.01in}\hat{n}_{-}+\hspace{0.01in}k)}D_{q^{4}%
,\hspace{0.01in}x^{+}}^{k}D_{q^{4},\hspace{0.01in}x^{-}}^{k}f.
\label{UmOrdInvOpeDreQuan}%
\end{equation}

\section{Derivatives and Integrals on $q$-De\-formed Eu\-clidean
Space\label{KapParDer}}

Partial derivatives can be consistently defined for $q$-de\-formed space-time
coordinates \cite{CarowWatamura:1990zp,Wess:1990vh}. These $q$%
\textit{-de\-formed partial derivatives} obey the same commutation relations
as the covariant coordinate generators $X_{i}$:%
\begin{gather}
\partial_{0}\hspace{0.01in}\partial_{+}=\hspace{0.01in}\partial_{+}%
\hspace{0.01in}\partial_{0},\quad\partial_{0}\hspace{0.01in}\partial
_{-}=\hspace{0.01in}\partial_{-}\hspace{0.01in}\partial_{0},\quad\partial
_{0}\hspace{0.01in}\partial_{\hspace{0.01in}3}=\partial_{\hspace{0.01in}%
3}\hspace{0.01in}\partial_{0},\nonumber\\
\partial_{+}\hspace{0.01in}\partial_{\hspace{0.01in}3}=q^{2}\partial
_{\hspace{0.01in}3}\hspace{0.01in}\partial_{+},\quad\partial_{\hspace
{0.01in}3}\hspace{0.01in}\partial_{-}=\hspace{0.01in}q^{2}\partial_{-}%
\hspace{0.01in}\partial_{\hspace{0.01in}3},\nonumber\\
\partial_{+}\hspace{0.01in}\partial_{-}-\partial_{-}\hspace{0.01in}%
\partial_{+}=\hspace{0.01in}\lambda\hspace{0.01in}\partial_{\hspace{0.01in}%
3}\hspace{0.01in}\partial_{\hspace{0.01in}3}.
\end{gather}

There are two consistent ways to commute $q$-de\-formed partial derivatives
with $q$-de\-formed coordinates. The corresponding $q$\textit{-de\-formed
Leibniz rules} take the form
\cite{CarowWatamura:1990zp,Wess:1990vh,Wachter:2020A}:%
\begin{gather}
\partial_{B}X^{A}=\delta_{B}^{A}+q^{4}\hat{R}{^{AC}}_{BD}\,X^{D}\partial
_{C},\nonumber\\
\partial_{A}X^{0}=X^{0}\hspace{0.01in}\partial_{A},\quad\partial_{0}%
\hspace{0.01in}X^{A}=X^{A}\hspace{0.01in}\partial_{0},\nonumber\\
\partial_{0}\hspace{0.01in}X^{0}=1+X^{0}\hspace{0.01in}\partial_{0}.
\label{DifKalExtEukQuaDreUnk}%
\end{gather}
Here $\hat{R}{^{AC}}_{BD}$ denotes the vector representation of the R-ma\-trix
associated with three-di\-men\-sion\-al $q$-de\-formed Euclidean space
\cite{Lorek:1997eh}. By redefining $\hat{\partial}_{A}=q^{6}\partial_{A}$ and
$\hat{\partial}_{0}=\partial_{0}$, the\textit{ }Leibniz rules of the second
differential calculus can be written as%
\begin{gather}
\hat{\partial}_{B}\hspace{0.01in}X^{A}=\delta_{B}^{A}+q^{-4}(\hat{R}%
^{-1}){^{AC}}_{BD}\,X^{D}\hat{\partial}_{C},\nonumber\\
\hat{\partial}_{A}\hspace{0.01in}X^{0}=X^{0}\hspace{0.01in}\hat{\partial}%
_{A},\quad\hat{\partial}_{0}\hspace{0.01in}X^{A}=X^{A}\hspace{0.01in}%
\hat{\partial}_{0},\nonumber\\
\hat{\partial}_{0}\hspace{0.01in}X^{0}=1+X^{0}\hspace{0.01in}\hat{\partial
}_{0}. \label{DifKalExtEukQuaDreKon}%
\end{gather}

Using either set of $q$-de\-formed Leibniz rules, one can compute the action
of $q$-de\-formed partial derivatives on nor\-mal-or\-dered monomials of
noncommutative coordinates. This action is then extended to commutative
coordinates via the Moyal-Weyl map $\mathcal{W}$:%
\begin{equation}
\partial_{i}\triangleright(x^{+})^{n_{+}}(x^{3})^{n_{3}}(x^{-})^{n_{-}%
}t^{\hspace{0.01in}n_{0}}=\mathcal{W}^{\hspace{0.01in}-1}\big (\partial
_{i}\triangleright(X^{+})^{n_{+}}(X^{3})^{n_{3}}(X^{-})^{n_{-}}(X^{0})^{n_{0}%
}\big ).
\end{equation}
Since the Mo\-yal-Weyl map is linear, this prescription naturally extends to
power series in commutative space-time coordinates:%
\begin{equation}
\partial_{i}\triangleright f(\mathbf{x},t)=\mathcal{W}^{\hspace{0.01in}%
-1}\big (\partial_{i}\triangleright\mathcal{W}(f(\mathbf{x},t))\big ).
\end{equation}

For nor\-mal-or\-dered monomials as defined in Eq.~(\ref{StePro0}), the
Leibniz rules in Eq.~(\ref{DifKalExtEukQuaDreUnk})\ yield explicit operator
representations \cite{Bauer:2003}:%
\begin{align}
\partial_{+}\triangleright f(\mathbf{x},t)  &  =D_{q^{4},\hspace{0.01in}x^{+}%
}f(\mathbf{x},t),\nonumber\\
\partial_{\hspace{0.01in}3}\triangleright f(\mathbf{x},t)  &  =D_{q^{2}%
,\hspace{0.01in}x^{3}}f(q^{2}x^{+},x^{3},x^{-},t),\nonumber\\
\partial_{-}\triangleright f(\mathbf{x},t)  &  =D_{q^{4},\hspace{0.01in}x^{-}%
}f(x^{+},q^{2}x^{3},x^{-},t)+\lambda\hspace{0.01in}x^{+}D_{q^{2}%
,\hspace{0.01in}x^{3}}^{2}f(\mathbf{x},t). \label{UnkOpeDarAbl}%
\end{align}
The derivative $\partial_{0}$ is represented on the commutative space-time
algebra by the standard time derivative:%
\begin{equation}
\partial_{0}\triangleright\hspace{-0.01in}f(\mathbf{x},t)=\frac{\partial
f(\mathbf{x},t)}{\partial t}. \label{OpeDarZeiAblExtQuaEuk}%
\end{equation}

Certain calculations require evaluating the action of higher powers of
$q$-de\-formed partial derivatives on a power series in commutative space-time
coordinates. Using the operator representations given in
Eq.~(\ref{UnkOpeDarAbl}), we obtain (for $m_{+},m_{3}\in\mathbb{N}_{0}$)%
\begin{align}
(\partial_{+})^{m_{+}}\triangleright f  &  =D_{q^{4},\hspace{0.01in}x^{+}%
}^{m_{+}}f,\nonumber\\
(\partial_{\hspace{0.01in}3})^{m_{3}}\triangleright f  &  =D_{q^{2}%
,\hspace{0.01in}x^{3}}^{m_{3}}f(q^{2m_{3}}x^{+}), \label{MehAbl3+Dim3}%
\end{align}
and (for $m_{-}\in\mathbb{N}_{0}$)%
\begin{align}
(\partial_{-})^{m_{-}}\triangleright f  &  =\left[  \hspace{0.01in}q^{2\hat
{n}_{3}}D_{q^{4},\hspace{0.01in}x^{-}}+\hspace{0.01in}\lambda\hspace
{0.01in}x^{+}D_{q^{2},\hspace{0.01in}x^{3}}^{2}\right]  ^{m_{-}}f\nonumber\\
&  =\sum_{i\hspace{0.01in}=\hspace{0.01in}0}^{m_{-}}\hspace{0.01in}\lambda^{i}%
\genfrac{[}{]}{0pt}{}{m_{-}}{i}%
_{q^{-4}}(x^{+})^{i}D_{q^{2},\hspace{0.01in}x^{3}}^{2i}\big (D_{q^{4}%
,\hspace{0.01in}x^{-}}q^{2\hat{n}_{3}}\big )^{m_{-}-\hspace{0.01in}%
i}f\nonumber\\
&  =\sum_{i\hspace{0.01in}=\hspace{0.01in}0}^{m_{-}}\hspace{0.01in}\lambda^{i}%
\genfrac{[}{]}{0pt}{}{m_{-}}{i}%
_{q^{-4}}(x^{+})^{i}D_{q^{2},\hspace{0.01in}x^{3}}^{2i}D_{q^{4},\hspace
{0.01in}x^{-}}^{m_{-}-\hspace{0.01in}i}f(q^{2(m_{-}-\hspace{0.01in}i)}x^{3}).
\label{MehAblMinDim3}%
\end{align}
The second identity in the calculation above follows from the commutation
relation%
\begin{equation}
\big (D_{q^{4},\hspace{0.01in}x^{-}}q^{2\hat{n}_{3}}\big )\big (\lambda
\hspace{0.01in}x^{+}D_{q^{2},\hspace{0.01in}x^{3}}^{2}\big )=q^{-4}%
\big (\lambda\hspace{0.01in}x^{+}D_{q^{2},\hspace{0.01in}x^{3}}^{2}%
\big )\big (D_{q^{4},\hspace{0.01in}x^{-}}q^{2\hat{n}_{3}}\big ),
\end{equation}
together with the $q$-bi\-no\-mi\-al theorem \cite{Klimyk:1997eb}:%
\begin{equation}
(x+a)^{n}=\sum_{k\hspace{0.01in}=\hspace{0.01in}0}^{n}%
\genfrac{[}{]}{0pt}{}{n}{k}%
_{q}x^{k}a^{n-k}\quad\text{if}\quad a\hspace{0.01in}x=q\hspace{0.01in}xa.
\end{equation}
Combining both results in Eq.~(\ref{MehAbl3+Dim3}), one finds%
\begin{align}
(\partial_{\hspace{0.01in}3})^{m_{3}}(\partial_{+})^{m_{+}}\triangleright f
&  =D_{q^{2},\hspace{0.01in}x^{3}}^{m_{3}}\hspace{0.01in}q^{2m_{3}\hat{n}_{+}%
}D_{q^{4},\hspace{0.01in}x^{+}}^{m_{+}}f\nonumber\\
&  =q^{-2m_{3}m_{+}}D_{q^{2},\hspace{0.01in}x^{3}}^{m_{3}}\hspace
{0.01in}D_{q^{4},\hspace{0.01in}x^{+}}^{m_{+}}\hspace{0.01in}q^{2m_{3}\hat
{n}_{+}}f\nonumber\\
&  =q^{-2m_{3}m_{+}}D_{q^{2},\hspace{0.01in}x^{3}}^{m_{3}}\hspace
{0.01in}D_{q^{4},\hspace{0.01in}x^{+}}^{m_{+}}f(q^{2m_{3}}x^{+}).
\label{WirPotAblm3mMi}%
\end{align}
Using the results from Eqs.~(\ref{MehAblMinDim3}) and (\ref{WirPotAblm3mMi}),
the combined action is given by%
\begin{align}
&  (\partial_{-})^{m_{-}}(\partial_{\hspace{0.01in}3})^{m_{3}}(\partial
_{+})^{m_{+}}\triangleright f=\sum_{i\hspace{0.01in}=\hspace{0.01in}0}^{m_{-}%
}\hspace{0.01in}\lambda^{i}%
\genfrac{[}{]}{0pt}{}{m_{-}}{i}%
_{q^{-4}}q^{-2m_{3}(m_{+}+\hspace{0.01in}m_{-}-\hspace{0.01in}i)}(x^{+}%
)^{i}\nonumber\\
&  \qquad\qquad\qquad\qquad\qquad\qquad\times D_{q^{4},\hspace{0.01in}x^{-}%
}^{m_{-}-\hspace{0.01in}i}D_{q^{2},\hspace{0.01in}x^{3}}^{m_{3}+2i}%
D_{q^{4},\hspace{0.01in}x^{+}}^{m_{+}}f(q^{2m_{3}}x^{+},q^{2(m_{-}%
-\hspace{0.01in}i)}x^{3})\nonumber\\[0.03in]
&  =q^{-2m_{3}(m_{+}+\hspace{0.01in}m_{-})}D_{q^{4},\hspace{0.01in}x^{-}%
}^{m_{-}}D_{q^{2},\hspace{0.01in}x^{3}}^{m_{3}}\hspace{0.01in}D_{q^{4}%
,\hspace{0.01in}x^{+}}^{m_{+}}f(q^{2m_{3}}x^{+},q^{2m_{-}}x^{3})+\mathcal{O}%
(h). \label{MehAblDreDim}%
\end{align}

The operator representations for the partial derivatives $\hat{\partial}_{i}$
can be derived from the Leibniz rules in Eq.$~$(\ref{DifKalExtEukQuaDreKon}).
The Leibniz rules in Eqs.$~$(\ref{DifKalExtEukQuaDreUnk}) and
(\ref{DifKalExtEukQuaDreKon}) are related by the substitutions%
\begin{gather}
q\rightarrow q^{-1},\quad X^{-}\rightarrow X^{+},\quad X^{+}\rightarrow
X^{-},\nonumber\\
\partial_{+}\rightarrow\hat{\partial}_{-},\quad\partial_{-}\rightarrow
\hat{\partial}_{+},\quad\partial_{\hspace{0.01in}3}\rightarrow\hat{\partial
}_{\hspace{0.01in}3},\quad\partial_{0}\rightarrow\hat{\partial}_{0}%
.\label{UebRegGedUngAblDreQua}%
\end{gather}
Thus, the operator representations of the partial derivatives $\hat{\partial
}_{A}$ can be obtained from those of $\partial_{A}$ [cf.
Eq.~(\ref{UnkOpeDarAbl})] by replacing $q$ with $q^{-1}$ and interchanging the
indices $+$ and $-$:%
\begin{align}
\hat{\partial}_{-}\,\bar{\triangleright}\,f(\mathbf{x},t) &  =D_{q^{-4}%
,\hspace{0.01in}x^{-}}f(\mathbf{x},t),\nonumber\\
\hat{\partial}_{\hspace{0.01in}3}\,\bar{\triangleright}\,f(\mathbf{x},t) &
=D_{q^{-2},\hspace{0.01in}x^{3}}f(q^{-2}x^{-},x^{3},x^{+},t),\nonumber\\
\hat{\partial}_{+}\,\bar{\triangleright}\,f(\mathbf{x},t) &  =D_{q^{-4}%
,\hspace{0.01in}x^{+}}f(x^{-},q^{-2}x^{3},x^{+},t)-\lambda\hspace{0.01in}%
x^{-}D_{q^{-2},\hspace{0.01in}x^{3}}^{2}f(\mathbf{x},t).\label{KonOpeDarAbl}%
\end{align}
Finally, the time derivative $\hat{\partial}_{0}$ acts on the commutative
space-time algebra as the usual derivative with respect to $t$:%
\begin{equation}
\hat{\partial}_{0}\,\bar{\triangleright}\,f(\mathbf{x},t)=\frac{\partial
f(\mathbf{x},t)}{\partial t}.\label{OpeDarZeiAblExtQuaEukKon}%
\end{equation}
We note that the operator representations in Eq.~(\ref{KonOpeDarAbl}) are tied
to the normal ordering defined in Eq.~(\ref{InvNorOrd}) of the previous chapter.

We can also shift $q$-de\-formed partial derivatives from the \textit{right}
side of a nor\-mal-or\-dered monomial to the left side using the Leibniz
rules. This procedure leads to\ the so-called \textit{right}%
-re\-pre\-sen\-ta\-tions of partial derivatives, which are denoted by
$f\,\bar{\triangleleft}\,\partial_{i}$ or $f\triangleleft\hat{\partial}_{i}$
\cite{Bauer:2003}.

The operator representations in Eqs.~(\ref{UnkOpeDarAbl}) and
(\ref{KonOpeDarAbl}) split into two parts%
\begin{equation}
\partial_{A}\triangleright F=\left(  \partial_{A,\operatorname*{cla}}%
+\partial_{A,\operatorname*{cor}}\right)  \triangleright F.
\end{equation}
In the limit $q\rightarrow1$, $\partial_{A,\operatorname*{cla}}$ reduces to
the usual partial derivative, while $\partial_{A,\operatorname*{cor}}$
vanishes. A solution to the difference equation $\partial^{A}\triangleright
F=f$ can be written as \cite{Wachter:2004A}:%
\begin{align}
F  &  =(\partial_{A})^{-1}\triangleright f=\left(  \partial
_{A,\operatorname*{cla}}+\partial_{A,\operatorname*{cor}}\right)
^{-1}\triangleright f\nonumber\\
&  =\sum_{k\hspace{0.01in}=\hspace{0.01in}0}^{\infty}\left[  -(\partial
_{A,\operatorname*{cla}})^{-1}\partial_{A,\operatorname*{cor}}\right]
^{k}(\partial_{A,\operatorname*{cla}})^{-1}\triangleright f.
\end{align}
Applying this formula to the operator representations in
Eq.~(\ref{UnkOpeDarAbl}) yields%
\begin{align}
(\partial_{+})^{-1}\triangleright f(\mathbf{x},t)  &  =D_{q^{4},\hspace
{0.01in}x^{+}}^{-1}f(\mathbf{x},t),\nonumber\\
(\partial_{\hspace{0.01in}3})^{-1}\triangleright f(\mathbf{x},t)  &
=D_{q^{2},\hspace{0.01in}x^{3}}^{-1}f(q^{-2}x^{+},x^{3},x^{-},t),
\label{InvParAbl1}%
\end{align}
and%
\begin{gather}
(\partial_{-})^{-1}\triangleright f(\mathbf{x},t)=\nonumber\\
=\sum_{k\hspace{0.01in}=\hspace{0.01in}0}^{\infty}q^{2k\left(  k\hspace
{0.01in}+1\right)  }\left(  -\lambda\,x^{+}D_{q^{4},\hspace{0.01in}x^{-}}%
^{-1}D_{q^{2},\hspace{0.01in}x^{3}}^{2}\right)  ^{k}D_{q^{4},\hspace
{0.01in}x^{-}}^{-1}f(x^{+},q^{-2\left(  k\hspace{0.01in}+1\right)  }%
x^{3},x^{-},t). \label{InvParAbl2}%
\end{gather}
Here, $D_{q,\hspace{0.01in}x}^{-1}$ denotes a Jackson integral with respect to
the variable $x$. The explicit form of this Jackson integral depends on its
integration limits and the value of the deformation parameter $q$ [cf.
Eqs.(\ref{QInt0klQkl1})-(\ref{UneJackIntAll}) in Chap.~\ref{KapQIntTrig}]. The
time coordinate integral remains unaffected by $q$-de\-for\-ma\-tion [cf.
Eq.~(\ref{OpeDarZeiAblExtQuaEuk})]:%
\begin{equation}
(\partial_{0})^{-1}\triangleright f(\mathbf{x},t)\hspace{0.01in}=\int
\text{d}t\,f(\mathbf{x},t).
\end{equation}

The considerations above extend directly to the conjugate set of partial
derivatives $\hat{\partial}_{i}$ with minor adjustments. Recall that the
representations of $\hat{\partial}_{i}$ can be derived from those of
$\partial_{i}$ by replacing $q$ with $q^{-1}$ and exchanging the indices $+$
and $-$. Applying these substitutions to Eqs.~(\ref{InvParAbl1}) and
(\ref{InvParAbl2}) yields the corresponding results for the conjugate partial
derivatives $\hat{\partial}_{i}$.

By successively applying the integral operators defined in
Eqs.~(\ref{InvParAbl1}) and (\ref{InvParAbl2}), one can define an
\textit{integration over the entire }$q$\textit{-de\-formed Euclidean space}
\cite{Wachter:2004A,Wachter:2007A}:%
\begin{equation}
\int_{-\infty}^{+\infty}\text{d}_{q}^{3}x\,f(x^{+},x^{3},x^{-})=(\partial
_{-})^{-1}\big |_{-\infty}^{+\infty}\,(\partial_{\hspace{0.01in}3}%
)^{-1}\big |_{-\infty}^{+\infty}\,(\partial_{+})^{-1}\big |_{-\infty}%
^{+\infty}\triangleright f.\label{DefIntSpa}%
\end{equation}
On the right-hand side of the equation above, the integral operators
$(\partial_{A})^{-1}$ reduce to improper Jackson integrals
\cite{Wachter:2004A,Jambor:2004ph}:%
\begin{equation}
\int\text{d}_{q}^{3}x\,f=\int_{-\infty}^{+\infty}\text{d}_{q}^{3}%
x\,f(\mathbf{x})=\int_{-\infty}^{+\infty}\text{d}_{q^{2}}x^{-}\int_{-\infty
}^{+\infty}\text{d}_{q}x^{3}\int_{-\infty}^{+\infty}\text{d}_{q^{2}}%
x^{+}\,f(\mathbf{x}).\label{VerIntEntSpa}%
\end{equation}
These improper Jackson integrals are defined on a smaller $q$-lat\-tice so
that the full space integral has trivial braiding \cite{Kempf:1994yd}.
Moreover, the $q$-in\-te\-gral over all space satisfies a $q$%
-\textit{ana\-logue of Stokes' theorem} \cite{Wachter:2007A,Jambor:2004ph}:%
\begin{align}
\int_{-\infty}^{+\infty}\text{d}_{q}^{3}x\,\partial^{A}\triangleright f &
=\int_{-\infty}^{+\infty}\text{d}_{q}^{3}x\,f\,\bar{\triangleleft}%
\,\partial^{A}=0,\nonumber\\
\int_{-\infty}^{+\infty}\text{d}_{q}^{3}x\,\hat{\partial}^{A}\,\bar
{\triangleright}\,f &  =\int_{-\infty}^{+\infty}\text{d}_{q}^{3}%
x\,f\triangleleft\hat{\partial}^{A}=0.\label{StoThe}%
\end{align}

\section{Exponentials and Translations on $q$-De\-formed Eu\-clidean
Space\label{KapExp}}

\begin{figure}[ptb]
\centering
  \includegraphics[width=4.555in]{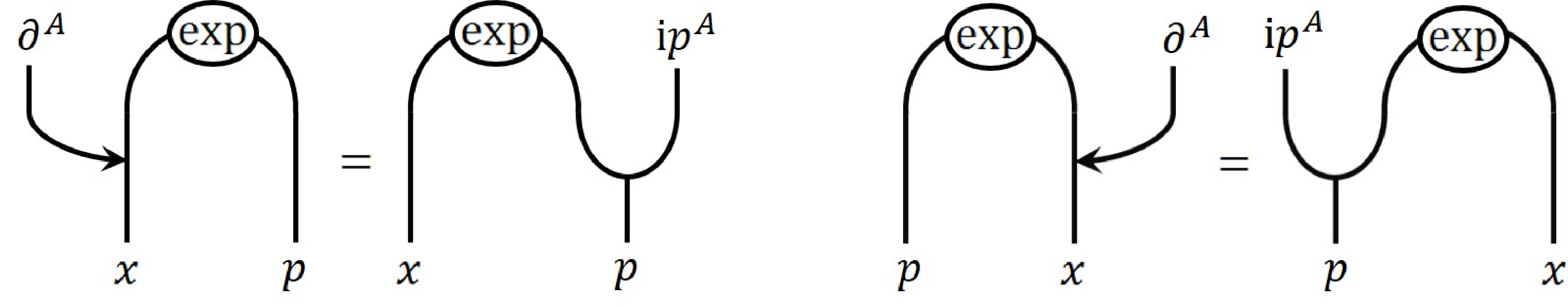}
  \caption{Eigenvalue equations of $q$-exponentials.}
  \label{Fig1}
\end{figure}

An \textit{exponential on a }$q$\textit{-de\-formed quantum space} is an
eigenfunction of all $q$-de\-formed partial derivatives
\cite{Majid:1993ud,Schirrmacher:1995,Wachter:2004ExpA}. In particular,
$q$-ex\-po\-nen\-tials may be viewed as eigenfunctions with respect to either
the left or right action of these derivatives:%
\begin{align}
\text{i}^{-1}\partial^{A}\triangleright\exp_{q}(\mathbf{x}|\text{i}\mathbf{p})
&  =\exp_{q}(\mathbf{x}|\text{i}\mathbf{p})\circledast p^{A},\nonumber\\
\exp_{q}(\text{i}^{-1}\mathbf{p}|\hspace{0.01in}\mathbf{x})\,\bar
{\triangleleft}\,\partial^{A}\text{i}^{-1} &  =p^{A}\circledast\exp
_{q}(\text{i}^{-1}\mathbf{p}|\hspace{0.01in}\mathbf{x}).\label{EigGl1N}%
\end{align}
These eigenvalue equations are illustrated in Fig.~\ref{Fig1}. The
$q$-ex\-po\-nen\-tials are uniquely determined by their eigenvalue equations
together with the normalization conditions%
\begin{align}
\exp_{q}(\mathbf{x}|\text{i}\mathbf{p})|_{x\hspace{0.01in}=\hspace{0.01in}0}
&  =\exp_{q}(\mathbf{x}|\text{i}\mathbf{p})|_{p\hspace{0.01in}=\hspace
{0.01in}0}=1,\nonumber\\
\exp_{q}(\text{i}^{-1}\mathbf{p}|\hspace{0.01in}\mathbf{x})|_{x\hspace
{0.01in}=\hspace{0.01in}0} &  =\exp_{q}(\text{i}^{-1}\mathbf{p}|\hspace
{0.01in}\mathbf{x})|_{p\hspace{0.01in}=\hspace{0.01in}0}=1.\label{NorBedExp}%
\end{align}%

From the operator representation in Eq.~(\ref{UnkOpeDarAbl}) in the previous
chapter, we derived explicit series expansions for $q$-ex\-ponen\-tials on
$q$-deformed Eu\-clidean space \cite{Wachter:2004ExpA}:%
\begin{gather}
\exp_{q}(\text{i}^{-1}\mathbf{p}|\mathbf{x})=\nonumber\\
=\sum_{\underline{n}\,=\,0}^{\infty}\frac{(\text{i}^{-1}p^{+})^{n_{+}%
}(\text{i}^{-1}p^{3})^{n_{3}}(\text{i}^{-1}p^{-})^{n_{-}}(x_{-})^{n_{-}}%
(x_{3})^{n_{3}}(x_{+})^{n_{+}}}{[[\hspace{0.01in}n_{+}]]_{q^{4}}%
!\,[[\hspace{0.01in}n_{3}]]_{q^{2}}!\,[[\hspace{0.01in}n_{-}]]_{q^{4}}!},
\label{ExpEukExp}%
\end{gather}
and%
\begin{gather}
\exp_{q}(\mathbf{x}|\text{i}\mathbf{p})=\nonumber\\
=\sum_{\underline{n}\,=\,0}^{\infty}\frac{(x^{+})^{n_{+}}(x^{3})^{n_{3}}%
(x^{-})^{n_{-}}(\text{i}p_{-})^{n_{-}}(\text{i}p_{3})^{n_{3}}(\text{i}%
p_{+})^{n_{+}}}{[[\hspace{0.01in}n_{+}]]_{q^{4}}!\,[[\hspace{0.01in}%
n_{3}]]_{q^{2}}!\,[[\hspace{0.01in}n_{-}]]_{q^{4}}!}. \label{ExpEukExp2}%
\end{gather}

By replacing $q$ with $q^{-1}$ in Eqs.~(\ref{ExpEukExp}) and (\ref{ExpEukExp2}%
), one obtains two further $q$-ex\-ponen\-tials, denoted by $\overline{\exp
}_{q}(x|$i$\mathbf{p})$ and $\overline{\exp}_{q}($i$^{-1}\mathbf{p}|x)$. Their
eigenvalue equations and normalization conditions follow directly from
Eqs.~(\ref{EigGl1N}) and (\ref{NorBedExp}) upon applying the substitutions%
\begin{equation}
\exp_{q}\rightarrow\hspace{0.01in}\overline{\exp}_{q},\qquad\triangleright
\,\rightarrow\,\bar{\triangleright},\qquad\bar{\triangleleft}\,\rightarrow
\,\triangleleft,\qquad\partial^{A}\rightarrow\hat{\partial}^{A}%
.\label{ErsRegQExp}%
\end{equation}

We can employ $q$-ex\-ponen\-tials to calculate $q$-trans\-la\-tions
\cite{Chryssomalakos:1993zm}. Substituting momentum variables by derivatives
in the $q$-ex\-ponen\-tials yields
\cite{Carnovale:1999,Majid:1993ud,Wachter:2007A}:%
\begin{align}
\exp_{q}(\mathbf{x}|\partial_{y})\triangleright g(\hspace{0.01in}\mathbf{y})
&  =g(\mathbf{x}\,\bar{\oplus}\,\mathbf{y}),\nonumber\\
\overline{\exp}_{q}(\mathbf{x}|\hat{\partial}_{y})\,\bar{\triangleright
}\,g(\hspace{0.01in}\mathbf{y})  &  =g(\mathbf{x}\oplus\mathbf{y}),
\label{q-TayN}%
\end{align}
and%
\begin{align}
g(\hspace{0.01in}\mathbf{y})\,\bar{\triangleleft}\,\exp_{q}(-\hspace
{0.01in}\partial_{y}|\hspace{0.01in}\mathbf{x})  &  =g(\hspace{0.01in}%
\mathbf{y}\,\bar{\oplus}\,\mathbf{x}),\nonumber\\
g(\hspace{0.01in}\mathbf{y})\triangleleft\hspace{0.01in}\overline{\exp}%
_{q}(-\hspace{0.01in}\hat{\partial}_{y}|\hspace{0.01in}\mathbf{x})  &
=g(\hspace{0.01in}\mathbf{y}\oplus\mathbf{x}). \label{q-TayRecN}%
\end{align}
For three-di\-men\-sion\-al $q$-de\-formed Euclidean space, one obtains, for
example, the following formula for calculating $q$-trans\-la\-tions
\cite{Wachter:2004ph}:%
\begin{gather}
f(\mathbf{x}\oplus\mathbf{y})=\nonumber\\
=\sum_{i_{+}=\hspace{0.01in}0}^{\infty}\sum_{i_{3}=\hspace{0.01in}0}^{\infty
}\sum_{i_{-}=\hspace{0.01in}0}^{\infty}\sum_{k\hspace{0.01in}=\hspace
{0.01in}0}^{i_{3}}\frac{(-q^{-1}\lambda\lambda_{+})^{k}}{[[2k]]_{q^{-2}}%
!!}\frac{(x^{-})^{i_{-}}(x^{3})^{i_{3}-\hspace{0.01in}k}(x^{+})^{i_{+}%
+\hspace{0.01in}k}\,(\hspace{0.01in}y^{-})^{k}}{[[i_{-}]]_{q^{-4}}%
!\,[[i_{3}-k]]_{q^{-2}}!\,[[i_{+}]]_{q^{-4}}!}\nonumber\\
\times\big (D_{q^{-4},\hspace{0.01in}y^{-}}^{i_{-}}D_{q^{-2},\hspace
{0.01in}y^{3}}^{i_{3}+\hspace{0.01in}k}\hspace{0.01in}D_{q^{-4},\hspace
{0.01in}y^{+}}^{i_{+}}f\big )(q^{2(k\hspace{0.01in}-\hspace{0.01in}i_{3}%
)}y^{-},q^{-2i_{+}}y^{3}). \label{ForQTra}%
\end{gather}

\begin{figure}[ptb]
  \centering
  \includegraphics[width=1.8827in]{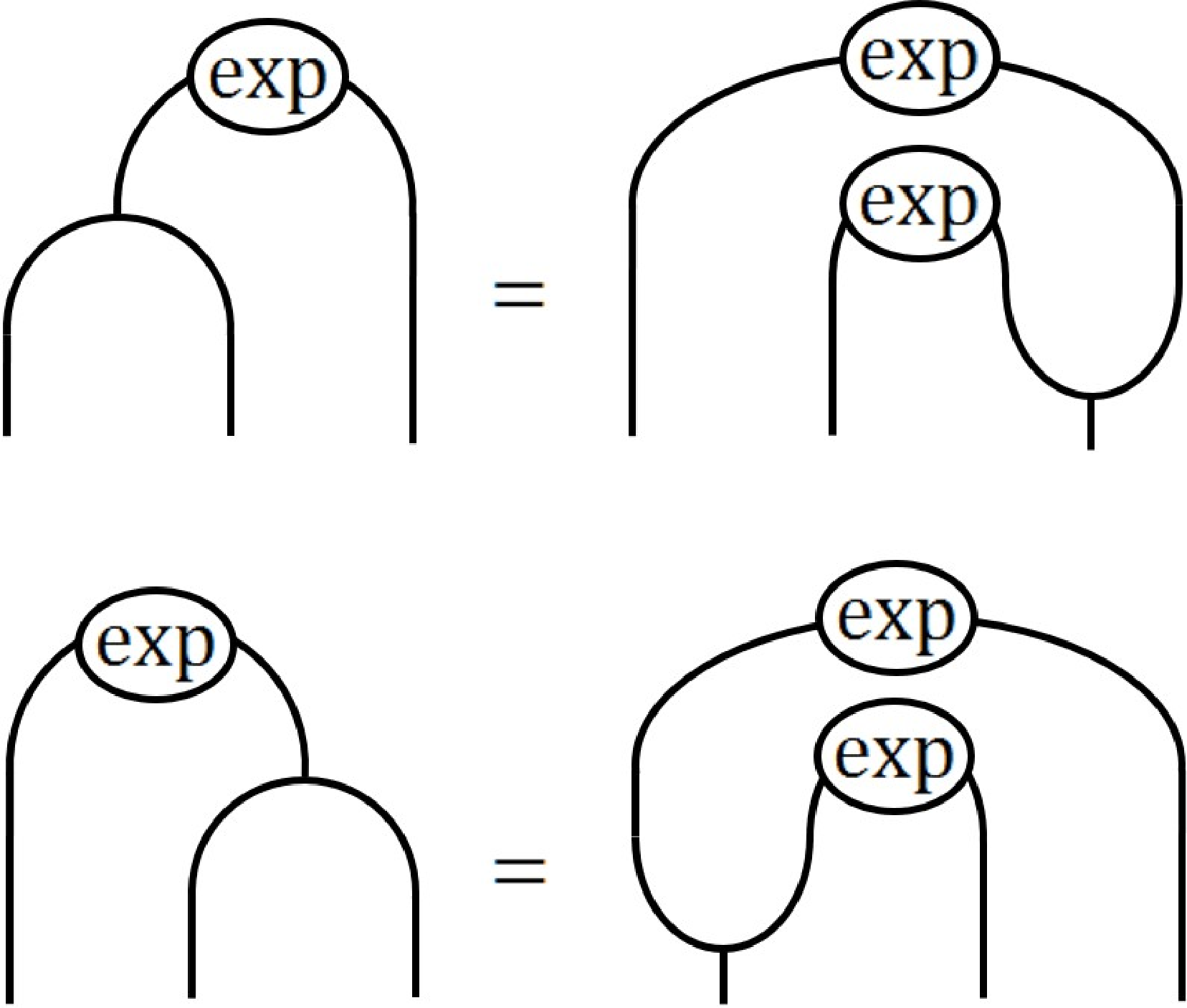}
  \caption{Addition theorems for $q$-exponentials.}
  \label{Fig2}
\end{figure}

In analogy with the undeformed case, $q$-ex\-ponen\-tials satisfy
\textit{addition theorems} \cite{Majid:1993ud,Schirrmacher:1995,Wachter:2007A}%
\begin{align}
\exp_{q}(\mathbf{x}\,\bar{\oplus}\,\mathbf{y}|\text{i}\mathbf{p})  &
=\exp_{q}(\mathbf{x}|\exp_{q}(\hspace{0.01in}\mathbf{y}|\text{i}%
\mathbf{p})\circledast\text{i}\mathbf{p}),\nonumber\\
\exp_{q}(\text{i}\mathbf{x}|\mathbf{p}\,\bar{\oplus}\,\mathbf{p}^{\prime})  &
=\exp_{q}(\mathbf{x}\circledast\exp_{q}(\mathbf{x}|\hspace{0.01in}%
\text{i}\mathbf{p})|\hspace{0.01in}\text{i}\mathbf{p}^{\prime}),
\label{AddTheExp}%
\end{align}
and%
\begin{align}
\overline{\exp}_{q}(\mathbf{x}\oplus\mathbf{y}|\text{i}\mathbf{p})  &
=\overline{\exp}_{q}(\mathbf{x}|\overline{\exp}_{q}(\hspace{0.01in}%
\mathbf{y}|\text{i}\mathbf{p})\circledast\text{i}\mathbf{p}),\nonumber\\
\overline{\exp}_{q}(\text{i}\mathbf{x}|\mathbf{p}\oplus\mathbf{p}^{\prime})
&  =\overline{\exp}_{q}(\mathbf{x}\circledast\overline{\exp}_{q}%
(\mathbf{x}|\text{i}\mathbf{p})|\hspace{0.01in}\text{i}\mathbf{p}^{\prime}).
\end{align}
Additional addition theorems can be derived by interchanging position and
momentum variables. Graphical representations of the two addition theorems in
Eq.~(\ref{AddTheExp}) are shown in Fig.~\ref{Fig2}.%

The $q$-de\-formed quantum spaces provide natural examples of \textit{braided
Hopf algebras} \cite{Majid:1996kd}. Within this framework, the two distinct
forms of $q$-trans\-lations arise from representations of two braided
coproducts, denoted $\underline{\Delta}$ and $\underline{\bar{\Delta}}$,
acting on the associated commutative coordinate algebras \cite{Wachter:2007A}:%
\begin{align}
f(\mathbf{x}\oplus\mathbf{y})  &  =((\mathcal{W}^{\hspace{0.01in}-1}%
\otimes\mathcal{W}^{\hspace{0.01in}-1})\circ\underline{\Delta})(\mathcal{W}%
(f)),\nonumber\\[0.08in]
f(\mathbf{x}\,\bar{\oplus}\,\mathbf{y})  &  =((\mathcal{W}^{\hspace{0.01in}%
-1}\otimes\mathcal{W}^{-1})\circ\underline{\bar{\Delta}})(\mathcal{W}(f)).
\label{KonReaBraCop}%
\end{align}
Braided Hopf algebras admit braided antipodes, denoted $\underline{S}$ and
$\underline{\bar{S}}$. These antipodes induce representations on the
commutative algebras:%
\begin{align}
f(\ominus\,\mathbf{x})  &  =(\mathcal{W}^{\hspace{0.01in}-1}\circ\underline
{S}\hspace{0.01in})(\mathcal{W}(f)),\nonumber\\[0.08in]
f(\bar{\ominus}\,\mathbf{x})  &  =(\mathcal{W}^{\hspace{0.01in}-1}%
\circ\underline{\bar{S}}\hspace{0.01in})(\mathcal{W}(f)). \label{qInvDef}%
\end{align}
We refer to the operations in Eq.~(\ref{qInvDef})\ as $q$%
\textit{-in\-ver\-sions}. For the $q$-de\-formed Euclidean space, one finds
the following operator representation \cite{Wachter:2004ph}:%
\begin{gather}
\hat{U}^{-1}f(\ominus\,\mathbf{x})=\nonumber\\
=\sum_{i=0}^{\infty}(-\hspace{0.01in}q\lambda\lambda_{+})^{i}\,\frac
{(x^{+}x^{-})^{i}}{[[2i]]_{q^{-2}}!!}\,q^{-2\hat{n}_{+}(\hat{n}_{+}%
+\hspace{0.01in}\hat{n}_{3})-2\hat{n}_{-}(\hat{n}_{-}+\hspace{0.01in}\hat
{n}_{3})-\hat{n}_{3}\hat{n}_{3}}\nonumber\\
\times D_{q^{-2},\hspace{0.01in}x^{3}}^{\hspace{0.01in}2i}\,f(-\hspace
{0.01in}q^{2-4i}x^{-},-\hspace{0.01in}q^{1-2i}x^{3},-\hspace{0.01in}%
q^{2-4i}x^{+}). \label{AntUnKonMon3dim}%
\end{gather}

The braided coproducts and braided antipodes satisfy the \textit{Hopf algebra
axioms} (see Ref.~\cite{Majid:1996kd}):%
\begin{align}
m\circ(\underline{S}\otimes\operatorname*{id})\circ\underline{\Delta}  &
=m\circ(\operatorname*{id}\otimes\,\underline{S}\hspace{0.01in})\circ
\underline{\Delta}=\underline{\varepsilon},\nonumber\\
m\circ(\underline{\bar{S}}\otimes\operatorname*{id})\circ\underline
{\bar{\Delta}}  &  =m\circ(\operatorname*{id}\otimes\,\underline{\bar{S}%
}\hspace{0.01in})\circ\underline{\bar{\Delta}}=\underline{\bar{\varepsilon}},
\label{HopfVerAnfN}%
\end{align}
and%
\begin{align}
(\operatorname*{id}\otimes\,\underline{\varepsilon})\circ\underline{\Delta}
&  =\operatorname*{id}=(\underline{\varepsilon}\otimes\operatorname*{id}%
)\circ\underline{\Delta},\nonumber\\
(\operatorname*{id}\otimes\,\underline{\bar{\varepsilon}})\circ\underline
{\bar{\Delta}}  &  =\operatorname*{id}=(\underline{\bar{\varepsilon}}%
\otimes\operatorname*{id})\circ\underline{\bar{\Delta}}. \label{HopfAxi2}%
\end{align}
Here, $m$ denotes multiplication in the braided Hopf algebra, while
$\underline{\varepsilon}$ or $\underline{\bar{\varepsilon}}$ are the
corresponding counits. These counits act as linear maps annihilating the
coordinate generators:%
\begin{equation}
\varepsilon(X^{i})=\underline{\bar{\varepsilon}}(X^{i})=0.
\end{equation}
Accordingly, on a commutative coordinate algebra they are represented by
evaluation at the origin:%
\begin{equation}
\underline{\varepsilon}(\mathcal{W}(f))=\underline{\bar{\varepsilon}%
}(\mathcal{W}(f))=\left.  f(\mathbf{x})\right\vert _{x\hspace{0.01in}%
=\hspace{0.01in}0}=f(0). \label{ReaVerZopNeuEleKomAlg}%
\end{equation}
Within this setting, the Hopf algebra axioms (\ref{HopfVerAnfN}) and
(\ref{HopfAxi2}) naturally translate into the following rules for
$q$-trans\-la\-tions and $q$-in\-ver\-sions \cite{Wachter:2007A}:%
\begin{align}
f((\ominus\,\mathbf{x})\oplus\mathbf{x})  &  =f(\mathbf{x}\oplus
(\ominus\,\mathbf{x}))=f(0),\nonumber\\
f((\bar{\ominus}\,\mathbf{x})\,\bar{\oplus}\,\mathbf{x})  &  =f(\mathbf{x}%
\,\bar{\oplus}\,(\bar{\ominus}\,\mathbf{x}))=f(0), \label{qAddN}%
\end{align}
and%
\begin{align}
f(\mathbf{x}\oplus\mathbf{y})|_{y\hspace{0.01in}=\hspace{0.01in}0}  &
=f(\mathbf{x})=f(\mathbf{y}\oplus\mathbf{x})|_{y\hspace{0.01in}=\hspace
{0.01in}0},\nonumber\\
f(\mathbf{x}\,\bar{\oplus}\,\mathbf{y})|_{y\hspace{0.01in}=\hspace{0.01in}0}
&  =f(\mathbf{x})=f(\mathbf{y}\,\bar{\oplus}\,\mathbf{x})|_{y\hspace
{0.01in}=\hspace{0.01in}0}. \label{qNeuEle}%
\end{align}

\begin{figure}[ptb]
  \centering
  \includegraphics[width=2.5754in]{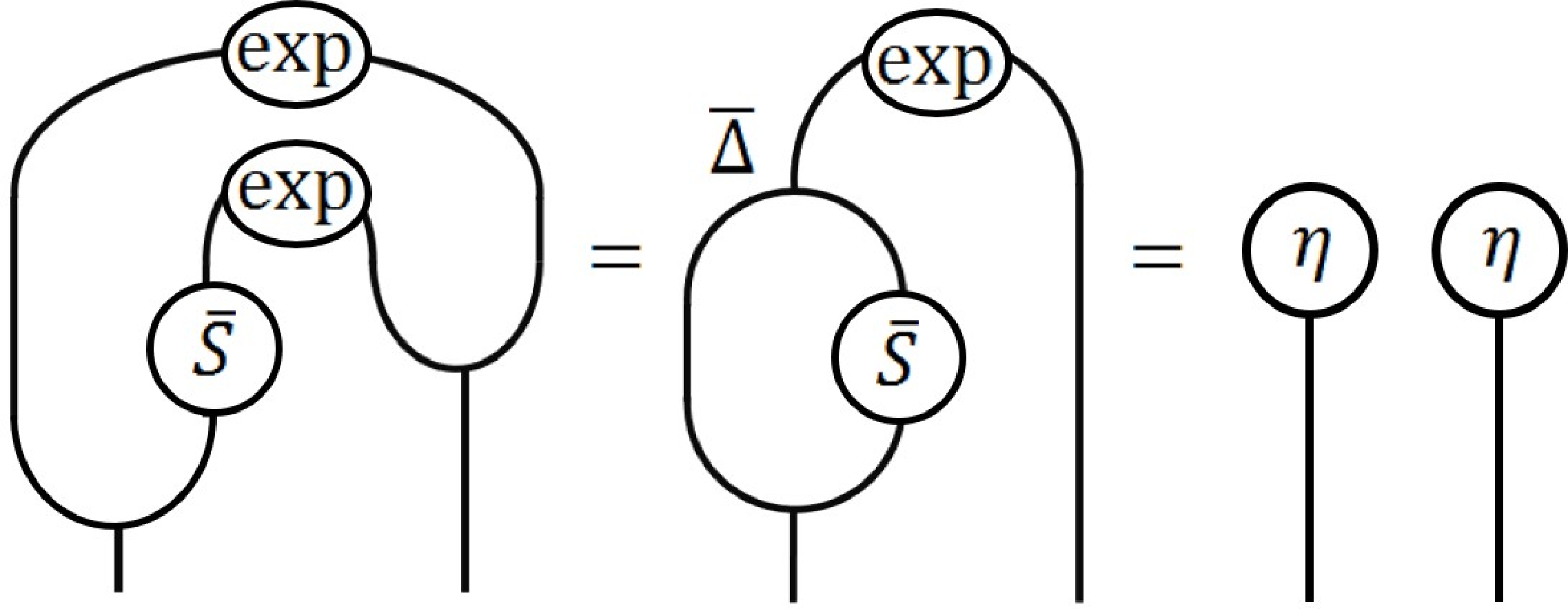}
  \caption{Invertibility of $q$-exponentials.}
  \label{Fig3}
\end{figure}

Based on $q$-in\-ver\-sions, we define \textit{inverse }$q$%
\textit{-ex\-po\-nen\-tials} as%
\begin{equation}
\exp_{q}(\bar{\ominus}\,\mathbf{x}|\text{i}\mathbf{p})=\exp_{q}(\text{i}%
\mathbf{x}|\text{{}}\bar{\ominus}\,\mathbf{p}). \label{InvExpAlgDefKom}%
\end{equation}
By employing the addition theorems, the identities in Eq.~(\ref{qAddN}), and
the normalization conditions of our $q$-ex\-po\-nen\-tials, one obtains:%
\begin{equation}
\exp_{q}(\text{i}\mathbf{x}\circledast\exp_{q}(\bar{\ominus}\,\mathbf{x}%
|\hspace{0.01in}\text{i}\mathbf{p})\circledast\mathbf{p})=\exp_{q}%
(\mathbf{x}\,\bar{\oplus}\,(\bar{\ominus}\,\mathbf{x})|\hspace{0.01in}%
\text{i}\mathbf{p})=\exp_{q}(\mathbf{x}|\text{i}\mathbf{p})|_{x=0}=1.
\end{equation}
Graphical representations of these identities are shown in Fig.~\ref{Fig3}%
.\footnote{Further details of these graphical calculations can be found in
Ref.~\cite{Majid:2002kd}.}%

\begin{figure}[ptb]
  \centering
  \includegraphics[width=1.817in]{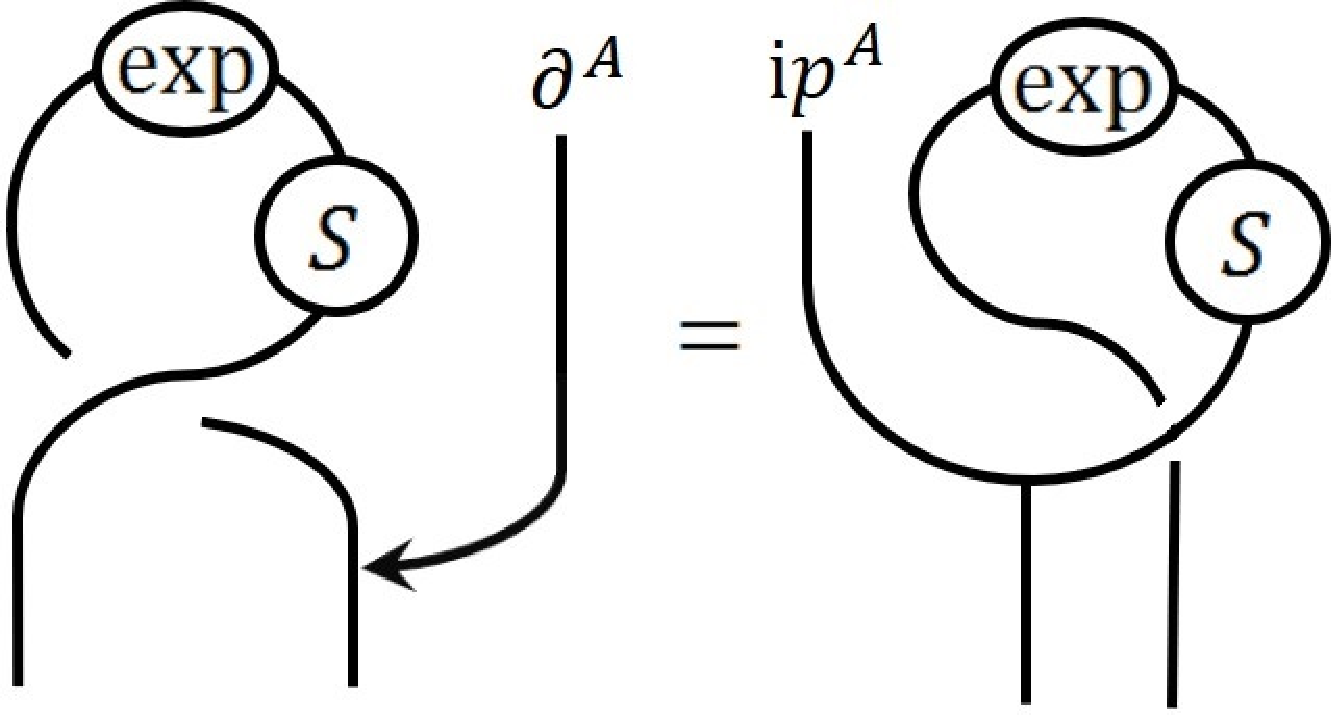}
  \caption{Eigenvalue equation of twisted $q$-exponential.}
  \label{Fig4}
\end{figure}

It is often convenient to exchange the tensor factors in the inverse
$q$-ex\-ponen\-tials of Eq.~(\ref{InvExpAlgDefKom}) using the inverse of the
universal R-ma\-trix (see the graphical representation in Fig.~\ref{Fig4}):%
\begin{align}
\exp_{q}^{\ast}(\text{i}\mathbf{p}|\hspace{0.01in}\mathbf{x})  &  =\tau
\circ\lbrack(\mathcal{R}_{[2]}^{-1}\otimes\mathcal{R}_{[1]}^{-1}%
)\triangleright\exp_{q}(\text{i}\mathbf{x}|\hspace{-0.03in}\ominus
\hspace{-0.01in}\mathbf{p})],\nonumber\\
\exp_{q}^{\ast}(\mathbf{x}|\text{i}\mathbf{p})  &  =\tau\circ\lbrack
(\mathcal{R}_{[2]}^{-1}\otimes\mathcal{R}_{[1]}^{-1})\triangleright\exp
_{q}(\ominus\hspace{0.02in}\mathbf{p}|\hspace{0.01in}\text{i}\mathbf{x})].
\label{DuaExp2}%
\end{align}
Here, $\tau$ denotes the ordinary twist operator. These \textit{twisted }%
$q$\textit{-ex\-ponen\-tials} satisfy the following eigenvalue equations (cf.
Fig.~\ref{Fig4}):%
\begin{align}
\exp_{q}^{\ast}(\text{i}\mathbf{p}|\hspace{0.01in}\mathbf{x})\triangleleft
\partial^{A}  &  =\text{i}p^{A}\circledast\exp_{q}^{\ast}(\text{i}%
\mathbf{p}|\hspace{0.01in}\mathbf{x}),\nonumber\\
\partial^{A}\,\bar{\triangleright}\,\exp_{q}^{\ast}(\mathbf{x}|\text{i}%
^{-1}\mathbf{p})  &  =\exp_{q}^{\ast}(\mathbf{x}|\text{i}^{-1}\mathbf{p}%
)\circledast\text{i}p^{A}. \label{EigGleExpQueAbl}%
\end{align}
%

\section{$q$-Deformed Position and Momentum
Eigenfunctions\label{KapVolRelImpOrt}}

Using $q$-ex\textit{\-}po\textit{\-}nen\textit{\-}tials, we define
$q$\textit{-de\-formed momentum eigenfunctions} \cite{Wachter:2019A}:%
\begin{equation}
u_{\hspace{0.01in}\mathbf{p}}(\mathbf{x})=\operatorname*{vol}\nolimits^{-1/2}%
\exp_{q}(\mathbf{x}|\text{i}\mathbf{p}),\qquad u^{\mathbf{p}}(\mathbf{x}%
)=\operatorname*{vol}\nolimits^{-1/2}\exp_{q}(\text{i}^{-1}\mathbf{p}%
|\hspace{0.01in}\mathbf{x}). \label{ImpEigFktqDefN}%
\end{equation}
The volume element $\operatorname*{vol}$ is specified in Eq.~(\ref{VolEleDef})
of Chap.~\ref{DefQDelFktKap}. These $q$-de\-formed momentum eigenfunctions
satisfy [cf. Eq.~(\ref{EigGl1N}) in App.~\ref{KapExp}]%
\begin{align}
\text{i}^{-1}\partial^{A}\triangleright u_{\hspace{0.01in}\mathbf{p}%
}(\mathbf{x})  &  =u_{\hspace{0.01in}\mathbf{p}}(\mathbf{x})\circledast
p^{A},\nonumber\\
u^{\mathbf{p}}(\mathbf{x})\,\bar{\triangleleft}\,\partial^{A}\hspace
{0.01in}\text{i}^{-1}  &  =p^{A}\circledast u^{\mathbf{p}}(\mathbf{x}).
\label{EigGleImpOpeImpEigFkt0}%
\end{align}
We also introduce the dual\ momentum eigenfunctions [cf. Eq.~(\ref{DuaExp2})
in App.~\ref{KapExp}]:%
\begin{equation}
(u^{\ast})_{\mathbf{p}}(\mathbf{x})=\operatorname*{vol}\nolimits^{-1/2}%
\exp_{q}^{\ast}(\text{i}\mathbf{p}|\hspace{0.01in}\mathbf{x}),\qquad(u^{\ast
})^{\mathbf{p}}(\mathbf{x})=\operatorname*{vol}\nolimits^{-1/2}\exp_{q}^{\ast
}(\mathbf{x}|\text{i}^{-1}\mathbf{p}). \label{DefDuaImpEigFktWdh}%
\end{equation}
Their eigenvalue equations read [cf. Eq.~(\ref{EigGleExpQueAbl}) in
App.~\ref{KapExp}]:%
\begin{align}
(u^{\ast})_{\mathbf{p}}(\mathbf{x})\triangleleft\partial^{A}\hspace
{0.01in}\text{i}^{-1}\hspace{-0.01in}  &  =p^{A}\circledast(u^{\ast
})_{\mathbf{p}}(\mathbf{x}),\nonumber\\
\text{i}^{-1}\partial^{A}\,\bar{\triangleright}\,(u^{\ast})^{\mathbf{p}%
}(\mathbf{x})  &  =(u^{\ast})^{\mathbf{p}}(\mathbf{x})\circledast p^{A}.
\label{ImpEigFktqDef2}%
\end{align}
The $q$-de\-formed momentum eigenfunctions satisfy the \textit{completeness
relations} \cite{Kempf:1994yd,Wachter:2019A}%
\begin{align}
\int\text{d}_{q}^{3}p\,u_{\hspace{0.01in}\mathbf{p}}(\mathbf{x})\circledast
(u^{\ast})_{\mathbf{p}}(\mathbf{y})  &  =\operatorname*{vol}\nolimits^{-1}%
\hspace{-0.01in}\delta_{q}^{3}(\mathbf{x}\oplus(\ominus\hspace{0.01in}%
\kappa^{-1}\mathbf{y})),\nonumber\\
\int\text{d}_{q}^{3}p\,(u^{\ast})^{\mathbf{p}}(\mathbf{y})\circledast
u^{\mathbf{p}}(\mathbf{x})  &  =\operatorname*{vol}\nolimits^{-1}%
\hspace{-0.01in}\delta_{q}^{3}((\ominus\hspace{0.01in}\kappa^{-1}%
\mathbf{y})\oplus\mathbf{x}), \label{VolRelZeiWelDreDim}%
\end{align}
where $\delta_{q}^{\hspace{0.01in}3}(\mathbf{x})$ denotes the $q$-de\-formed
version of the three-di\-men\-sion\-al delta function [cf.
Eq.~(\ref{DreDimDelLDef}) in Chap.~\ref{DefQDelFktKap}]. The corresponding
\textit{orthogonality relations} are:%
\begin{align}
\int\text{d}_{q}^{3}x\,(u^{\ast})_{\mathbf{p}}(\mathbf{x})\circledast
u_{\hspace{0.01in}\mathbf{p}^{\prime}}(\mathbf{x})  &  =\operatorname*{vol}%
\nolimits^{-1}\hspace{-0.01in}\delta_{q}^{\hspace{0.01in}3}((\ominus
\hspace{0.01in}\kappa^{-1}\mathbf{p})\oplus\mathbf{p}^{\prime})\nonumber\\
\int\text{d}_{q}^{3}x\,u^{\mathbf{p}}(\mathbf{x})\circledast(u^{\ast
})^{\mathbf{p}^{\prime}}(\mathbf{x})  &  =\operatorname*{vol}\nolimits^{-1}%
\hspace{-0.01in}\delta_{q}^{\hspace{0.01in}3}(\hspace{0.01in}\mathbf{p}%
\oplus(\ominus\hspace{0.01in}\kappa^{-1}\mathbf{p}^{\prime})).
\label{SkaProEbeDreExpWie0}%
\end{align}

The $q$\textit{-de\-formed position eigenfunctions} are defined in terms of
$q$-de\-formed delta functions \cite{Wachter:2019A}:%
\begin{align}
u_{\mathbf{y}}(\mathbf{x})  &  =\operatorname*{vol}\nolimits^{-1}%
\hspace{-0.01in}\delta_{q}^{3}(\mathbf{x}\oplus(\ominus\hspace{0.01in}%
\kappa^{-1}\mathbf{y}))=\operatorname*{vol}\nolimits^{-1}\hspace
{-0.01in}\delta_{q}^{3}((\ominus\hspace{0.01in}\kappa^{-1}\mathbf{x}%
)\oplus\mathbf{y})=(u^{\ast})^{\mathbf{y}}(\mathbf{x}),\nonumber\\
(u^{\ast})_{\mathbf{y}}(\mathbf{x})  &  =\operatorname*{vol}\nolimits^{-1}%
\hspace{-0.01in}\delta_{q}^{3}(\mathbf{y}\oplus(\ominus\hspace{0.01in}%
\kappa^{-1}\mathbf{x}))=\operatorname*{vol}\nolimits^{-1}\hspace
{-0.01in}\delta_{q}^{3}((\ominus\hspace{0.01in}\kappa^{-1}\mathbf{y}%
)\oplus\mathbf{x})=u^{\mathbf{y}}(\mathbf{x}). \label{DefEigPos1}%
\end{align}
The position operators act by multiplication operators in position space:%
\begin{equation}
X^{A}\triangleright f(x)=x^{A}\circledast f(x),\qquad f(x)\triangleleft
X^{A}=f(x)\circledast x^{A}. \label{DefWirOrtOpe1}%
\end{equation}
Hence, the position eigenfunctions\ satisfy:%
\begin{align}
X^{A}\triangleright u_{\mathbf{y}}(\mathbf{x})  &  =x^{A}\circledast
u_{\mathbf{y}}(\mathbf{x})=u_{\mathbf{y}}(\mathbf{x})\circledast
y^{A},\nonumber\\
(u^{\ast})_{\mathbf{y}}(\mathbf{x})\triangleleft X^{A}  &  =(u^{\ast
})_{\mathbf{y}}(\mathbf{x})\circledast x^{A}=y^{A}\circledast(u^{\ast
})_{\mathbf{y}}(\mathbf{x}). \label{DefGleOrtEigFkt}%
\end{align}
The \textit{orthogonality relations} for $q$-de\-formed position
eigenfunctions read%
\begin{gather}
\int\text{d}_{q}^{3}x\,u^{\mathbf{y}}(\mathbf{x})\circledast(u^{\ast
})^{\mathbf{y}^{\prime}}(\mathbf{x})=\int\text{d}_{q}^{3}x\,(u^{\ast
})_{\mathbf{y}}(\mathbf{x})\circledast u_{\mathbf{y}^{\prime}}(\mathbf{x}%
)\nonumber\\
=\operatorname*{vol}\nolimits^{-1}\hspace{-0.01in}\delta_{q}^{3}%
(\hspace{0.01in}\mathbf{y}\oplus(\ominus\hspace{0.01in}\kappa^{-1}%
\mathbf{y}^{\prime})). \label{OrtNorBedOrtEig2}%
\end{gather}
The corresponding \textit{completeness relations} are%
\begin{gather}
\int\text{d}_{q}^{3}\hspace{0.01in}y\,(u^{\ast})^{\mathbf{y}}(\mathbf{x}%
^{\prime})\circledast u^{\mathbf{y}}(\mathbf{x})=\int\text{d}_{q}^{3}%
\hspace{0.01in}y\,u_{\hspace{0.01in}\mathbf{y}}(\mathbf{x}^{\prime
})\circledast(u^{\ast})_{\mathbf{y}}(\mathbf{x})\nonumber\\
=\operatorname*{vol}\nolimits^{-1}\hspace{-0.01in}\delta_{q}^{3}%
(\mathbf{x}^{\prime}\oplus(\ominus\hspace{0.01in}\kappa^{-1}\mathbf{x})).
\label{VolRelOrtEigFkt}%
\end{gather}

\section{$q$-Deformed Klein-Gordon Equation\label{KapPlaWavSol}}

In Ref.~\cite{Wachter:2021B} we introduced a $q$\textit{-de\-formed analogue
of the Klein-Gor\-don equation} describing a spinless particle of rest mass
$m$ in three-di\-men\-sion\-al $q$-de\-formed Euclidean space:%
\begin{equation}
c^{-2}\partial_{t}^{\hspace{0.01in}2}\triangleright\varphi(\mathbf{x}%
,t)-\hspace{-0.01in}\nabla_{q}^{2}\triangleright\varphi(\mathbf{x}%
,t)+(m\hspace{0.01in}c)^{2}\hspace{0.01in}\varphi(\mathbf{x}%
,t)=0.\label{KleGorGleLin}%
\end{equation}
The $q$-de\-formed Laplacian $\nabla_{q}^{2}$ depends on the metric of
three-di\-men\-sion\-al $q$-de\-formed Euclidean space [cf.
Eq.~(\ref{MetDreiDim}) in App.~\ref{KapQuaZeiEle}]:%
\begin{equation}
\nabla_{q}^{2}=\Delta_{q}=\mathbf{\partial}\circ\mathbf{\partial}=\partial
^{A}\partial_{A}=g^{AB}\partial_{B}\hspace{0.01in}\partial_{A}.\label{qLapOpe}%
\end{equation}

The $q$-de\-formed Klein-Gor\-don equation in Eq.~(\ref{KleGorGleLin}) admits
\textit{plane wave solutions} of the form%
\begin{equation}
\varphi_{\mathbf{p}}(\mathbf{x},t)=\frac{c}{\sqrt{2}}\,u_{\hspace
{0.01in}\mathbf{p}}(\mathbf{x})\circledast\exp(-\text{i\hspace{0.01in}%
}tE_{\mathbf{p}})\circledast E_{\mathbf{p}}^{\hspace{0.01in}-1/2}%
,\label{EbeWelKGF}%
\end{equation}
with the time-de\-pen\-dent phase factor represented by%
\begin{equation}
\exp(-\hspace{0.01in}\text{i\hspace{0.01in}}tE_{\mathbf{p}})=\sum
_{n\hspace{0.01in}=\hspace{0.01in}0}^{\infty}\frac{(-\hspace{0.01in}%
\text{i\hspace{0.01in}}tE_{\mathbf{p}})^{n}}{n!}.
\end{equation}
Substituting Eq.~(\ref{EbeWelKGF}) into Eq.~(\ref{KleGorGleLin}) yields%
\begin{align}
0 &  =c^{-2}\partial_{t}^{\hspace{0.01in}2}\triangleright\varphi_{\mathbf{p}%
}-\nabla_{q}^{2}\triangleright\varphi_{\mathbf{p}}+(m\hspace{0.01in}%
c)^{2}\varphi_{\mathbf{p}}\nonumber\\
&  =\varphi_{\mathbf{p}}\circledast(\hspace{0.01in}p^{B}\hspace{-0.01in}%
\circledast p_{B}-c^{-2}E_{\mathbf{p}}\circledast E_{\mathbf{p}}%
+(m\hspace{0.01in}c)^{2}),\label{qKleGorGleImp1}%
\end{align}
provided that the following $q$\textit{-de\-formed dispersion relation} holds:%
\begin{equation}
c^{-2}E_{\mathbf{p}}\circledast E_{\mathbf{p}}=p^{B}\hspace{-0.01in}%
\circledast p_{B}+(m\hspace{0.01in}c)^{2}.
\end{equation}
Since $m^{2}$ and $\mathbf{p}^{2}=p^{B}\hspace{-0.01in}\circledast p_{B}$
commute, this relation can be solved formally for $E_{\mathbf{p}}$:%
\begin{align}
E_{\mathbf{p}}  & =c\,(\mathbf{p}^{2}+(m\hspace{0.01in}c)^{2})^{1/2}%
\nonumber\\
& =c\sum_{k\hspace{0.01in}=\hspace{0.01in}0}^{\infty}\binom{1/2}%
{k}\,\mathbf{p}^{2k}(m\hspace{0.01in}c)^{1-\hspace{0.01in}2k}%
,\label{EneKleGorSer}%
\end{align}
where the powers of $\mathbf{p}^{2}$ must be replaced by normal-ordered
expressions \cite{Wachter:2020B}:%
\begin{align}
\mathbf{p}^{2k} &  =\hspace{0.01in}\overset{k-\text{times}}{\overbrace
{\mathbf{p}^{2}\circledast\ldots\circledast\mathbf{p}^{2}}}\nonumber\\
&  =\sum_{l\hspace{0.01in}=\hspace{0.01in}0}^{k}\hspace{0.01in}q^{-2l}%
(-q-q^{-1})^{k-l}%
\genfrac{[}{]}{0pt}{}{k}{l}%
_{q^{4}}\,(\hspace{0.01in}p_{-})^{k-l}(\hspace{0.01in}p_{3})^{2l}%
(\hspace{0.01in}p_{+})^{k-l}.\label{EntPotP}%
\end{align}

\section{Derivation of Auxiliary Formulas\label{AnhHerFor}}

In this appendix, we derive several auxiliary formulas needed in
Chap.~\ref{KapVacEneTot}\ for evaluating the expectation value $\langle
H^{2}\rangle_{S_{m}}$ [cf. Eq.~(\ref{ErwHQua}) in Chap.~\ref{KapVacEneTot}].
Inspection of Eqs.~(\ref{MehSum01}), (\ref{MehSum2}), (\ref{MehSum30}) and
(\ref{MehSum3}) in Chap.~\ref{KapVacEneTot}, together with
Eq.~(\ref{WirImpPotDreQDelEin}) from Chap.~\ref{VacEneQEuc}, shows that
expressions of the form below occur repeatedly ($m,n\in\mathbb{Z}$):%
\begin{align}
&  \sum_{\varepsilon,\hspace{0.01in}\varepsilon^{\prime}=\hspace{0.01in}\pm
}\,\sum_{k,\hspace{0.01in}l\hspace{0.01in}=\hspace{0.01in}-\infty}^{m}\left.
\left\vert (1-q)\hspace{0.01in}x\right\vert \,\phi_{q}(q^{n}x,y)\right\vert
_{_{x\hspace{0.01in}=\hspace{0.01in}\varepsilon\hspace{0.01in}\alpha
\hspace{0.01in}q^{k},\,y\hspace{0.01in}=\hspace{0.01in}\varepsilon^{\prime
}\alpha\hspace{0.01in}q^{l}}}=\nonumber\\
&  \qquad=\sum_{\varepsilon,\hspace{0.01in}\varepsilon^{\prime}=\hspace
{0.01in}\pm}\,\sum_{k,\hspace{0.01in}l\hspace{0.01in}=\hspace{0.01in}-\infty
}^{m}\left\vert (1-q)\hspace{0.01in}\varepsilon\hspace{0.01in}\alpha
\hspace{0.01in}q^{k}\right\vert \,\phi_{q}(\varepsilon\hspace{0.01in}%
\alpha\hspace{0.01in}q^{k+n},\varepsilon^{\hspace{0.01in}\prime}\alpha
\hspace{0.01in}q^{l})\nonumber\\
&  \qquad=\sum_{\varepsilon,\hspace{0.01in}\varepsilon^{\prime}=\hspace
{0.01in}\pm}\,\sum_{k,\hspace{0.01in}l\hspace{0.01in}=\hspace{0.01in}-\infty
}^{m}(q-1)\hspace{0.01in}\alpha\hspace{0.01in}q^{k}\hspace{0.01in}%
\delta_{\varepsilon,\varepsilon^{\prime}}\hspace{0.01in}\delta_{k+n,l}%
\nonumber\\
&  \qquad=2(q-1)\hspace{0.01in}\alpha\sum_{k,\hspace{0.01in}l\hspace
{0.01in}=\hspace{0.01in}-\infty}^{m}q^{k}\hspace{0.01in}\delta_{k+n,l}%
.\label{HerForErg1}%
\end{align}
In this calculation, we first substituted the $q$-lat\-tice values for $x$ and
$y$, and then used the properties of the function $\phi_{q}$ [cf.
Eqs.~(\ref{ChaIdePhi1}) and (\ref{ChaIdePhi2}) in Chap.~\ref{KapQDisEin}]. To
ensure convergence of the infinite series, we assume $q>1$:%
\begin{equation}
\sum_{k,\hspace{0.01in}l\hspace{0.01in}=-\infty}^{m}q^{k}\hspace{0.01in}%
\delta_{k+n,l}=\left\{
\begin{array}
[c]{c}%
\sum\limits_{k\hspace{0.01in}=\hspace{0.01in}-\infty}^{m-n}q^{k}\quad
\text{if}\quad n>0,\\
\sum\limits_{k\hspace{0.01in}=\hspace{0.01in}-\infty}^{m}q^{k}\quad
\text{if}\quad n\leq0
\end{array}
\right.  =\frac{q^{m-\Theta(n)\cdot n+1}}{q-1},\label{HerForErg2}%
\end{equation}
where $\Theta(x)$ denotes the Heaviside step function. 
Eq.~(\ref{HerForErg2}) follows directly from the formula for a geometric
series:%
\begin{equation}
\sum_{k\hspace{0.01in}=\hspace{0.01in}-\infty}^{m-\Theta(n)\cdot n}%
q^{k}=q^{m-\Theta(n)\cdot n}\sum_{k\hspace{0.01in}=\hspace{0.01in}-\infty}%
^{0}q^{k}=q^{m-\Theta(n)\cdot n}\sum_{k\hspace{0.01in}=\hspace{0.01in}%
0}^{\infty}q^{-k}=\frac{q^{m-\Theta(n)\cdot n+1}}{q-1}.
\end{equation}
Substituting Eq.~(\ref{HerForErg2}) into Eq.~(\ref{HerForErg1}), we obtain%
\begin{equation}
\sum_{\varepsilon,\hspace{0.01in}\varepsilon^{\prime}=\hspace{0.01in}\pm
}\,\sum_{k,\hspace{0.01in}l\hspace{0.01in}=\hspace{0.01in}-\infty}^{m}\left.
\left\vert (1-q)\hspace{0.01in}x\right\vert \,\phi_{q}(q^{n}x,y)\right\vert
_{_{\substack{x\hspace{0.01in}=\hspace{0.01in}\varepsilon\hspace{0.01in}%
\alpha\hspace{0.01in}q^{k}\\y\hspace{0.01in}=\hspace{0.01in}\varepsilon
^{\prime}\alpha\hspace{0.01in}q^{l}}}}=2\alpha\hspace{0.01in}q^{m-\Theta
(n)\cdot n+1}.\label{MehSumPhi}%
\end{equation}

In the evaluation of $\langle H^{2}\rangle_{S_{m}}$ one also encounters sums
of the type below [cf. Eq.~(\ref{MehSum1}) in Chap.~\ref{KapVacEneTot}
together with Eq.~(\ref{WirImpPotDreQDelEin}) in Chap.~\ref{VacEneQEuc}]:%
\begin{align}
&  \sum_{\varepsilon,\hspace{0.01in}\varepsilon^{\prime}=\hspace{0.01in}\pm
}\,\sum_{k,\hspace{0.01in}l\hspace{0.01in}=\hspace{0.01in}-\infty}^{m}\left.
\left\vert (1-q)\hspace{0.01in}x\right\vert \,D_{q^{2},x}\hspace{0.01in}%
\phi_{q}(q^{n}x,y)\right\vert _{_{x\hspace{0.01in}=\hspace{0.01in}%
\varepsilon\hspace{0.01in}\alpha\hspace{0.01in}q^{k},\,y\hspace{0.01in}%
=\hspace{0.01in}\varepsilon^{\prime}\alpha\hspace{0.01in}q^{l}}}=\nonumber\\
&  \qquad=\sum_{\varepsilon,\hspace{0.01in}\varepsilon^{\prime}=\pm}%
\,\sum_{k,\hspace{0.01in}l\hspace{0.01in}=\hspace{0.01in}-\infty}^{m}\left.
\left\vert (1-q)\hspace{0.01in}x\right\vert \,\frac{\phi_{q}(q^{n+2}%
x,y)-\phi_{q}(q^{n}x,y)}{(q^{2}-1)\hspace{0.01in}x}\right\vert
_{\substack{x\hspace{0.01in}=\hspace{0.01in}\varepsilon\hspace{0.01in}%
\alpha\hspace{0.01in}q^{k}\\y\hspace{0.01in}=\hspace{0.01in}\varepsilon
^{\prime}\alpha\hspace{0.01in}q^{l}}}\nonumber\\
&  \qquad=\sum_{\varepsilon,\hspace{0.01in}\varepsilon^{\prime}=\hspace
{0.01in}\pm}\,\sum_{k,\hspace{0.01in}l\hspace{0.01in}=\hspace{0.01in}-\infty
}^{m}\left\vert (1-q)\hspace{0.01in}\varepsilon\hspace{0.01in}\alpha
\hspace{0.01in}q^{k}\right\vert \,\frac{\phi_{q}(\varepsilon\hspace
{0.01in}\alpha\hspace{0.01in}q^{k+n+2},\varepsilon^{\hspace{0.01in}\prime
}\alpha\hspace{0.01in}q^{l})-\phi_{q}(\varepsilon\hspace{0.01in}\alpha
\hspace{0.01in}q^{k+n},\varepsilon^{\hspace{0.01in}\prime}\alpha
\hspace{0.01in}q^{l})}{\varepsilon\hspace{0.01in}\alpha\hspace{0.01in}%
q^{k}(q^{2}-1)}\nonumber\\
&  \qquad=\sum_{\varepsilon,\hspace{0.01in}\varepsilon^{\prime}=\hspace
{0.01in}\pm}\,\sum_{k,\hspace{0.01in}l\hspace{0.01in}=\hspace{0.01in}-\infty
}^{m}\frac{\delta_{\varepsilon,\varepsilon^{\prime}}(\delta_{k+n+2,l}%
-\delta_{k+n,l})}{\varepsilon\hspace{0.01in}(q+1)}=0.
\label{MehSumAblPhi}%
\end{align}
In the above derivation, we expanded the Jackson derivative [cf.
Eq.~(\ref{DefJacAbl}) in Chap.~\ref{KapqAnaExpTriFkt}], substituted the
$q$-lat\-tice values for both coordinates, and applied the properties of the
function $\phi_{q}$\ [cf. Eqs.~(\ref{ChaIdePhi1}) and (\ref{ChaIdePhi2}) in
Chap.~\ref{KapQDisEin}]. The final equality reflects the cancellation between
the contributions for $\varepsilon=\pm$.

Finally, we consider [cf. Eq.~(\ref{MehSum2}) in Chap.~\ref{KapVacEneTot} and
Eq.~(\ref{WirImpPotDreQDelEin}) in Chap.~\ref{VacEneQEuc}]:%
\begin{align}
&  \sum_{\varepsilon,\varepsilon^{\prime}=\hspace{0.01in}\pm}\,\sum
_{k,\hspace{0.01in}l\hspace{0.01in}=\hspace{0.01in}-\infty}^{m}\left.
\left\vert (1-q)\hspace{0.01in}x\right\vert \,D_{q^{2},x}^{2}\hspace
{0.01in}\phi_{q}(q^{-2}x,y)\right\vert _{_{x\hspace{0.01in}=\hspace
{0.01in}\varepsilon\hspace{0.01in}\alpha\hspace{0.01in}q^{k},\,y\hspace
{0.01in}=\hspace{0.01in}\varepsilon^{\prime}\alpha\hspace{0.01in}q^{l}}%
}=\nonumber\\
&  \qquad=\sum_{\varepsilon,\hspace{0.01in}\varepsilon^{\prime}=\hspace
{0.01in}\pm}\,\sum_{k,\hspace{0.01in}l\hspace{0.01in}=-\infty}^{m}%
\frac{(q-1)\hspace{0.01in}\alpha\hspace{0.01in}q^{k}}{[(1-q^{2})(\varepsilon
\hspace{0.01in}\alpha\hspace{0.01in}q^{k})]^{2}}\,\Big [\phi_{q}%
(\varepsilon\hspace{0.01in}\alpha\hspace{0.01in}q^{k-2},\varepsilon
^{\hspace{0.01in}\prime}\alpha\hspace{0.01in}q^{l})\nonumber\\
&  \qquad\qquad\qquad\qquad-(1+q^{-2})\hspace{0.01in}\phi_{q}(\varepsilon
\hspace{0.01in}\alpha\hspace{0.01in}q^{k},\varepsilon^{\hspace{0.01in}\prime
}\alpha\hspace{0.01in}q^{l})+q^{-2}\phi_{q}(\varepsilon\hspace{0.01in}%
\alpha\hspace{0.01in}q^{k\hspace{0.01in}+2},\varepsilon^{\hspace{0.01in}%
\prime}\alpha\hspace{0.01in}q^{l})\Big ]\nonumber\\
&  \qquad=\frac{1}{(q^{2}-1)(q+1)\hspace{0.01in}\alpha}\sum_{\varepsilon
,\hspace{0.01in}\varepsilon^{\prime}=\hspace{0.01in}\pm}\delta_{\varepsilon
,\hspace{0.01in}\varepsilon^{\prime}}\Big [\sum_{k,\hspace{0.01in}%
l\hspace{0.01in}=\hspace{0.01in}-\infty}^{m}q^{-k}\,(\delta_{k-2,l}%
-\delta_{k,l})\nonumber\\
&  \qquad\qquad\qquad\qquad+q^{-2}\sum_{k,\hspace{0.01in}l\hspace
{0.01in}=\hspace{0.01in}-\infty}^{m}q^{-k}\,(\delta_{k\hspace{0.01in}%
+2,l}-\delta_{k,l})\Big ]\nonumber\\
&  \qquad=\frac{2}{(q^{2}-1)(q+1)\hspace{0.01in}\alpha}(-q^{-2})(q^{-m}%
+q^{-(m-1)})=-\frac{2\hspace{0.01in}q^{-2}}{(q^{2}-1)\hspace{0.01in}%
\alpha\hspace{0.01in}q^{m}}.\label{MehSumAbl2Phi}%
\end{align}
In this derivation, we expanded the second Jackson derivative [cf.
Eq.~(\ref{DefJacAbl}) in Chap.~\ref{KapqAnaExpTriFkt}], substituted the
$q$-lat\-tice values for the coordinates, and applied the properties of the
function $\phi_{q}$\ [cf. Eqs.~(\ref{ChaIdePhi1}) and (\ref{ChaIdePhi2}) in
Chap.~\ref{KapQDisEin}]. Here, the sums over $\varepsilon$ and $\varepsilon
^{\hspace{0.01in}\prime}$ yield a factor of two, while the double sums over
$k$ and $l$ reduce to telescoping series: the first vanishes, and the second
collapses to the two terms $-q^{-m}$ and $-q^{-(m-1)}$.

{\normalsize
\bibliographystyle{unsrt}
\bibliography{book,habil}
}

\end{document}